%% file: main.tex
\documentclass{article}
\usepackage{microtype}
\usepackage{graphicx}
\usepackage{booktabs}
\usepackage{hyperref}

\usepackage[preprint]{paperstyle}
\usepackage{amsmath,amssymb,amsthm}
\usepackage{multirow,wrapfig,tikz,longtable,placeins,needspace}
\usepackage{enumitem}
\usetikzlibrary{arrows.meta,positioning,fit,backgrounds}

\definecolor{ccblue}{HTML}{2A78D6}
\definecolor{ccorange}{HTML}{EB6834}
\definecolor{ccgray}{HTML}{8F8E88}
\definecolor{ccred}{HTML}{D62F2F}
\definecolor{ccviolet}{HTML}{6250D6}
\definecolor{ccink}{HTML}{1F1F1F}

\newtheorem{theorem}{Theorem}
\newtheorem{lemma}{Lemma}
\newtheorem{proposition}{Proposition}
\newtheorem{corollary}{Corollary}
\theoremstyle{definition}
\newtheorem{definition}{Definition}
\newtheorem{example}{Example}

\papertitlerunning{Towards Certified Model-Claim Verification for Coding-Agent Services}

\begin{document}

\twocolumn[
  \papertitle{CertClaim: Towards Certified Model-Claim Verification\texorpdfstring{\\}{ }for Coding-Agent Services}

  \begin{paperauthorlist}
    \paperauthor{Jiacheng Lu}{sjtu,stepfun}
    \paperauthor{Qixuan Yang}{sjtu,stepfun}
    \paperauthor{Tao Song}{sjtu}
    \paperauthor{Daxin Jiang}{stepfun}
    \paperauthor{Jing Li}{stepfun}
    \paperauthor{Zuo Bai}{stepfun}
    \paperauthor{Haibing Guan}{sjtu}
  \end{paperauthorlist}

  \paperaffiliation{sjtu}{Shanghai Jiao Tong University, Shanghai, China}
  \paperaffiliation{stepfun}{StepFun, Shanghai, China}

  \papercorrespondingauthor{Tao Song}{}
  \papercorrespondingauthor{Zuo Bai}{}

  \paperkeywords{Model Auditing, Model Substitution, Coding Agents, Certified Verification, Hypothesis Testing}

  \vskip 0.3in
]

\printAffiliationsAndNotice{}

\begin{abstract}
Coding-agent services are priced by the model they claim, making the verification of such model claims crucial for their users. In this paper, we reveal that existing model auditing methods rely on a latent assumption that the deployment is static, i.e., the claimed model is always served in one reference configuration. This assumption does not hold for coding-agent services, whose providers may legitimately change the sampling preset, the reasoning mode or the system prompt without changing the model. As a result, existing methods either falsely accuse honest providers or miss cheap derivatives of the claimed model. To address this problem, we propose a certified model-claim verification method (dubbed CertClaim). Instead of testing the service against one reference deployment, we make the authorized deployments part of the claim and test whether some authorized deployment explains all the evidence of the service, by calibrating multiple evidence channels in every authorized deployment and fusing them within each deployment. We provide theoretical analyses of its false-accusation rate and detection condition. Extensive experiments on eight claims, two agent harnesses and three task suites verify the effectiveness of CertClaim, which is valid in every setting while separating substitutes far better than existing methods, as well as its resistance to potential adaptive attacks.
\end{abstract}

\input{sections/intro}
\input{sections/related}
\input{sections/revisit}
\input{sections/method}
\input{sections/theory}
\input{sections/experiments}
\input{sections/conclusion}

\bibliography{refs}
\bibliographystyle{paperstyle}

\clearpage
\appendix
\onecolumn
\input{sections/appendix}

\end{document}

%% file: sections/intro.tex
\section{Introduction}
\label{sec:intro}

Recently, large language models (LLMs) have been widely deployed as coding agents (e.g., Claude Code, Codex and OpenCode) that read, edit and execute code on behalf of developers~\citep{yang2024sweagent,wu2026responsible}. The capability of such an agent is mostly determined by the model behind it, so coding-agent services (e.g., first-party subscriptions, model gateways and resellers) are usually priced by the model they claim.

\begin{figure*}[t]
\centering
\includegraphics[width=\textwidth]{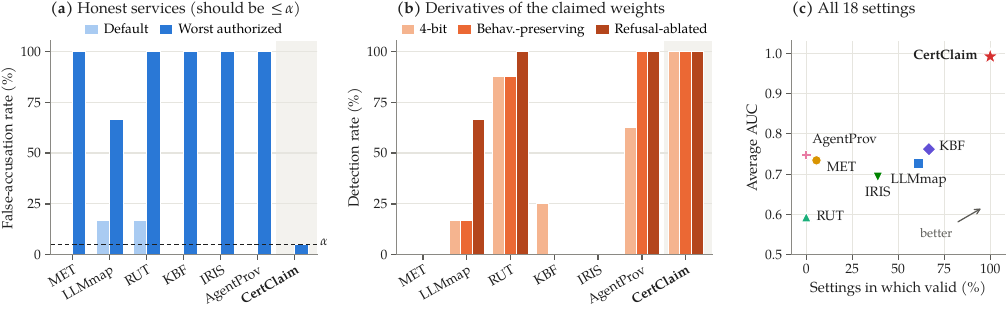}
\vspace{-1.5em}
\caption{Existing model auditing methods test a coding-agent service against one static reference deployment of the claimed model. (a) The false-accusation rate of honest services that claim Qwen3.5-27B~\citep{qwen35_27b}, in the default deployment and in their worst authorized deployment (e.g., with reasoning switched off). (b) The detection rate of three derivatives of Qwen3.5-27B, i.e., its official 4-bit checkpoint, a behavior-preserving edit and a refusal-ablated model. (c) The proportion of the 18 settings of our experiments in which each method is valid at $\alpha=5\%$, against its average AUC. Existing methods either falsely accuse honest providers or miss derivatives of the claimed model, whereas CertClaim is valid in every setting and detects all three derivatives.}
\label{fig:landscape}
\end{figure*}

However, users can hardly verify whether the claimed model is actually behind a service. A provider has a direct economic incentive to serve a cheaper substitute (e.g., a quantized, distilled or fine-tuned derivative or a smaller sibling of the claimed model)~\citep{cai2025substitution}, and silent model substitutions have already been observed on commercial LLM API gateways~\citep{gatescope2026}. In this paper, we study model-claim verification for coding-agent services, i.e., deciding whether a service is serving its claimed model when the user can only interact with it through the agent (i.e., under the black-box setting). Classical model protection, which requires either the cooperation of the provider (e.g., trusted execution environments~\citep{xie2026tee} and verifiable inference~\citep{toploc2025,liu2026sae}) or access to the model parameters (e.g., white-box fingerprinting~\citep{shao2025sok}), cannot be adopted by users in this task. To the best of our knowledge, black-box model auditing is currently the only feasible approach.

In general, existing model auditing methods can be divided into three categories based on the evidence they compare, i.e., equality tests of the output distribution (e.g., MET~\citep{gao2025met}), fingerprints that identify the model (e.g., LLMmap~\citep{pasquini2025llmmap}), and behavioral probes of agentic tool use (e.g., AgentProv~\citep{agentprov2026}). They share a two-stage pipeline: the auditor first collects the responses of the claimed model to a set of queries (i.e., probes) in a reference configuration, and then examines whether the responses of the suspicious service follow the reference distribution. We argue that such a verifier is practical if and only if it is both valid and complete. Specifically, validity requires that an honest provider is falsely accused with a probability of at most a prescribed level in any deployment it is authorized to use and under any audit budget, while completeness requires that cheap substitutes, including derivatives of the claimed weights, are detected.

In this paper, we reveal that existing model auditing methods rely on a latent assumption that the deployment is static, i.e., the claimed model is always served in one reference configuration. This assumption is reasonable for the model APIs that most of them were designed for, where the client sets the sampling parameters and the prompt. However, it does not hold for coding-agent services, where the provider controls the deployment and may legitimately change it without changing the model (e.g., switching the reasoning mode, adding system prompts or serving an official low-precision checkpoint). As a result, existing methods either falsely accuse honest providers that change an authorized setting, which a larger audit budget makes worse, or miss derivatives of the claimed model (see Figure~\ref{fig:landscape}), since an authorized change may move the behavior of an agent more than a derivative does. Simply calibrating existing tests over the authorized deployments restores their validity, but no single evidence channel detects every cheap substitute on every claim. Accordingly, an intriguing and critical question arises: \emph{Could we design valid and complete model-claim verification for coding-agent services?}

The answer to the aforementioned question is in the positive, though achieving it is non-trivial. First, authorized deployments are discrete and semantic changes of the serving configuration that may move the behavior of an agent more than a derivative does, so no tolerance around one reference deployment can separate them (dubbed `Challenge 1'). Second, since no single channel suffices on every claim, multiple channels have to be combined, whereas a provider may let the channels that are calibrated separately be explained by different authorized deployments (dubbed `Challenge 2').

Motivated by this understanding, we propose CertClaim, a certified model-claim verification method for coding-agent services. Instead of testing the service against one reference deployment, we make the authorized deployments part of the claim (dubbed `deployment contract') and test whether some authorized deployment explains all the evidence of the service. In general, CertClaim consists of three main steps. First, we run a few batches of coding tasks through each authorized deployment (dubbed `cell') and compare an audit with every cell via stratified permutation tests, so that the false-accusation rate is controlled within every cell however far apart the cells are (i.e., Challenge 1). Second, we audit the service through an agent harness controlled by the auditor and score it on multiple evidence channels, including in-harness channels and out-of-session probes (i.e., Challenge 2). Third, we fuse all channels within each cell before taking the maximum over cells (dubbed `cell-coherent fusion'), so that the claim survives only if a single authorized deployment explains all the evidence (i.e., Challenge 2). Our theoretical analyses show that CertClaim controls the false-accusation rate under every authorized deployment and every audit budget, and characterize when a substitute is detected.

We evaluate CertClaim on eight claims, two agent harnesses and three task suites, involving 44 models. In the metrics of a recent model-identity benchmark~\citep{shao2025sok}, CertClaim is valid in every setting with an average AUC above 0.99, whereas every existing method is invalid under at least one type of authorized deployment change and stays below 0.77 on average. Besides, CertClaim resists potential adaptive attacks, such as serving a substitute in an authorized configuration and routing part of the sessions to a substitute.

In summary, our main contributions are four-fold: (1) We study model-claim verification for coding-agent services and reveal that existing model auditing methods rely on a latent assumption that the deployment is static. (2) Based on our analyses, we design a certified model-claim verification method (i.e., CertClaim), which makes the authorized deployments part of the claim and fuses multiple evidence channels within each authorized deployment. (3) We provide theoretical analyses of the false-accusation rate and the detection condition of CertClaim. (4) We conduct extensive experiments on open-weight and frontier claims, agent harnesses and task suites, which verify the effectiveness of CertClaim as well as its resistance to potential adaptive attacks.

%% file: sections/related.tex
\section{Related Work}
\label{sec:related}

\subsection{Model Auditing for LLM Services}
\label{sec:related-auditing}

Model auditing verifies which model is behind a black-box LLM service from its responses to auditor-specified queries (i.e., probes)~\citep{cai2025substitution}. Existing methods can be divided into three categories based on the evidence they compare, as follows.

\vspace{0.3em}
\noindent\textbf{Equality Tests of Output Distributions.} MET~\citep{gao2025met} compares completions of Wikipedia prompts via a maximum mean discrepancy test, RUT~\citep{rut2026} examines the ranks of the responses of a service among samples drawn from the claimed weights, and IRIS~\citep{iris2026} compares the outputs of random-generation challenges with a reference window. However, they compare the service with the claimed model in one reference configuration, so a deployment change of an honest provider looks like a substitution to them.

\vspace{0.3em}
\noindent\textbf{Fingerprints that Identify the Model.} LLMmap~\citep{pasquini2025llmmap} classifies the responses to a few crafted queries with a pretrained open-set model, and KBF~\citep{kbf2026} uses the repeatable wrong answers of a model near its knowledge boundary. Recent studies also fingerprint agents from their trajectories and actions~\citep{procgrep2026,white2026forensics,lugoloobi2026actions,gupta2026dissecting}, including the concurrent LIDAR~\citep{lidar2026} for coding agents. However, every model is enrolled in one configuration, so an authorized variant of the claimed model may be nearer to another model, while a derivative of the claimed model is nearest to the claim itself.

\vspace{0.3em}
\noindent\textbf{Behavioral Probes of Agentic Tool Use.} AgentProv~\citep{agentprov2026} audits agentic APIs through their choices among functionally equivalent tools. However, its probes are still compared with the claimed model in one reference configuration, and a provider that changes its system prompt or sampling preset may change these choices as well.

\subsection{Classical Model Protection and Certified Verification}
\label{sec:related-protection}

Classical model protection relies on either the cooperation of the provider, e.g., trusted execution environments that attest the serving stack~\citep{xie2026tee} and verifiable inference that commits to the computation of every request~\citep{toploc2025,liu2026sae}, or access to the model, e.g., white-box fingerprints of weights and internal representations~\citep{shao2025sok}. However, users of a coding-agent service have neither of them. Besides, conformal prediction~\citep{vovk2005alrw,bates2023outliers} and exact randomization tests~\citep{hemerik2018random} turn exchangeable observations into valid $p$-values, and certified verification exploits such calibration to obtain verdicts that hold for every perturbation within a region, e.g., randomized smoothing~\citep{cohen2019smoothing} and certified dataset watermarks~\citep{qiao2025certdw}. However, the regions they consider are norm balls of the inputs, whereas the deployment changes of a coding-agent service are discrete and semantic changes of the serving configuration.

%% file: sections/revisit.tex
\section{Revisiting Model Auditing for Coding-agent Services}
\label{sec:revisit}

In this section, we revisit existing model auditing methods in the context of coding-agent services.

\vspace{0.3em}
\noindent\textbf{The Main Pipeline of Existing Model Auditing Methods.} Let $M_c$ denote the claimed model, $g_0$ its reference configuration (e.g., the default setting of the model card), and $P_{M,g}$ the distribution of the responses of model $M$ served in configuration $g$. In the first stage (i.e., reference collection), the auditor sends a set of probes to $M_c$ served in $g_0$ and collects its responses. In the second stage (i.e., claim verification), the auditor sends the same probes to the suspicious service, tests the simple null hypothesis $H_0:P_S=P_{M_c,g_0}$, and regards the service as not serving $M_c$ if $p\le\alpha$. In other words, these methods rely on the latent assumption that the deployment is static. We examine whether authorized deployment changes break them, and whether the two most straightforward remedies fix this problem.

\subsection{Impact of Authorized Deployment Changes}
\label{sec:revisit-drift}

\begin{figure*}[t]
\centering
\includegraphics[width=\textwidth]{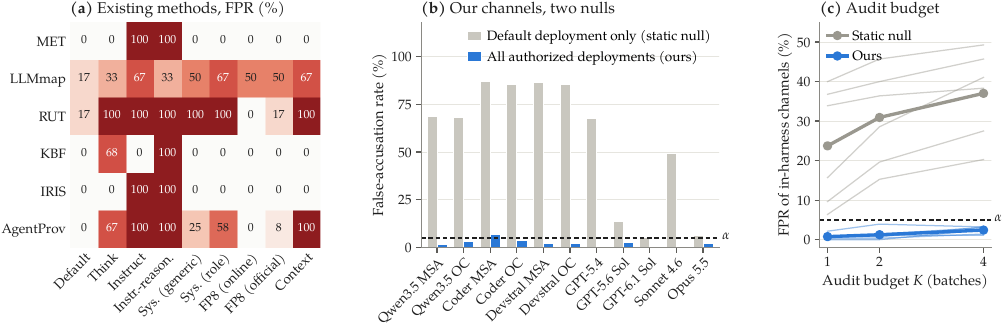}
\vspace{-1.6em}
\caption{False accusations under authorized deployment changes. (a) The false-positive rate (\%) of existing methods on honest services that claim Qwen3.5-27B in each authorized configuration, with their published protocols and budgets. (b) The false-positive rate of the same evidence channels of CertClaim (one batch of tasks) tested against the default deployment only (i.e., the static null) and against all authorized deployments, on every setting of the authored suite. (c) The false-positive rate of our in-harness channels as the audit budget grows, on the six open-weight settings (thin lines) and on average (thick lines). The dashed line marks $\alpha=5\%$.}
\label{fig:static}
\end{figure*}

\noindent\textbf{Settings.} We audit a coding-agent service that claims Qwen3.5-27B~\citep{qwen35_27b} with six representative methods, i.e., MET~\citep{gao2025met}, LLMmap~\citep{pasquini2025llmmap}, RUT~\citep{rut2026}, KBF~\citep{kbf2026}, IRIS~\citep{iris2026} and AgentProv~\citep{agentprov2026}, each with its published protocol and budget at $\alpha=0.05$. The honest service serves Qwen3.5-27B in nine configurations from the model card and common serving practice (e.g., its thinking and non-thinking presets, system prompts, FP8 versions and an injected project memory). To isolate the role of the null hypothesis, we also test our own evidence channels (Section~\ref{sec:method}) on the same audits, against the default configuration only (i.e., the static null) and against all nine configurations.

\vspace{0.3em}
\noindent\textbf{Results.} As shown in Figure~\ref{fig:static}(a), every existing method falsely accuses honest providers in some authorized configuration. In particular, MET, RUT and IRIS reject every audit of the services that switch reasoning off, and the methods designed to be robust are robust only to the changes they anticipated (e.g., KBF accuses the thinking preset in 67.5\% of the audits). Moreover, this failure does not stem from the evidence of these methods. As shown in Figure~\ref{fig:static}(b), our own channels falsely accuse 68.6\% of the honest audits of Qwen3.5-27B under the static null (5.6\%--86.8\% on the other seven claims), compared with 1.4\% (0\%--7.1\%) under all authorized configurations. Besides, as shown in Figure~\ref{fig:static}(c), the false-positive rate of our in-harness channels under the static null grows with the audit budget in every open-weight setting (from 23.7\% to 37.0\% on average as the budget grows from one to four batches of tasks), whereas it stays at most 3.9\% under all authorized configurations. These results verify that the false accusations stem from the static null hypothesis itself, and that a larger audit budget makes them worse (Proposition~\ref{prop:paradox}).

\subsection{Can We Simply Widen the Tolerance or Calibrate Existing Tests over the Contract?}
\label{sec:revisit-remedies}

\noindent\textbf{Settings.} We examine two remedies. For the first one (dubbed `Tol-T'), we accept a service whenever the standardized separation of its in-harness evidence from the default deployment stays below a tolerance $\tau$. We choose $\tau$ as an oracle from the honest audits (i.e., the smallest $\tau$ under which every authorized configuration passes at level $\alpha$), which favors this remedy. For the second one (dubbed `-C'), we test the evidence of every existing method (and each of our channels) against honest audits of every authorized configuration, and reject the claim only if every configuration is rejected.

\begin{figure*}[t]
\centering
\includegraphics[width=\textwidth]{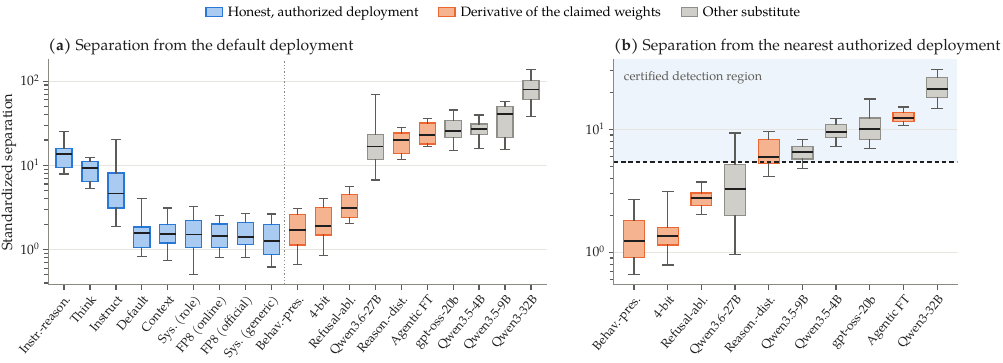}
\vspace{-1.6em}
\caption{Authorized deployment changes move the behavior of a coding agent more than derivatives of the claimed weights do. Each box is the standardized separation of the in-harness evidence of 40 audits of a service that claims Qwen3.5-27B (mini-swe-agent, one batch of tasks; whiskers: 5th to 95th percentiles), (a) from the default deployment and (b) from the nearest authorized deployment. The dashed line is the threshold of our detection condition under Gaussian scores (Example~\ref{ex:gaussian}).}
\label{fig:gap}
\end{figure*}

\input{tables/calibrated}

\vspace{0.3em}
\noindent\textbf{Results.} As shown in Figure~\ref{fig:gap}, honest services in three authorized deployments are farther from the default deployment than three derivatives of the claimed weights (with median separations of 4.6--13.5 vs.\ 1.7--3.1). In other words, authorized deployment changes move the behavior of the agent more than these derivatives do. Accordingly, the oracle tolerance detects none of the three derivatives and only 26.3\% of the audits of all substitutes on average. This is probably because authorized changes act on the sampling and the prompts that shape the whole trajectory, whereas a derivative perturbs the weights only slightly. Moreover, as shown in Figure~\ref{fig:gap}(b), these three derivatives stay below the threshold of our detection condition even against their nearest authorized deployment, i.e., the behavior of the agent alone hardly separates them from the claimed model. On the other hand, as shown in Table~\ref{tab:calibrated}, calibrating existing tests over the contract restores their validity on most claims, but every single channel misses some substitute on some claim. For example, MET-C separates every substitute of the five frontier claims but separates the refusal-ablated derivative of Qwen3.5-27B worse than chance, while AgentProv-C separates every substitute of Qwen3.5-27B but is not valid on three frontier claims. Which channel works depends on the claim, which the auditor cannot know before the audit.

Taken together, these results reveal that the failures of existing model auditing methods share one root cause: they rely on a latent assumption that the deployment is static, whereas the provider of a coding-agent service controls its deployment and may change it within what it is authorized to do. Accordingly, valid and complete verification can be achieved neither by widening the tolerance around one reference deployment nor by calibrating a single existing test over the contract.

%% file: tables/calibrated.tex
\begin{table*}[t]
\centering
\caption{Simply calibrating existing tests over the authorized deployments (dubbed `-C') restores their validity but not their completeness. Each cell is the AUC against the hardest substitute of the claim (worst case over its substitutes; mini-swe-agent, authored task suite). Values of at least 0.95 are in boldface; red marks a channel that is not valid at $\alpha=5\%$ on that claim.}
\label{tab:calibrated}
\scriptsize
\setlength{\tabcolsep}{2.6pt}
\begin{tabular}{lcccccccc}
\toprule
Channel & Qwen3.5-27B & Qwen3-Coder & Devstral & GPT-5.4 & GPT-5.6 Sol & GPT-6.1 Sol & Sonnet 4.6 & Opus 5.5 \\
\midrule
MET-C & 0.37 & \textbf{0.99} & 0.32 & \textbf{1.00} & \textbf{1.00} & \textbf{1.00} & \textbf{1.00} & \textbf{1.00} \\
LLMmap-C & 0.57 & \textbf{1.00} & 0.47 & 0.91 & \textbf{1.00} & \textbf{1.00} & \textcolor{red}{0.80} & \textbf{1.00} \\
KBF-C & 0.04 & 0.75 & 0.06 & 0.37 & 0.00 & \textbf{1.00} & 0.11 & \textbf{0.99} \\
IRIS-C & 0.34 & 0.44 & 0.35 & 0.82 & \textbf{0.97} & \textbf{1.00} & 0.06 & \textbf{1.00} \\
AgentProv-C & \textbf{1.00} & \textbf{1.00} & 0.74 & \textcolor{red}{0.68} & \textbf{1.00} & \textbf{0.98} & \textcolor{red}{0.68} & \textcolor{red}{0.88} \\
R (ours) & \textbf{1.00} & \textbf{1.00} & 0.48 & -- & -- & -- & -- & -- \\
In-harness (ours) & 0.55 & 0.72 & 0.35 & \textbf{1.00} & 0.89 & \textbf{1.00} & \textbf{1.00} & 0.46 \\
\bottomrule
\end{tabular}
\end{table*}

%% file: sections/method.tex
\section{Methodology}
\label{sec:method}

As we demonstrated in Section~\ref{sec:revisit}, existing model auditing methods either falsely accuse honest providers under authorized deployment changes or miss derivatives of the claimed model, and neither widening the tolerance nor calibrating a single existing test over the authorized deployments fixes both. We argue that this is mostly because they rely on a latent assumption that the deployment is static. Motivated by this understanding, we propose a certified model-claim verification method (dubbed CertClaim). Instead of testing the service against one reference deployment, we make the authorized deployments part of the claim and test whether some authorized deployment explains all the evidence of the service.

\subsection{Preliminaries}
\label{sec:method-prelim}

\noindent\textbf{Problem Setting.} We consider two parties, i.e., the provider and the auditor. The provider sells a coding-agent service that claims a model $M_c$ and controls its serving stack, including the weights, the sampling configuration, the reasoning mode, the system prompts and context added on the provider side, and the numerical format. The auditor (e.g., a user, a marketplace or a regulator) controls the client, including the agent harness, the tasks and every request, and only observes what any client receives. This assumption is realistic because coding-agent services only expose their APIs to users. The auditor also obtains honest behavior of $M_c$ for calibration from the public checkpoint (open-weight claims) or the first-party endpoint (closed claims), which is reasonable because the claim refers to exactly these weights or this endpoint. An honest provider serves $M_c$ in one authorized configuration throughout an audit, whereas a dishonest provider may serve any substitute, for all sessions or for a fraction of them, and may adapt to the auditor.

\vspace{0.3em}
\noindent\textbf{Desired Properties.} Consistent with Section~\ref{sec:intro}, we require a model-claim verifier to be both valid and complete. Validity requires that an honest provider is falsely accused with a probability of at most $\alpha$ in every authorized deployment and under every audit budget, while completeness requires that a substitute is rejected with high probability once it is sufficiently separated from every authorized deployment. We formalize both properties in Section~\ref{sec:theory}.

\subsection{Design Requirements and Overall Pipeline}
\label{sec:method-overview}

\begin{figure*}[t]
\centering
\resizebox{\textwidth}{!}{%
\begin{tikzpicture}[every node/.style={font=\fontfamily{ppl}\selectfont\scriptsize, text=ccink},
  box/.style={draw, rounded corners=2pt, align=center, minimum height=9mm, inner sep=3pt, line width=0.5pt},
  arrow/.style={-{Latex[length=1.8mm]}, line width=0.6pt, draw=ccink!75}]
\node[box, fill=ccblue!8, draw=ccblue]       (claim)    at (0,0)       {\textbf{Claim}: model $M_c$\\+ contract $\mathcal{G}$\\(authorized cells)};
\node[box, fill=ccgray!12, draw=ccgray]      (audit)    at (3.0,0)     {In-harness audit\\$K$ batches $\times$ 3 task pairs\\fresh sessions};
\node[box, fill=ccorange!10, draw=ccorange]  (channels) at (6.15,0)    {Channel scores\\per task pair\\A\; U\; Uv\; C\; L\; B};
\node[box, fill=ccblue!8, draw=ccblue]       (cells)    at (9.45,0)    {For every cell $g$:\\stratified permutation,\\joint test $\to p^{\mathrm{in}}_g$};
\node[box, fill=ccblue!3, draw=ccblue!60, dashed] (cal) at (6.15,-1.25) {\textbf{Step I}: calibration batches of every cell\\(public checkpoint or first-party endpoint)};
\node[box, fill=ccgray!12, draw=ccgray]      (probe)    at (3.0,-2.5)  {Probe audit\\token ranks / existing probes};
\node[box, fill=ccblue!8, draw=ccblue]       (pcells)   at (9.45,-2.5) {For every cell $g$:\\probe $p$-value $p^{(e)}_g$};
\node[box, fill=ccred!8, draw=ccred]         (fuse)     at (12.85,-1.25) {\textbf{Step III}: cell-coherent fusion\\$p_g=(1{+}E)\min(p^{\mathrm{in}}_g,p^{(e)}_g)$\\reject iff $\max_g p_g\le\alpha$};
\node[font=\fontfamily{ppl}\selectfont\scriptsize\bfseries] at (4.6,0.9) {Step II: collecting multi-channel evidence};
\draw[arrow] (claim) -- (audit); \draw[arrow] (audit) -- (channels); \draw[arrow] (channels) -- (cells);
\draw[arrow] (claim.south) |- (probe.west); \draw[arrow] (probe) -- (pcells);
\draw[arrow] (cells.east) -| (fuse.north); \draw[arrow] (pcells.east) -| (fuse.south);
\draw[arrow, dashed, draw=ccblue!70] (cal.east) -| (cells.south); \draw[arrow, dashed, draw=ccblue!70] (cal.east) -| (pcells.north);
\end{tikzpicture}}
\caption{The main pipeline of CertClaim. Step I (R1): the claim specifies the deployments that the provider is authorized to use (dubbed `cells'), and we run a few calibration batches of coding tasks through every cell. Step II (R2): we audit the service with the same kind of tasks through an agent harness controlled by the auditor, score every pair of episodes on multiple in-harness channels, and add out-of-session probe channels. Step III (R3): we compare the audit with every cell, fuse all channels within each cell, and reject the claim only if no single cell explains all the evidence.}
\label{fig:pipeline}
\end{figure*}
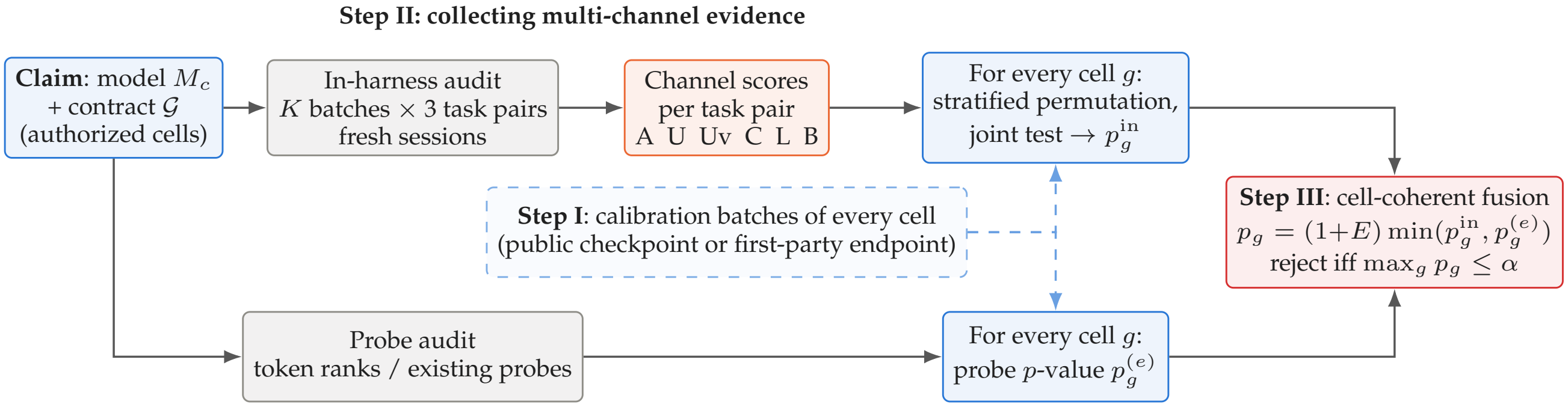

As we demonstrated in Section~\ref{sec:revisit-remedies}, widening the tolerance around one reference deployment misses the derivatives of the claimed model, and calibrating a single existing test leaves every channel blind to some substitute on some claim. These failures motivate three requirements. \textbf{R1} (Validity per deployment): the false-accusation rate is controlled within every authorized deployment and for every audit budget, so that validity comes from calibration rather than from evidence that is insensitive to the deployment. \textbf{R2} (Diverse evidence): the evidence includes several channels, including channels that are sensitive to the claimed weights, so that every cheap substitute is separated from every authorized deployment in some channel. \textbf{R3} (Coherence): the claim is accepted only if a single authorized deployment explains all channels at once, since otherwise different channels may be explained by different deployments. Accordingly, CertClaim consists of three components, each responding to one requirement, as shown in Figure~\ref{fig:pipeline}.

\subsection{Calibrating over the Deployment Contract}
\label{sec:method-claim}

In this step, we specify the deployments that the provider is authorized to use and calibrate the claimed model over them (i.e., R1), as follows.

\begin{definition}[Deployment Contract]
\label{def:contract}
A claim is a pair $c=(M_c,\mathcal{G})$ of a model $M_c$ and a finite set $\mathcal{G}=\{g_1,\dots,g_m\}$ of serving configurations (dubbed `cells'). Each cell fixes everything that the provider controls and the client does not, i.e., the sampling preset, the reasoning mode or effort, the provider-side system prompt and context, and the numerical format of the weights. A service is honest for $c$ if it serves $M_c$ in one cell $g^\star\in\mathcal{G}$ throughout an audit, where $g^\star$ is unknown to the auditor.
\end{definition}

In practice, we build contracts from public information (e.g., the sampling presets of the model card, the reasoning modes or efforts of the API, common system-prompt practices and official low-precision checkpoints), and a continuous parameter can be covered by a grid of cells (Appendix~\ref{app:theory}). For every cell, we run several calibration batches of freshly generated tasks, organized into strata $s\in\mathcal{S}$ of exchangeable units (Section~\ref{sec:method-evidence}).

\begin{definition}[Calibration Set]
\label{def:calibration}
For a cell $g\in\mathcal{G}$, a stratum $s\in\mathcal{S}$ and an evidence channel $\phi_j$ (i.e., a score of an audit unit, fixed before calibration), the calibration set is $C^{g,s}_j=\{\phi_j(X_i)\}_{i=1}^{n_{g,s}}$, where $X_1,\dots,X_{n_{g,s}}\overset{iid}{\sim}P^{s}_{M_c,g}$ are units of the claimed model served in cell $g$.
\end{definition}

Inspired by conformal prediction~\citep{vovk2005alrw}, we compare an audit with the calibration set of every cell by a permutation test that only exchanges units within the same stratum, so that the test is exact within every cell however far apart the cells are.

\begin{definition}[Stratified Cell-wise Permutation $p$-value]
\label{def:pvalue}
Let $T^{s}_{j}$ be the $K$ audit scores of channel $j$ in stratum $s$, and for a cell $g$ let $D_{g,j}=\bigl|\sum_{s\in\mathcal{S}}\bigl(\overline{T^{s}_{j}}-\overline{C^{g,s}_{j}}\bigr)\bigr|$. Let $\Pi$ be the group that permutes the pooled $K+n_{g,s}$ units independently within every stratum, let $\pi_1,\dots,\pi_B$ be i.i.d.\ uniform draws from $\Pi$ shared by all channels, and put $\pi_0=\mathrm{id}$. The $p$-value of channel $j$ in cell $g$ is $p_{g,j}=\bigl(1+\sum_{b=1}^{B}\mathbb{1}\{D_{g,j}^{\pi_b}\ge D_{g,j}\}\bigr)/(B+1)$.
\end{definition}

Since the channels are strongly correlated, a Bonferroni correction over them is overly conservative. Accordingly, we combine them within the same permutation test into a joint $p$-value $p^{\mathrm{in}}_g$ of every cell, which takes the better of a min-$p$ and a Fisher combination of the channels over the same permutations (Definition~\ref{def:joint} in Appendix~\ref{app:details}). The min-$p$ part is powerful when one channel carries the evidence, and the Fisher part when the evidence is spread over several channels.

\subsection{Collecting Multi-channel Evidence}
\label{sec:method-evidence}

In this step, we audit the suspicious service through an agent harness controlled by the auditor and score the audit on multiple evidence channels (i.e., R2). An audit consists of $K$ batches. A batch contains one coding task for each of three task axes, and every task is run in two paired variants that differ in a single controlled factor: whether the agent is asked to check its work (\emph{verify}), whether a transient failure is injected (\emph{recover}), and whether a visible test conflicts with the specification (\emph{resolve}). Every episode runs in a fresh sandbox and a fresh provider session through the harness of the auditor (e.g., mini-swe-agent~\citep{minisweagent2025} or OpenCode~\citep{opencode2026}), so every request of the audit has the format of ordinary agent traffic.

\begin{definition}[Audit Unit and Evidence Channel]
\label{def:channel}
An audit unit $X$ is everything that the auditor observes in the two episodes of one task, and $P^{s}_{M,g}$ denotes its law in stratum (i.e., task axis) $s$ when model $M$ is served in configuration $g$. An evidence channel is a measurable score $\phi_j(X)\in\mathbb{R}$, $j\in\{1,\dots,J\}$, fixed before calibration, where the score of a unit is the mean of the scores of its two episodes.
\end{definition}

We treat the pair of episodes of a task as the exchangeable unit, since the two variants share the difficulty of the task. Specifically, we design the following in-harness channels, all fitted on data that are disjoint from calibration and audits (Appendix~\ref{app:details}): \textbf{A}, the likelihood of the workflow of the agent (e.g., transitions between inspecting, editing, testing and recovering); \textbf{U} and \textbf{Uv}, the usage (e.g., the number of calls and the characters per call); \textbf{C}, the response to an optional in-session request for about 300 integers; \textbf{L}, the likelihood ratio of behavioral fields between the best-matching enrolled public model and the claimed model; and \textbf{B} (open-weight claims), the likelihood of the generated tokens under the claimed weights. Besides, we add out-of-session probe channels $e\in\{1,\dots,E\}$, each returning a $p$-value $p^{(e)}_g$ for every cell. For open-weight claims, we use token-rank probes (dubbed \textbf{R}), whose responses receive exactly uniform randomized ranks against locally simulated responses of the same cell (Theorem~\ref{thm:pivot}). Since our verifier is valid for any channel that is valid within every cell, we also reuse the probes of existing methods calibrated over the contract, and admit such a channel only if its own-cell $p$-values on held-out honest audits pass a validity check (Appendix~\ref{app:details}). Arguably, these channels complement each other: the in-harness channels see the behavior of the agent, while the probes see the weights more directly.

\subsection{Claim Verification via Cell-coherent Fusion}
\label{sec:method-test}

In this step, we fuse all evidence within each cell and reject the claim only if every cell is rejected (i.e., R3), as follows.

\begin{proposition}[Claim Verification via Cell-coherent Fusion]
\label{def:verifier}
For every cell $g\in\mathcal{G}$, let $p_g=\min\bigl\{1,(1+E)\min\bigl(p^{\mathrm{in}}_g,p^{(1)}_g,\dots,p^{(E)}_g\bigr)\bigr\}$. We claim that the suspicious service is not serving the claimed model $M_c$ (in any authorized deployment) if and only if $\max_{g\in\mathcal{G}}p_g\le\alpha$, where $\alpha$ (e.g., $0.05$) is the significance level. We denote this verifier by $V_\alpha$.
\end{proposition}

In other words, the claim is rejected if and only if, for every cell, some channel refutes that cell at level $\alpha/(1+E)$. In particular, we fuse the channels within each cell before maximizing over cells (dubbed `cell-coherent fusion'), instead of letting each channel choose its own best cell. As such, a substitute whose channels are explained by different cells (e.g., a quantized model served with the sampling preset that best matches the agent behavior of the claimed model) is still rejected (Proposition~\ref{prop:coherent}).

%% file: sections/theory.tex
\section{Theoretical Analyses}
\label{sec:theory}

In this section, we provide theoretical analyses of CertClaim. Further results (e.g., exact nulls from open weights, mixed routing and continuous deployment parameters) and all proofs are in Appendices~\ref{app:theory} and~\ref{app:proofs}.

\subsection{Certified Model-Claim Verification}
\label{sec:theory-def}

Let $\mathcal{F}_{\mathcal{G}}(M_c)=\{\bigotimes_{k\le K,\,s\in\mathcal{S}}P^{s}_{M_c,g}\mid g\in\mathcal{G}\}$ be the laws of an audit that an honest provider can produce. Model-claim verification tests the composite null hypothesis $H_0$ that the audit follows a law in $\mathcal{F}_{\mathcal{G}}(M_c)$ against the alternative $H_1$ that it follows the law of a substitute $(S,\eta)$, i.e., a model $S\ne M_c$ served in some configuration $\eta$. Since a false accusation may end a contract or trigger legal forensics, we require the type-I error to be controlled for every authorized deployment, as follows.

\begin{definition}[$(\alpha,\mathcal{G})$-Certified Model-Claim Verification]
\label{def:certified}
A verifier is $(\alpha,\mathcal{G})$-certified for claim $c=(M_c,\mathcal{G})$ if and only if, for every calibration size, every audit budget $K$ and every honest service, $\sup_{g\in\mathcal{G}}\Pr_{M_c,g}(\text{reject})\le\alpha$.
\end{definition}

Definition~\ref{def:certified} is the counterpart of certified dataset watermarks~\citep{qiao2025certdw}: the verdict `the claim holds' must survive every deployment in the contract, instead of every perturbation in a bounded neighborhood. Following certified verification~\citep{qiao2025certdw,cohen2019smoothing}, we characterize completeness by the condition in Theorem~\ref{thm:power}.

\subsection{Validity of CertClaim and Failures of a Static Null}
\label{sec:theory-validity}

\begin{theorem}[Certified False-accusation Control]
\label{thm:validity}
For any score functions $\phi_1,\dots,\phi_J$ fixed before calibration, any probe channels that return valid $p$-values within every cell, any calibration sizes $n_{g,s}\ge1$, any budget $K\ge1$, any number of permutations $B\ge1$ and any distributions $P^{s}_{M_c,g}$, the verifier $V_\alpha$ in Proposition~\ref{def:verifier} is $(\alpha,\mathcal{G})$-certified.
\end{theorem}

In general, Theorem~\ref{thm:validity} indicates that the false-accusation rate is controlled without any distributional assumption, under arbitrary dependence between channels and for every audit budget, provided that the calibration data of every cell are exchangeable with the audit of an honest service in that cell. In contrast, the following proposition explains the two failures in Section~\ref{sec:revisit}.

\begin{proposition}[Two Failures of a Static Null]
\label{prop:paradox}
(1) (Budget paradox.) Let $g\in\mathcal{G}$ satisfy $P_{M_c,g}\ne P_{M_c,g_0}$. If level-$\alpha$ tests $(\psi_N)_{N\ge1}$ of $H_0:Q=P_{M_c,g_0}$ from $N$ observations are consistent against $P_{M_c,g}$, the honest service in cell $g$ is falsely accused with probability tending to one as $N\to\infty$. (2) (Blindness of deployment-invariant evidence.) If a verifier depends on the audit only through a statistic $\rho$, and a substitute $(S,\eta)$ satisfies $\rho_{\#}P_{S,\eta}=\rho_{\#}P_{M_c,g}$ for some $g\in\mathcal{G}$, then any $(\alpha,\mathcal{G})$-certified verifier based on $\rho$ rejects $S$ with probability at most $\alpha$.
\end{proposition}

In other words, a test of the static null buys power with false accusations, while restricting the evidence to what is insensitive to the deployment makes a method blind to any substitute that preserves this evidence. Calibrating over the contract removes this dilemma, since validity then comes from the calibration of every cell, so the evidence may be as sensitive as we like.

\subsection{Detection Condition and Cell-coherent Fusion}
\label{sec:theory-power}

\begin{theorem}[General Condition of Certified Detection]
\label{thm:power}
Suppose that in-harness channel $j$ takes values in an interval of length $c_j$, and for a substitute service $(S,\eta)$ let $\Delta_j(g)=\bigl|\sum_{s}(\mu^{s}_{S,j}-\mu^{s}_{g,j})\bigr|$ be its separation from cell $g$ in channel $j$, where $\mu^{s}_{g,j}$ and $\mu^{s}_{S,j}$ are the expected scores of the claimed model in cell $g$ and of the substitute in stratum $s$. Let $n=\min_{g,s}n_{g,s}$, $J'=2(1+E)J$ and $\beta\in(0,1)$. If $B\ge\frac{J'}{\alpha}\max\{8,3\log\frac{4m}{\beta}\}$ and for every cell $g\in\mathcal{G}$ some channel $j$ satisfies $\Delta_j(g)\ge r_j(K,n)$, where
\[
\begin{aligned}
r_j(K,n)=c_j\sqrt{|\mathcal{S}|}\,\Bigl[&\sqrt{\tfrac{1}{2}\bigl(\tfrac1K+\tfrac1n\bigr)\bigl(1+\tfrac1n\bigr)\log\tfrac{8J'}{\alpha}}\\
&+\sqrt{\tfrac{1}{2K}\log\tfrac{8J}{\beta}}+\sqrt{\tfrac{1}{2n}\log\tfrac{8m}{\beta}}\Bigr],
\end{aligned}
\]
then $V_\alpha$ rejects with probability at least $1-\beta$.
\end{theorem}

In general, Theorem~\ref{thm:power} indicates that a substitute is detected once every authorized cell is refuted, in some channel, by a margin that shrinks as $K^{-1/2}$ until the calibration size dominates and grows only logarithmically with the numbers of cells and channels. Besides, cell-coherent fusion never loses power to the alternative that lets each channel choose its own best cell, as follows.

\begin{proposition}[Cell-coherent Fusion Dominates]
\label{prop:coherent}
(1) For every cell, $p^{\mathrm{in}}_g\le2J\min_{j\le J}p_{g,j}$ deterministically. (2) For any array of $p$-values $(q_{g,c})$ and weights $w_c>0$, $\max_g\min_c q_{g,c}/w_c\le\min_c\max_gq_{g,c}/w_c$. Hence, with $w_j=1/(2(1+E)J)$ for in-harness channels and $w_e=1/(1+E)$ for probe channels, $V_\alpha$ rejects whenever the channel-first rule ``some channel $c$ refutes every cell at level $w_c\alpha$'' rejects. (3) The inequality in (2) is strict under \emph{masking}: if channel $1$ refutes every cell except $g_1$ and channel $2$ every cell except $g_2\ne g_1$, the channel-first rule cannot reject while $V_\alpha$ can.
\end{proposition}

Since $B=1{,}999$ in our experiments is below the number required by Theorem~\ref{thm:power}, we use it to explain which substitutes are detectable and evaluate detection empirically.

%% file: sections/experiments.tex
\section{Experiments}
\label{sec:experiments}

\subsection{Main Settings}
\label{sec:exp-setup}

\noindent\textbf{Claims and Models.} We conduct experiments on eight claims, including three open-weight claims (i.e., Qwen3.5-27B~\citep{qwen35_27b}, Qwen3-Coder-30B-A3B-Instruct~\citep{qwen3coder30b} (dubbed Qwen3-Coder) and Devstral-Small-2507~\citep{devstral2507} (dubbed Devstral)) and five frontier claims served through a commercial model gateway (i.e., GPT-5.4~\citep{openai_gpt54}, GPT-5.6 Sol~\citep{openai_gpt56sol}, GPT-6.1 Sol~\citep{openai_gpt61sol}, Claude Sonnet~4.6~\citep{anthropic_sonnet46} and Claude Opus~5.5~\citep{anthropic_opus55}). Each contract has six to nine cells built from public information, covering five types of authorized changes (i.e., sampling, reasoning, system prompt, project context and, for open-weight claims, FP8 precision).

\vspace{0.3em}
\noindent\textbf{Substitutes and Precision Variants.} Following the model-identity benchmark LeaFBench~\citep{shao2025sok}, we adopt substitutes of five types, i.e., models of another family, models of the same developer, post-trained derivatives of the claimed weights (e.g., a refusal-ablated model~\citep{arditi2024refusal}), quantized derivatives (4-bit and 6-bit), and merged derivatives $(1-\lambda)\theta_c+\lambda\theta'$ of the claimed weights $\theta_c$ and a sibling $\theta'$ with $\lambda\in\{0.1,0.25,0.5\}$. The low-precision releases include Qwen3.5-27B-FP8~\citep{qwen35_27b_fp8} and Qwen3-Coder-30B-A3B-Instruct-FP8~\citep{qwen3coder30b_fp8} as authorized deployments, and Qwen3.5-27B-GPTQ-Int4~\citep{qwen35_27b_gptq} as a substitute. In total, our experiments involve 44 models, including 15 derivatives of the claimed weights; Appendix~\ref{app:model-sources} provides the derivative constructions and the complete source inventory (Table~\ref{tab:model-citations}).

\vspace{0.3em}
\noindent\textbf{Agent Harnesses and Task Suites.} Following previous works on coding agents~\citep{yang2024sweagent}, we audit services through two agent harnesses, i.e., mini-swe-agent (MSA)~\citep{minisweagent2025} and OpenCode (OC)~\citep{opencode2026}, where the frontier claims are audited through MSA. We adopt three task suites, including an authored suite of repository tasks with hidden tests (Section~\ref{sec:method-evidence}), a suite built from EvalPlus~\citep{liu2023evalplus}, and a suite of 506 real issues from 11 projects of SWE-bench~\citep{jimenez2024swebench}. We use the authored suite for all claims, the EvalPlus suite for the open-weight claims under both harnesses, and the real-issue suite for Qwen3.5-27B under MSA, which leads to 18 settings and about 37k agent episodes in total.

\vspace{0.3em}
\noindent\textbf{Baseline Selection.} We compare CertClaim with six representative methods covering all three categories in Section~\ref{sec:related-auditing}, including (1) MET~\citep{gao2025met}, (2) LLMmap~\citep{pasquini2025llmmap}, (3) RUT~\citep{rut2026}, (4) KBF~\citep{kbf2026}, (5) IRIS~\citep{iris2026} and (6) AgentProv~\citep{agentprov2026}. We use the official code of MET, LLMmap, KBF and IRIS with their published protocols and budgets, and re-implement RUT (open-weight claims only, since it requires the claimed weights) and AgentProv following their papers. Their contract-calibrated versions are analyzed in Section~\ref{sec:revisit-remedies}, and the concurrent LIDAR~\citep{lidar2026} (code not released) is re-implemented in Appendix~\ref{app:lidar}.

\vspace{0.3em}
\noindent\textbf{Settings for Claim Verification.} We use $B=1{,}999$ permutations, $\alpha=0.05$ and $K=1$ batch of tasks per audit unless otherwise specified, and the audited substitute is never enrolled in the identity channel L. The probe channels are R and the contract-calibrated KBF probe for the open-weight claims, and the contract-calibrated KBF, MET and IRIS probes for the frontier claims. Every cell has 4--16 calibration batches, and we re-split the honest batches of every cell into calibration and audit batches ten times at random (Appendix~\ref{app:details}).

\vspace{0.3em}
\noindent\textbf{Evaluation Metrics.} Following LeaFBench~\citep{shao2025sok}, we adopt the AUC, the standardized partial AUC over $\mathrm{FPR}\in[0,0.05]$ (pAUC), the Mahalanobis distance (MD), the accuracy (ACC) of the decision of each method at $\alpha$, and the TPR at 1\% FPR, where the positives are the audits of substitutes in their default configuration, the negatives are the honest audits in all authorized cells, and the score of an audit is $-\log_{10}p$ (or the score of the method itself). The higher these metrics, the better the verification. Besides, a method is valid in a setting if the lower end of the 95\% Clopper--Pearson interval of its false-positive rate (FPR) is at most $\alpha$.

\input{tables/overall}
\input{tables/per_setting_main}

\subsection{Main Results}
\label{sec:exp-main}

\begin{figure*}[t]
\centering
\includegraphics[width=\textwidth]{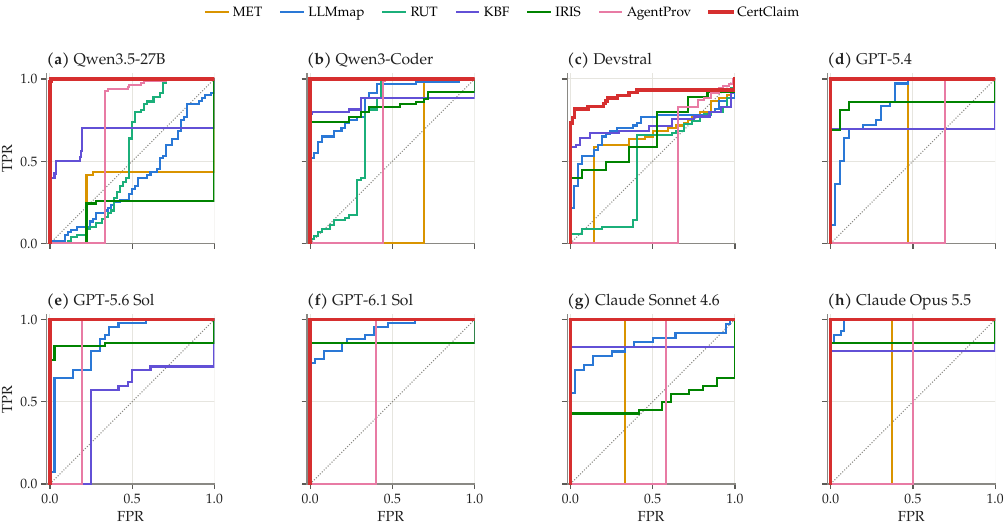}
\vspace{-1.6em}
\caption{ROC curves of model-claim verification on the eight claims (authored task suite, mini-swe-agent, one batch of tasks for CertClaim), where the positives are the audits of substitutes and the negatives are the honest audits in all authorized deployments. RUT is not applicable to the frontier claims. Curves are threshold-free, so they do not show whether a method is valid at $\alpha$ (Table~\ref{tab:main-leaf}).}
\label{fig:roc}
\end{figure*}

As shown in Table~\ref{tab:overall-leaf}, CertClaim is valid in all 18 settings and achieves an average AUC of 0.992 and TPR@1\%FPR of 0.970. In contrast, every existing method is invalid in at least one-third of the settings, and none of them achieves an average AUC above 0.77. As shown in Table~\ref{tab:main-leaf}, CertClaim achieves the best (or tied-best) results in every setting of the authored suite, with an AUC of 1.000 except on the Devstral claim (0.920 and 0.929). Existing methods fail in two ways: none of them is valid on the Qwen3.5-27B claim, whose contract includes switching reasoning on and off, while the valid ones on the other claims achieve an AUC of 0.70--0.99 (except MET on GPT-6.1 Sol), i.e., they miss part of the substitutes. As shown in Figure~\ref{fig:roc}, the ROC curve of CertClaim lies on or above those of all existing methods at every FPR on seven of the eight claims (and at every FPR below 0.9 on Devstral-Small-2507), and it achieves a TPR of at least 0.98 at zero FPR on all claims except Devstral-Small-2507 (0.73), whose misses are mostly its merged derivatives. On the EvalPlus and real-issue suites, CertClaim achieves an AUC of 1.000 in all seven settings, whereas the best valid existing method achieves at most 0.900 (Figure~\ref{fig:roc-more} and Table~\ref{tab:suites-leaf} in Appendix~\ref{app:results}).

\input{tables/breakdown}

As shown in Table~\ref{tab:breakdown-leaf}, every existing method is invalid under at least one type of authorized deployment change, and MET, LLMmap and AgentProv falsely accuse 14.6\%, 14.6\% and 20.0\% of the honest services even in the default deployment, probably because their decision rules are not calibrated against the variation of the claimed model across sessions. In contrast, CertClaim is valid under every type of change with an AUC of at least 0.986. As for the types of substitution, CertClaim achieves an AUC of 1.000 on models of another family and of the same developer and on post-trained derivatives, and 0.998 on quantized derivatives (vs.\ at most 0.889 of existing methods). Although CertClaim misses part of the merged derivatives (with an AUC of 0.877), it still outperforms all existing methods (at most 0.728). It is mostly because merging Devstral-Small-2507 with its predecessor at $\lambda\le0.25$ moves the agent less than the authorized changes do (i.e., below the radius of Theorem~\ref{thm:power} in every channel), on which all methods fail (with a pAUC of at most 0.58). These results verify the effectiveness of CertClaim.

\subsection{Analyzing the Detection Condition of CertClaim}
\label{sec:exp-region}

In this section, we visualize the detection condition of CertClaim (Theorem~\ref{thm:power}) on the in-harness channels. For every audit, we compute its standardized separation from every authorized cell (i.e., the largest one over the in-harness channels) and take the minimum over the cells, which exceeds the threshold of Example~\ref{ex:gaussian} if and only if the audit lies in the certified detection region.

\begin{figure*}[t]
\centering
\includegraphics[width=\textwidth]{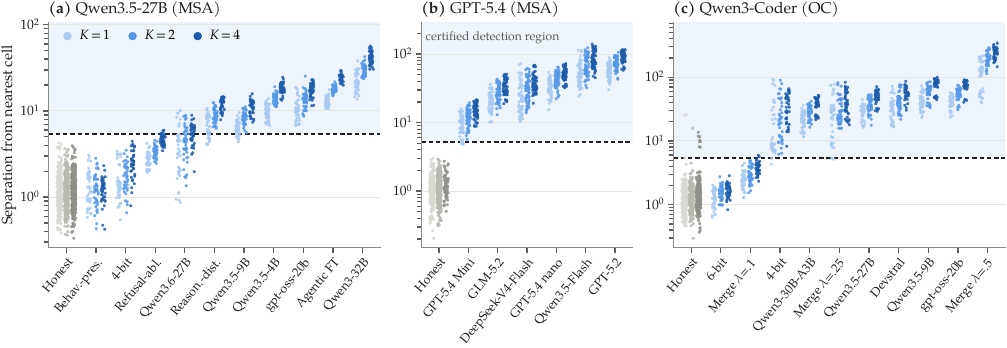}
\vspace{-1.6em}
\caption{The certified detection region of CertClaim on the in-harness channels. Each point is an audit with $K=1$, $2$ or $4$ batches of tasks, placed at its standardized separation from the nearest authorized cell; the dashed line is the threshold of the detection condition under Gaussian scores (Example~\ref{ex:gaussian}), above which an audit is certified to be detected. Results on the other eight settings of the authored suite are in Figure~\ref{fig:certified-more} (Appendix~\ref{app:results}).}
\label{fig:certified}
\end{figure*}

\vspace{0.3em}
\noindent\textbf{Results.} As shown in Figure~\ref{fig:certified}, the separation of honest services stays around one for every audit budget, whereas that of a substitute grows with the budget, roughly as $\sqrt{K}$ (e.g., from 21.2 to 40.3 for Qwen3-32B~\citep{qwen3_32b} under the Qwen3.5-27B claim as $K$ grows from one to four), which is consistent with Theorem~\ref{thm:power}. With one batch of tasks, every substitute of GPT-5.4 and most substitutes of Qwen3.5-27B and Qwen3-Coder-30B-A3B-Instruct lie in the certified detection region. The substitutes below the threshold are the nearest derivatives of the claimed weights (e.g., the 4-bit and the behavior-preserving derivatives of Qwen3.5-27B and the 6-bit derivative of Qwen3-Coder-30B-A3B-Instruct), whose separation barely grows with the budget. These derivatives are detected by the probe channels instead (Section~\ref{sec:exp-ablation}), which is why CertClaim fuses both kinds of evidence.

\subsection{Ablation Study}
\label{sec:exp-ablation}

We discuss the effects of the three components of CertClaim (Section~\ref{sec:method-overview}) and of the budgets, evaluating all variants on the same audits.

\begin{figure*}[t]
\centering
\includegraphics[width=\textwidth]{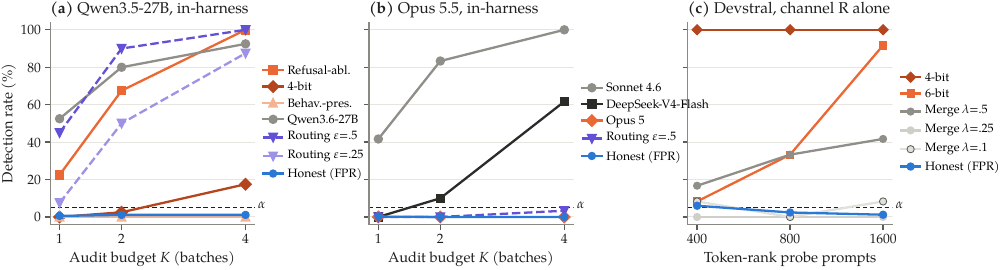}
\vspace{-1.6em}
\caption{The effects of the audit and probe budgets. (a, b) The detection rate of the in-harness channels alone as the number $K$ of batches of tasks grows, on the Qwen3.5-27B and Claude Opus~5.5 claims (mini-swe-agent), for the substitutes and mixed-routing services that are hardest to separate; the full verifier detects every substitute shown here with $K=1$. (c) The detection rate of the token-rank channel R alone on the Devstral-Small-2507 claim as the number of probe prompts grows. Honest: the false-positive rate on honest services in all authorized deployments.}
\label{fig:budgets}
\end{figure*}

\input{tables/ablation}

\begin{figure*}[t]
\centering
\includegraphics[width=\textwidth]{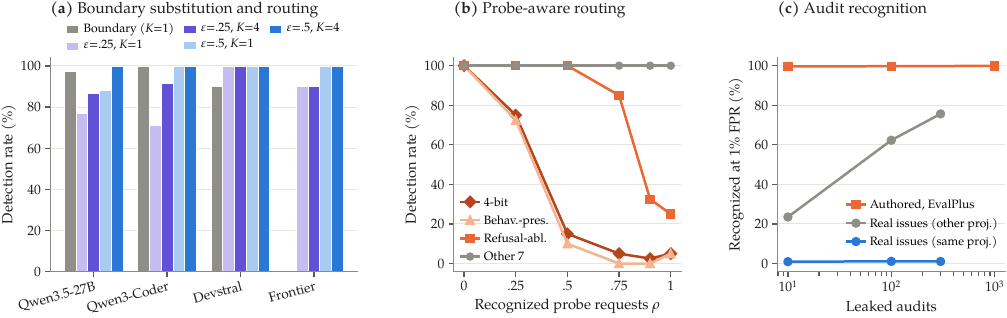}
\vspace{-1.6em}
\caption{The resistance of CertClaim to potential adaptive attacks. (a) The detection rate of substitutes served in an authorized configuration (Boundary) and of mixed routing, where a fraction $\varepsilon$ of the sessions (open-weight claims) or of the calls (frontier claims, with $\varepsilon=0.3$ instead of $0.25$) is served by a substitute, averaged over the harnesses and task suites of each claim. (b) The detection rate on the Qwen3.5-27B claim when a fraction $\rho$ of the probe requests is answered by the claimed model (one batch of tasks; Other~7: the mean over the other seven substitutes). (c) The proportion of audits that a character $n$-gram classifier recognizes at 1\% FPR, given leaked audits of the synthetic suites (minimum over the authored and EvalPlus suites and both harnesses) or of the real-issue suite (negatives: other issues of the same projects or of other projects).}
\label{fig:adaptive}
\end{figure*}

\vspace{0.3em}
\noindent\textbf{Effects of Contract Calibration.} We calibrate the same channels on the default deployment only (dubbed `static null'). As shown in Table~\ref{tab:ablation}, the static null is invalid in 16 of the 18 settings (with an FPR of up to 87.9\%), and its high detection rates of adaptive services are paid for with false accusations. This verifies that the validity of CertClaim comes from its calibration over the contract (i.e., R1).

\vspace{0.3em}
\noindent\textbf{Effects of Multi-channel Evidence.} With the in-harness channels only, CertClaim remains valid but misses the nearest derivatives (e.g., with an AUC of 0.656 and 0.578 on the 4-bit and the behavior-preserving derivatives of Qwen3.5-27B under MSA) and most of the mixed routing. With the probe channels only, it detects fewer adaptive services and relies entirely on requests that a provider may recognize (Section~\ref{sec:exp-adaptive}). Combining both kinds of channels (i.e., R2) achieves the best results.

\vspace{0.3em}
\noindent\textbf{Effects of Cell-coherent Fusion.} We replace cell-coherent fusion with channel-first fusion, which lets every channel choose its own best cell. The two rules perform similarly on substitutes in their default configuration, but channel-first fusion detects fewer adaptive services, e.g., 66.7\% vs.\ 100\% of the 4-bit derivative of Qwen3.5-27B and 33.3\% vs.\ 100\% of the 6-bit derivative of Devstral served in an authorized configuration. It is mostly because different channels of these services are explained by different cells (i.e., masking), which only cell-coherent fusion (i.e., R3) refutes. Its cost is a slightly higher FPR in a few settings (e.g., 7.1\% vs.\ 2.4\% on Qwen3-Coder under MSA), which remains valid.

\vspace{0.3em}
\noindent\textbf{Effects of the Audit and Probe Budgets.} As shown in Figure~\ref{fig:budgets}(a) and (b), the detection rate of the in-harness channels alone grows with the audit budget for the substitutes near the threshold, e.g., from 22.5\% to 100\% for the refusal-ablated derivative of Qwen3.5-27B and from 41.7\% to 100\% for Claude Sonnet~4.6 under the Claude Opus~5.5 claim, while their false-positive rate stays at most 1.1\%. However, the in-harness channels detect at most 17.5\% of the 4-bit and the behavior-preserving derivatives of Qwen3.5-27B and none of Claude Opus~5~\citep{anthropic_opus5} even with four batches, whereas the full verifier detects all of them with one batch through the probe channels. Overall, increasing the audit budget from one to four batches raises the average worst-substitute AUC of CertClaim from 0.954 to 0.964, while CertClaim remains valid in all settings. As shown in Figure~\ref{fig:budgets}(c), increasing the number of token-rank probes from 400 to 1{,}600 raises the detection rate of the 6-bit derivative of Devstral by channel R alone from 8.3\% to 91.7\%, while the false-positive rate of channel R decreases from 6.0\% to 1.2\%. In contrast, the detection rate of the merges at $\lambda\le0.25$ stays at most 8.3\% at every budget, which is consistent with their small separation from the claimed model (Appendix~\ref{app:results}).

\subsection{The Resistance to Potential Adaptive Attacks}
\label{sec:exp-adaptive}

We evaluate CertClaim against a provider that knows our method and adapts its substitute to pass the audit.

\begin{figure*}[t]
\centering
\includegraphics[width=\textwidth]{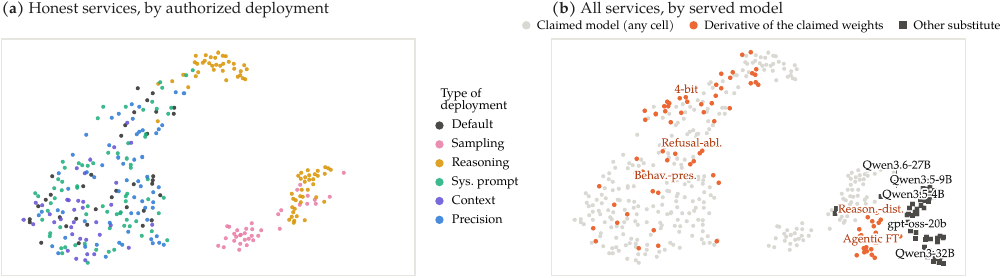}
\vspace{-1.6em}
\caption{The t-SNE visualization of the in-harness evidence of services that claim Qwen3.5-27B (mini-swe-agent). Each point is the mean of the in-harness channel scores of one batch of tasks. (a) Honest batches, colored by the type of their authorized deployment. (b) All batches, colored by the served model. Results on GPT-5.4 and Qwen3-Coder-30B-A3B-Instruct are in Figure~\ref{fig:tsne-more} (Appendix~\ref{app:results}).}
\label{fig:tsne}
\end{figure*}

\begin{figure}[t]
\centering
\includegraphics[width=\columnwidth]{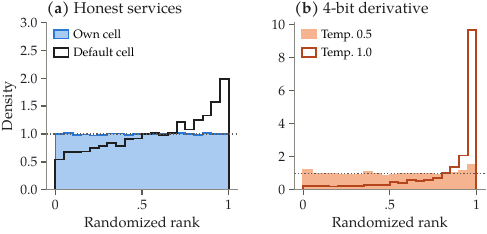}
\vspace{-1.8em}
\caption{The randomized ranks of the token-rank probe R on the Qwen3.5-27B claim (dotted: the uniform density). (a) Honest services against the local simulations of their own cell and the non-thinking preset against the default cell. (b) The 4-bit derivative against the non-thinking preset at the two client temperatures.}
\label{fig:pit}
\end{figure}

\vspace{0.3em}
\noindent\textbf{The Resistance to Contract-boundary Substitution.} A provider may serve its substitute in the authorized configuration that brings it closest to the claimed model. As shown in Figure~\ref{fig:adaptive}(a) (Table~\ref{tab:adaptive-all} in Appendix~\ref{app:results} for every setting), CertClaim detects 97.5\% and 100\% of these services on the Qwen3.5-27B and Qwen3-Coder claims, and 90.0\% on the Devstral claim, where all misses are merged derivatives. This is because a substitute in an authorized configuration is still compared with the claimed model in the same configuration.

\vspace{0.3em}
\noindent\textbf{The Resistance to Mixed Routing.} A provider may serve only a fraction $\varepsilon$ of the sessions (open-weight claims) or of the calls (frontier claims) with a cheaper model. As shown in Figure~\ref{fig:adaptive}(a), CertClaim detects 88.3\%--100\% of the audits of services that route half of the traffic with one batch of tasks and all of them with four batches, and 70.8\%--100\% and 86.7\%--100\% of those that route a quarter, respectively, which is consistent with our analysis of mixed routing (Theorem~\ref{thm:sequential} in Appendix~\ref{app:theory}).

\vspace{0.3em}
\noindent\textbf{The Resistance to Probe-aware Routing.} A provider may recognize the out-of-session probes, which are single-turn requests without the tools of the harness, and answer them with the claimed model. We simulate this attack on the Qwen3.5-27B claim by answering a fraction $\rho$ of the token-rank probe requests with the claimed model in its default cell. As shown in Figure~\ref{fig:adaptive}(b) (Table~\ref{tab:probe-aware} in Appendix~\ref{app:results}), CertClaim still detects all audits of the substitutes that the in-harness channels separate from the claimed model (e.g., the agentic fine-tune and Qwen3.5-9B~\citep{qwen35_9b}) for every $\rho$, whereas the detection rate of the 4-bit and the behavior-preserving derivatives falls to 15\% and 10\% at $\rho=0.5$. In other words, these two derivatives evade CertClaim only if the provider recognizes most of the probe requests.

\vspace{0.3em}
\noindent\textbf{The Resistance to Audit-aware Routing.} A provider may also recognize the audit sessions themselves. Given ten leaked audits, a character $n$-gram classifier recognizes at least 99.6\% of the audits of the authored and EvalPlus suites at 1\% FPR, i.e., synthetic tasks are recognizable from their wording. In contrast, as shown in Figure~\ref{fig:adaptive}(c), on the real-issue suite, whose audit tasks are new issues of the same projects as the requests of other users, it recognizes at most 1.1\% of the audits at 1\% FPR, and its AUC is only 0.518--0.536 even with 300 leaked audits. Since CertClaim also achieves an AUC of 1.000 on this suite, an auditor can resist this attack by drawing audit tasks from the projects of the audited traffic.

\subsection{A Closer Look at the Effectiveness of CertClaim}
\label{sec:exp-closer}

In this section, we visualize the evidence of CertClaim to further explain why it is valid and complete.

\vspace{0.3em}
\noindent\textbf{Visualizing the In-harness Evidence.} As shown in Figure~\ref{fig:tsne}(a), honest batches in the thinking preset and with reasoning switched off form their own clusters far from the default deployment, whereas the other authorized changes (e.g., system prompts and FP8 checkpoints) overlap with it. This is why a static null falsely accuses the former and why validity has to come from calibrating every cell. As shown in Figure~\ref{fig:tsne}(b), most substitutes form clusters outside all authorized deployments, whereas the 4-bit, the behavior-preserving and the refusal-ablated derivatives lie inside the cloud of honest batches, which is consistent with Figure~\ref{fig:gap}(b) and explains why CertClaim also needs evidence channels that see the weights more directly.

\vspace{0.3em}
\noindent\textbf{The Exact Null of the Token-rank Probe.} As shown in Figure~\ref{fig:pit}(a), the randomized ranks of honest services against the local simulations of their own cell are uniform, which verifies Theorem~\ref{thm:pivot}(1), whereas those of the non-thinking preset against the default cell are not, i.e., channel R also has to be calibrated in every cell. As shown in Figure~\ref{fig:pit}(b), the ranks of the 4-bit derivative against the non-thinking preset are close to uniform at client temperature 0.5 (13.3\% of them exceed 0.9), i.e., this authorized cell masks the derivative at this temperature, whereas 58.9\% of them exceed 0.9 at temperature 1.0. Since channel R combines both temperatures within each cell, a masking cell has to explain both of them, which is why channel R alone detects this derivative in every audit.

%% file: tables/overall.tex
\begin{table}[t]
\centering
\caption{Overall performance averaged over all 18 claim--harness--suite settings, in the metrics of LeaFBench, and the number of settings in which each method is valid at $\alpha=5\%$. RUT needs the weights of the claimed model, so it is evaluated on the 13 open-weight settings only.}
\label{tab:overall-leaf}
\scriptsize
\setlength{\tabcolsep}{2.4pt}
\begin{tabular}{lccccccc}
\toprule
Metric & MET & LLMmap & RUT & KBF & IRIS & AgentProv & CertClaim \\
\midrule
AUC$\uparrow$ & 0.734 & 0.726 & 0.593 & 0.762 & 0.693 & 0.747 & \textbf{0.992} \\
pAUC$\uparrow$ & 0.576 & 0.676 & 0.503 & 0.821 & 0.748 & 0.516 & \textbf{0.987} \\
MD$\uparrow$ & 1.57 & 1.15 & 0.81 & 2.10 & 1.01 & 1.17 & \textbf{9.82} \\
ACC$\uparrow$ & 0.693 & 0.672 & 0.671 & 0.828 & 0.668 & 0.639 & \textbf{0.973} \\
TPR@1\%FPR$\uparrow$ & 0.111 & 0.322 & 0.029 & 0.636 & 0.500 & 0.000 & \textbf{0.970} \\
\midrule
Valid at $\alpha$ & 1/18 & 11/18 & 0/13 & 12/18 & 7/18 & 0/18 & 18/18 \\
\bottomrule
\end{tabular}
\end{table}

%% file: tables/per_setting_main.tex
\begin{table*}[t]
\centering
\caption{The performance of model-claim verification on the authored task suite, with the open-weight claims under two agent harnesses (MSA: mini-swe-agent; OC: OpenCode) and the frontier claims under mini-swe-agent. AUC and standardized pAUC over FPR $\in[0,0.05]$ of LeaFBench between the scores of substitutes (positives) and of honest services in all authorized deployments (negatives); the other metrics are in Appendix~\ref{app:results}. The best result in each column is in boldface; results of methods that are not valid at $\alpha=5\%$ are in red. RUT is not applicable to the frontier claims (--), since it needs the weights of the claimed model to sample its reference responses and to score every token.}
\label{tab:main-leaf}
\scriptsize
\setlength{\tabcolsep}{3.0pt}
\begin{tabular}{llccccccccccc}
\toprule
 & & \multicolumn{2}{c}{Qwen3.5-27B} & \multicolumn{2}{c}{Qwen3-Coder} & \multicolumn{2}{c}{Devstral} & \multicolumn{5}{c}{Frontier} \\
\cmidrule(lr){3-4} \cmidrule(lr){5-6} \cmidrule(lr){7-8} \cmidrule(lr){9-13}
Method & Metric & MSA & OC & MSA & OC & MSA & OC & GPT-5.4 & GPT-5.6 Sol & GPT-6.1 Sol & Sonnet 4.6 & Opus 5.5 \\
\midrule
\multirow{2}{*}{MET} & AUC$\uparrow$ & \textcolor{red}{0.591} & \textcolor{red}{0.618} & \textcolor{red}{0.655} & \textcolor{red}{0.655} & \textcolor{red}{0.647} & \textcolor{red}{0.647} & \textcolor{red}{0.764} & \textcolor{red}{\textbf{1.000}} & \textbf{1.000} & \textcolor{red}{0.833} & \textcolor{red}{0.812} \\
 & pAUC$\uparrow$ & \textcolor{red}{0.510} & \textcolor{red}{0.513} & \textcolor{red}{0.506} & \textcolor{red}{0.506} & \textcolor{red}{0.534} & \textcolor{red}{0.534} & \textcolor{red}{0.514} & \textcolor{red}{\textbf{1.000}} & \textbf{1.000} & \textcolor{red}{0.526} & \textcolor{red}{0.521} \\
\midrule
\multirow{2}{*}{LLMmap} & AUC$\uparrow$ & \textcolor{red}{0.390} & \textcolor{red}{0.427} & 0.873 & 0.873 & 0.712 & 0.712 & 0.863 & \textcolor{red}{0.877} & \textcolor{red}{0.928} & 0.848 & 0.992 \\
 & pAUC$\uparrow$ & \textcolor{red}{0.493} & \textcolor{red}{0.492} & 0.763 & 0.763 & 0.637 & 0.637 & 0.601 & \textcolor{red}{0.654} & \textcolor{red}{0.871} & 0.804 & 0.941 \\
\midrule
\multirow{2}{*}{RUT} & AUC$\uparrow$ & \textcolor{red}{0.541} & \textcolor{red}{0.539} & \textcolor{red}{0.701} & \textcolor{red}{0.701} & \textcolor{red}{0.472} & \textcolor{red}{0.472} & -- & -- & -- & -- & -- \\
 & pAUC$\uparrow$ & \textcolor{red}{0.487} & \textcolor{red}{0.487} & \textcolor{red}{0.506} & \textcolor{red}{0.506} & \textcolor{red}{0.516} & \textcolor{red}{0.516} & -- & -- & -- & -- & -- \\
\midrule
\multirow{2}{*}{KBF} & AUC$\uparrow$ & \textcolor{red}{0.663} & \textcolor{red}{0.693} & 0.864 & 0.864 & 0.730 & 0.730 & 0.699 & \textcolor{red}{0.571} & 0.857 & 0.833 & 0.812 \\
 & pAUC$\uparrow$ & \textcolor{red}{0.709} & \textcolor{red}{0.735} & 0.896 & 0.896 & 0.792 & 0.792 & 0.843 & \textcolor{red}{0.516} & 0.927 & 0.915 & 0.902 \\
\midrule
\multirow{2}{*}{IRIS} & AUC$\uparrow$ & \textcolor{red}{0.413} & \textcolor{red}{0.445} & \textcolor{red}{0.835} & \textcolor{red}{0.835} & 0.706 & 0.706 & 0.845 & \textcolor{red}{0.848} & 0.857 & \textcolor{red}{0.519} & 0.857 \\
 & pAUC$\uparrow$ & \textcolor{red}{0.491} & \textcolor{red}{0.495} & \textcolor{red}{0.866} & \textcolor{red}{0.866} & 0.701 & 0.701 & 0.841 & \textcolor{red}{0.893} & 0.927 & \textcolor{red}{0.707} & 0.927 \\
\midrule
\multirow{2}{*}{AgentProv} & AUC$\uparrow$ & \textcolor{red}{0.803} & \textcolor{red}{0.805} & \textcolor{red}{0.777} & \textcolor{red}{0.777} & \textcolor{red}{0.578} & \textcolor{red}{0.578} & \textcolor{red}{0.653} & \textcolor{red}{0.903} & \textcolor{red}{0.800} & \textcolor{red}{0.708} & \textcolor{red}{0.750} \\
 & pAUC$\uparrow$ & \textcolor{red}{0.522} & \textcolor{red}{0.522} & \textcolor{red}{0.516} & \textcolor{red}{0.516} & \textcolor{red}{0.503} & \textcolor{red}{0.503} & \textcolor{red}{0.506} & \textcolor{red}{0.553} & \textcolor{red}{0.519} & \textcolor{red}{0.509} & \textcolor{red}{0.513} \\
\midrule
\multirow{2}{*}{CertClaim} & AUC$\uparrow$ & \textbf{1.000} & \textbf{1.000} & \textbf{1.000} & \textbf{1.000} & \textbf{0.920} & \textbf{0.929} & \textbf{1.000} & \textbf{1.000} & \textbf{1.000} & \textbf{1.000} & \textbf{1.000} \\
 & pAUC$\uparrow$ & \textbf{0.999} & \textbf{0.999} & \textbf{1.000} & \textbf{1.000} & \textbf{0.890} & \textbf{0.886} & \textbf{1.000} & \textbf{1.000} & \textbf{1.000} & \textbf{1.000} & \textbf{1.000} \\
\bottomrule
\end{tabular}
\end{table*}

%% file: tables/breakdown.tex
\begin{table*}[t]
\centering
\caption{The AUC by type of authorized deployment change (left: negatives restricted to honest services in that type of deployment; positives are all substitutes) and by type of substitution (right: positives restricted to that type; negatives are all honest services), averaged over the settings in which the type occurs. A cell on the left is red when the method is not valid at $\alpha=5\%$ on the pooled honest services of that type.}
\label{tab:breakdown-leaf}
\scriptsize
\setlength{\tabcolsep}{3.2pt}
\begin{tabular}{lccccccccccc}
\toprule
 & \multicolumn{6}{c}{Authorized deployment change (honest services)} & \multicolumn{5}{c}{Substitution (substitutes)} \\
\cmidrule(lr){2-7}\cmidrule(lr){8-12}
Method & Default & Sampling & Reasoning & \begin{tabular}[b]{@{}c@{}}Sys.\\prompt\end{tabular} & Context & Precision & \begin{tabular}[b]{@{}c@{}}Other\\family\end{tabular} & \begin{tabular}[b]{@{}c@{}}Same\\developer\end{tabular} & \begin{tabular}[b]{@{}c@{}}Post-\\trained\end{tabular} & Quantized & Merged \\
\midrule
MET & \textcolor{red}{0.898} & \textcolor{red}{0.730} & \textcolor{red}{0.486} & \textcolor{red}{0.728} & \textcolor{red}{0.734} & \textcolor{red}{0.736} & 0.844 & 0.745 & 0.630 & 0.489 & 0.425 \\
LLMmap & \textcolor{red}{0.684} & 0.687 & \textcolor{red}{0.696} & \textcolor{red}{0.724} & \textcolor{red}{0.742} & \textcolor{red}{0.659} & 0.877 & 0.497 & 0.488 & 0.581 & 0.728 \\
RUT & 0.966 & \textcolor{red}{0.506} & \textcolor{red}{0.600} & \textcolor{red}{0.307} & \textcolor{red}{0.264} & \textcolor{red}{0.955} & 0.588 & 0.713 & 0.549 & 0.447 & 0.433 \\
KBF & 0.765 & \textcolor{red}{0.733} & \textcolor{red}{0.681} & 0.784 & 0.764 & 0.769 & 0.552 & 0.976 & 0.454 & 0.889 & 0.581 \\
IRIS & 0.732 & 0.748 & \textcolor{red}{0.408} & \textcolor{red}{0.676} & \textcolor{red}{0.720} & \textcolor{red}{0.704} & 0.794 & 0.606 & 0.500 & 0.483 & 0.698 \\
AgentProv & \textcolor{red}{0.964} & \textcolor{red}{0.845} & \textcolor{red}{0.636} & \textcolor{red}{0.678} & \textcolor{red}{0.482} & \textcolor{red}{0.965} & 0.766 & 0.766 & 0.833 & 0.555 & 0.626 \\
CertClaim & \textbf{0.992} & \textbf{0.986} & \textbf{1.000} & \textbf{0.994} & \textbf{0.993} & \textbf{0.988} & \textbf{1.000} & \textbf{1.000} & \textbf{1.000} & \textbf{0.998} & \textbf{0.877} \\
\bottomrule
\end{tabular}
\end{table*}

%% file: tables/ablation.tex
\begin{table}[t]
\centering
\caption{The effects of the components of CertClaim (one batch of tasks). Valid: the number of settings in which the variant is valid at $\alpha=5\%$. AUC, pAUC and Worst: the AUC, the pAUC and the AUC against the hardest substitute, averaged over all settings. Boundary and Routing: the detection rate (\%) at $\alpha$ of substitutes served in an authorized configuration and of mixed routing, averaged over the settings in which they are audited.}
\label{tab:ablation}
\scriptsize
\setlength{\tabcolsep}{1.6pt}
\begin{tabular}{lcccccc}
\toprule
Variant & Valid & AUC$\uparrow$ & pAUC$\uparrow$ & Worst$\uparrow$ & Boundary$\uparrow$ & Routing$\uparrow$ \\
\midrule
Static null (default only) & 2/18 & 0.822 & 0.793 & 0.467 & 100.0 & 91.7 \\
\midrule
In-harness only & 18/18 & 0.935 & 0.902 & 0.711 & 74.6 & 36.3 \\
Probes only & 18/18 & 0.988 & 0.984 & 0.948 & 94.2 & 86.1 \\
Channel-first fusion & 18/18 & 0.991 & 0.988 & 0.954 & 93.2 & 87.0 \\
CertClaim (cell-coherent) & 18/18 & 0.992 & 0.987 & 0.954 & 96.0 & 90.5 \\
\bottomrule
\end{tabular}
\end{table}

%% file: sections/conclusion.tex
\section{Conclusion}
\label{sec:conclusion}

In this paper, we studied model-claim verification for coding-agent services. We revealed that existing model auditing methods rely on a latent assumption that the deployment is static, so that they either falsely accuse honest providers that change an authorized setting or miss derivatives of the claimed model. Based on this understanding, we proposed a certified model-claim verification method (dubbed CertClaim). Instead of testing the service against one reference deployment, we made the authorized deployments part of the claim and tested whether some authorized deployment explains all the evidence of the service. We provided theoretical analyses of its false-accusation rate and detection condition, and conducted extensive experiments, which verified its effectiveness as well as its resistance to potential adaptive attacks. We hope that our work can provide a new angle on model auditing for agent services and facilitate the design of more trustworthy coding-agent services.

%% file: sections/appendix.tex

\section{Further Theoretical Results}
\label{app:theory}

In this section, we present further properties of our CertClaim, including its detection condition under Gaussian scores, its certified region between the cells of a continuous deployment parameter, the limit of any certified verifier, the fusion of evidence channels, exact null distributions from open weights, and sequential audits and mixed routing. Their proofs are in Appendix~\ref{app:proofs}.

\begin{example}[Detection Conditions under Gaussian Scores]
\label{ex:gaussian}
Let unit scores be $\mathcal{N}(\mu_g,\sigma^2)$ for the honest service in cell $g$ and $\mathcal{N}(\mu_S,\sigma^2)$ for the substitute, identically across strata. For large $B$, the cell-wise test of each channel is asymptotically the two-sample $z$-test, and the substitute is detected with probability at least $1-\beta$ if (asymptotically)
\[
\min_{g\in\mathcal{G}}|\mu_S-\mu_g|\ \ge\ \tau(K,n):=\sigma\bigl(z_{1-\alpha/(2J')}+z_{1-\beta/m}\bigr)\sqrt{\tfrac{1}{K|\mathcal{S}|}+\tfrac{1}{n|\mathcal{S}|}},
\]
where $z_{q}$ is the $q$-quantile of the standard Gaussian distribution.
\end{example}

In general, Example~\ref{ex:gaussian} indicates that a substitute with a larger separation from the nearest authorized cell is more likely to be detected under a fixed audit budget, and that a larger audit budget helps until the calibration size dominates.

\begin{corollary}[Certified Region for Continuous Deployment Parameters]
\label{cor:region}
Let the authorized set be $\Theta_c\subseteq\mathbb{R}^d$ and suppose that for every stratum $\mathrm{KL}\bigl(P^{s}_{M_c,\theta}\,\Vert\,P^{s}_{M_c,\theta'}\bigr)\le 2L^2\lVert\theta-\theta'\rVert^2$. If the grid $\mathcal{G}$ is an $h$-cover of $\Theta_c$, then for every $\theta\in\Theta_c$,
\[
\Pr_{M_c,\theta}(\text{reject})\ \le\ \alpha+\min\bigl\{\sqrt{K|\mathcal{S}|}\,L\,h,\ 1\bigr\}.
\]
\end{corollary}

\begin{example}[Temperature as a Contract Parameter]
\label{ex:temperature}
For softmax sampling with temperature $\tau$ from logits $z$, the per-token divergence satisfies $\mathrm{KL}(p_\tau\Vert p_{\tau'})\le\frac{(\tau-\tau')^2}{2\min(\tau,\tau')^4}\max_{\tilde\tau}\mathrm{Var}_{p_{\tilde\tau}}(z)$. With $\mathrm{Var}(z)\le v$, $\tau\ge\tau_{\min}$ and at most $T$ sampled tokens per unit, the chain rule gives the constant $L=\sqrt{Tv}/(2\tau_{\min}^2)$ in Corollary~\ref{cor:region}. A temperature grid of mesh $h$ therefore controls the false-accusation rate at $\alpha+\sqrt{K|\mathcal{S}|Tv}\,h/(2\tau_{\min}^2)$ over the whole authorized interval.
\end{example}

\begin{corollary}[Conditioning on an Observable Deployment Mode]
\label{cor:mondrian}
Let $\kappa(\cdot)$ be a statistic of the audit that is almost surely constant under each cell (e.g., whether the service bills hidden reasoning tokens). Taking the maximum in Proposition~\ref{def:verifier} only over $\{g:\kappa(g)=\kappa(\text{audit})\}$ preserves Theorem~\ref{thm:validity}.
\end{corollary}

\begin{theorem}[Indistinguishability Bound]
\label{thm:impossibility}
Let an audit comprise $N$ independent observations. For any $(\alpha,\mathcal{G})$-certified verifier and any substitute service $(S,\eta)$,
\[
\Pr_{S,\eta}(\text{reject})\ \le\ \alpha+\min_{g\in\mathcal{G}}\mathrm{TV}\bigl(P_{S,\eta}^{\otimes N},P_{M_c,g}^{\otimes N}\bigr)\ \le\ \alpha+\min_{g\in\mathcal{G}}\min\Bigl\{N\delta_{\mathrm{TV}}(g),\ \sqrt{\tfrac{N}{2}\,\delta_{\mathrm{KL}}(g)}\Bigr\},
\]
where $\delta_{\mathrm{TV}}(g)$ and $\delta_{\mathrm{KL}}(g)$ are the per-observation total-variation distance and Kullback--Leibler divergence between $(S,\eta)$ and $(M_c,g)$. Consequently, detecting $S$ with power $1-\beta$ requires $N\ge2(1-\alpha-\beta)^2/\min_g\delta_{\mathrm{KL}}(g)$, and enlarging the contract can only lower the attainable power.
\end{theorem}

In general, Theorem~\ref{thm:impossibility} indicates that no certified verifier can detect a substitute that is nearly indistinguishable from some authorized deployment on the audited workload, which is consistent with the failure of all methods on the merges of Devstral-Small-2507 at $\lambda\le0.25$ (Section~\ref{sec:exp-main}), and that the price of a larger contract is a lower attainable power.

\begin{theorem}[Fusion of Evidence Channels]
\label{thm:fusion}
Under arbitrary dependence between channels and probe audits, $V_\alpha$ is $(\alpha,\mathcal{G})$-certified, and for any substitute,
\[
\Pr(\text{reject})\ \ge\ \max\Bigl\{\max_{j\le J}\Pr\bigl(\textstyle\max_gp_{g,j}\le\tfrac{\alpha}{2(1+E)J}\bigr),\ \max_{e\le E}\Pr\bigl(\max_gp^{(e)}_g\le\tfrac{\alpha}{1+E}\bigr)\Bigr\}.
\]
Adding a channel therefore cannot reduce power below that of any single channel run at the corresponding level, and by Theorem~\ref{thm:power} it enlarges the leading term of the radius of a channel by at most the factor $\sqrt{\log(8J'/\alpha)/\log(8/\alpha)}$.
\end{theorem}

\begin{theorem}[Exact Nulls from Open Weights]
\label{thm:pivot}
Let the weights of the claimed model be available to the auditor.
(1) (Local simulation.) For every cell $g$, the auditor can draw the calibration data of Definition~\ref{def:calibration} from $P^{s}_{M_c,g}$ itself, so Theorem~\ref{thm:validity} holds without access to any trusted deployment. In particular, for a prompt $x$, a suspect response with score $t$ and $m$ locally simulated scores $t_1,\dots,t_m$ in cell $g^\star$, the randomized rank $\bigl(\#\{i:t_i<t\}+U\cdot(1+\#\{i:t_i=t\})\bigr)/(m+1)$ with $U\sim\mathcal{U}(0,1)$ is exactly uniform on $(0,1)$.
(2) (Rosenblatt pivot.) If the service samples token $y_t\sim q_t=\mathrm{Samp}_\theta\bigl(p_{M_c}(\cdot\mid h_t)\bigr)$ and the auditor knows the context $h_t$ and the sampling transformation, the randomized probability integral transforms $u_t=Q_t(y_t^-)+V_tq_t(y_t)$, $V_t\overset{iid}{\sim}\mathcal{U}(0,1)$, are i.i.d.\ $\mathcal{U}(0,1)$ for an honest service, with no calibration data at all. If the numerics of the provider differ from those of the auditor by at most $\varepsilon_t$ in total variation at step $t$, any level-$\alpha$ test of uniformity has a false-accusation rate of at most $\alpha+\sum_t\varepsilon_t$.
\end{theorem}

In general, Theorem~\ref{thm:pivot}(1) is the reason why the token-rank channel R is exact within every cell, provided that the local responses are drawn in the same way as the audited ones (i.e., one sample per request, Appendix~\ref{app:details}), and Theorem~\ref{thm:pivot}(2) bounds the effect of numerical differences between serving stacks.

\begin{theorem}[Sequential Audits and Mixed Routing]
\label{thm:sequential}
(1) (Anytime validity.) Let $p^{(r)}=\max_gp^{(r)}_g$ be the $p$-value of the verifier in round $r$, where successive rounds use fresh audit units and disjoint calibration data. For any $\kappa\in(0,1)$, $E_t=\prod_{r\le t}\kappa\,(p^{(r)})^{\kappa-1}$ is a non-negative supermartingale under every authorized cell, and rejecting at the first $t$ with $E_t\ge1/\alpha$ falsely accuses with probability at most $\alpha$, however the stopping time is chosen.
(2) (Mixed routing.) If each session is served by the substitute independently with probability $\varepsilon$ and by the claimed model in cell $g^\star$ otherwise, the separation of Theorem~\ref{thm:power} from cell $g^\star$ is at most $\varepsilon\Delta_j(g^\star)$ in every channel $j$, so certified detection requires $r_j(K,n)\le\varepsilon\Delta_j(g^\star)$ for some $j$, i.e., an audit budget $K=\Omega\bigl(c_j^2|\mathcal{S}|\log(J'm/\alpha\beta)/(\varepsilon\Delta_j(g^\star))^2\bigr)$. If routing is instead decided independently for each of the $N$ calls of an audit, no certified verifier can exceed power $\alpha+1-(1-\varepsilon)^{N}$.
\end{theorem}

In general, Theorem~\ref{thm:sequential}(2) indicates that the routed fraction that can be detected shrinks as $K^{-1/2}$, which is consistent with the increase of the detection rate of mixed routing from one to four batches of tasks in Figure~\ref{fig:adaptive}(a).

\input{sections/appendix_proofs}

\section{Detailed Settings}
\label{app:details}

\subsection{Contracts and Substitutes}

Table~\ref{tab:contracts} lists the deployment contract and the substitutes of every claim. The cells of the open-weight claims follow their model cards (e.g., the thinking preset of Qwen3.5-27B is temperature $1.0$, top-$p$ $0.95$, top-$k$ $20$ and presence penalty $1.5$, and its non-thinking presets switch thinking off with the general or the reasoning sampling preset), common serving practice (a generic and a role system prompt prepended by the provider, and an injected project memory) and official low-precision checkpoints. The cells of the frontier claims use the reasoning efforts or the temperatures that their APIs expose. The 4-bit derivative of Qwen3.5-27B is its official GPTQ checkpoint, the other post-trained derivatives of Qwen3.5-27B are public community or vendor checkpoints, and the quantized (round-to-nearest) and merged derivatives of the two coding claims are produced by us from the public weights. Devstral-Small-2507 is merged with its predecessor Devstral-Small-2505 and Qwen3-Coder-30B-A3B-Instruct with Qwen3-30B-A3B-Instruct-2507. Besides, every claim has a few benign deployment changes that are not authorized by its contract (e.g., an unauthorized temperature of $1.3$), which are not used in any metric. On the EvalPlus suite, the merged and quantized derivatives of the two coding claims are not audited.

\begin{table*}[t]
\centering
\caption{The deployment contract (authorized cells) and the substitutes of every claim.}
\label{tab:contracts}
\scriptsize
\setlength{\tabcolsep}{3pt}
\begin{tabular}{p{2.0cm}p{5.0cm}p{6.3cm}}
\toprule
Claim & Authorized cells & Substitutes \\
\midrule
Qwen3.5-27B & default; thinking preset; non-thinking presets (general, reasoning); generic and role system prompts; online FP8; official FP8 checkpoint; injected project memory & gpt-oss-20b, Qwen3-32B, Qwen3.5-9B, Qwen3.5-4B, Qwen3.6-27B, Devstral-Small-2507 (OpenCode and EvalPlus); agentic fine-tune, reasoning-distilled, refusal-ablated and behavior-preserving derivatives; official 4-bit (GPTQ) checkpoint \\
Qwen3-Coder-30B-A3B-Instruct & default; sampling preset of the model card; temperature $0.3$; generic and role system prompts; injected project memory; official FP8 checkpoint & Devstral-Small-2507, gpt-oss-20b, Qwen3-30B-A3B-Instruct-2507, Qwen3.5-27B, Qwen3.5-9B; 4-bit and 6-bit round-to-nearest; merges with Qwen3-30B-A3B-Instruct-2507 ($\lambda=0.1,0.25,0.5$) \\
Devstral-Small-2507 & default; temperatures $0.15$ and $0.7$; generic and role system prompts; injected project memory; FP8 & gpt-oss-20b, Mistral-Small-24B-Instruct-2501, Qwen3-Coder-30B-A3B-Instruct, Qwen3.5-27B, Qwen3.5-9B; 4-bit and 6-bit round-to-nearest; merges with Devstral-Small-2505 ($\lambda=0.1,0.25,0.5$) \\
GPT-5.4 & default; low and medium reasoning efforts; generic and role system prompts; injected project memory & DeepSeek-V4-Flash, GLM-5.2, GPT-5.2, GPT-5.4 Mini, GPT-5.4 nano, Qwen3.5-Flash \\
GPT-5.6 Sol & as GPT-5.4 & DeepSeek-V4-Flash, GLM-5.2, GPT-5.4, GPT-5.4 Mini, GPT-5.5, GPT-5.6 Terra, Qwen3.5-Flash \\
GPT-6.1 Sol & as GPT-5.4 & DeepSeek-V4-Flash, GLM-5.2, GPT-5.4, GPT-5.4 Mini, GPT-5.6 Sol, GPT-6 Sol, Qwen3.5-Flash \\
Claude Sonnet~4.6 & default; temperatures $0.7$ and $0.3$; generic and role system prompts; injected project memory & Claude Haiku~4.5, Claude Sonnet~4.5, DeepSeek-V4-Flash, GLM-5.2, Kimi K2.6, MiniMax-M2.5 \\
Claude Opus~5.5 & as Sonnet~4.6, plus adaptive thinking and low effort & Claude Haiku~4.5, Claude Opus~5, Claude Sonnet~4.6, DeepSeek-V4-Flash, GLM-5.2, Kimi K2.6, MiniMax-M2.5 \\
\bottomrule
\end{tabular}
\end{table*}

\FloatBarrier
\input{sections/appendix_models}

\subsection{Serving and Task Suites}

\noindent\textbf{Serving.} Open-weight models are served with vLLM~\citep{kwon2023vllm} on H800 GPUs, one model replica per GPU, behind a provider shim controlled by us. The shim applies the provider-side configuration of a cell (sampling overrides, reasoning mode, system prompt or context injection) to every request, opens one session per episode, and records what the harness sends and receives. API models are reached through a commercial gateway under their official API model IDs, except that GPT-6.1 Sol is served under the gateway alias \texttt{gpt-6.1-sol-coding}, and the shim translates between the chat and messages dialects. Every episode runs in a fresh sandbox.

\vspace{0.3em}
\noindent\textbf{Authored Suite.} A task schema defines a small repository (Python or JavaScript; single-file, multi-file, configuration and API-migration repairs) with a function to implement or repair, visible tests and hidden test cases. The \emph{verify} axis pairs a prompt without and with an explicit instruction to run the tests before finishing. The \emph{recover} axis pairs a clean environment with one in which the first test invocation fails with a transient, retry-safe error. The \emph{resolve} axis pairs consistent visible tests with a variant in which one visible test is stale and conflicts with the specification, which the agent is told to follow. Schemas are split into a pool for fitting the channels and enrolling public models and a disjoint pool for calibration and audits, and every batch draws one schema per axis at random with fresh constants and test cases. An episode has a wall-clock limit of 25 minutes and at most 50 model calls.

\vspace{0.3em}
\noindent\textbf{EvalPlus Suite.} The problems of HumanEval+ and MBPP+~\citep{liu2023evalplus} are turned into the same three paired axes: the agent completes a function stub in a repository with three visible test cases, and a hidden check runs all base and extended EvalPlus tests. A problem is kept only if its reference solution passes the hidden check and the visible tests, the stub fails both, and the stale visible test of the \emph{resolve} axis truly conflicts with the reference, which keeps 524 of the 542 problems.

\vspace{0.3em}
\noindent\textbf{Real-issue Suite.} We take the issues of 11 projects in the test split of SWE-bench~\citep{jimenez2024swebench} whose repository version has a source release, and use the source tree of that version with a fresh one-commit history as the fixture. The prompt is the issue text exactly as filed, so nothing about a task marks it as an audit. The strata are groups of projects (Django; scientific Python; tooling and web), and the two episodes of a unit are two different issues of the same stratum. An episode has at most 50 model calls and 1{,}500 seconds.

\subsection{Evidence Channels}

The workflow channel A is the negative log-likelihood of 101 descriptors of semantic actions (e.g., counts of and transitions between inspecting, editing, testing and recovering) under the claimed model. The usage channels U and Uv are the standardized distance of the number of calls, the visible characters per call and (for U only) the output tokens per call reported by the service from those of the claimed model, so that Uv cannot be moved by misreporting usage. The identity channel L is the likelihood ratio of 126 behavioral fields between the best-matching enrolled public model and the claimed model, with every public model of the run enrolled except the audited substitute. All scorers that use a categorical model discretize a field into at most three states (terciles of the fitting data, or its distinct values when there are at most eight) and use add-one-half smoothing. The in-session probe channel C is the surprisal of the response to an in-session request, which copies the conversation and asks for 290--330 integers between 1 and 355 (capped at 4{,}096 output tokens), under a reference distribution of the claimed model; it is not available for the Qwen3.5-27B claim on the authored suite, whose data were collected before the probe was added. The reference-likelihood channel B is the mean surprisal excess of all generated tokens (dubbed Ball) and of the action tokens inside tool calls (dubbed Bact), computed with the prompt log-probabilities of the claimed checkpoint over the conversation exactly as the harness rendered it, where tokens whose top-1 probability exceeds 0.9 are excluded as uninformative.

\subsection{Probe Channels and Their Validity Check}

\noindent\textbf{Token-rank Probe R.} For every cell, the auditor generates a local bank of 50 responses of 30 tokens to each of 400 WildChat prompts~\citep{zhao2024wildchat} (1{,}600 for Devstral) at two client temperatures ($0.5$ and $1.0$) from the public checkpoint, and an audit sends every prompt once at each temperature. The bank must be drawn in the same way as the audit (i.e., one sample per request): in our experiments, a bank drawn with many parallel samples per request broke the exact null of a near-greedy cell (temperature $0.15$) at 1{,}600 prompts, which drawing the bank with one sample per request restored.

\vspace{0.3em}
\noindent\textbf{Contract-calibrated Probes of Existing Methods.} A contract-calibrated probe (dubbed `-C') replaces the single reference of an existing method with banks of honest audits of every cell and takes the maximum $p$-value over cells. We admit such a probe as a channel of CertClaim only if, on held-out honest audits, the proportion of own-cell $p$-values at most $\alpha/(1+E)$ has a 95\% Clopper--Pearson lower bound of at most $\alpha/(1+E)$. Under this rule, KBF-C is admitted for every claim, and the official MET-C and IRIS-C are admitted for the frontier claims, whereas AgentProv-C is not admitted, since its sample-level test is anti-conservative under contract calibration (6.0\% of its own-cell $p$-values are at most $\alpha/(1+E)$). This rule was fixed after a diagnosis of the honest rejections of an earlier version of the verifier that fused AgentProv-C.

\subsection{Tests and Audits}

The joint $p$-value of a cell combines all in-harness channels within the same permutation test, as follows.

\begin{definition}[Joint Cell $p$-value]
\label{def:joint}
For $b\in\{0,\dots,B\}$, let $r^{(b)}_{g,j}=\#\{b'\le B:D^{\pi_{b'}}_{g,j}\ge D^{\pi_b}_{g,j}\}/(B+1)$ be the rank of draw $b$ in channel $j$, $M^{(b)}_g=\min_jr^{(b)}_{g,j}$ (min-$p$) and $S^{(b)}_g=\sum_j\log r^{(b)}_{g,j}$ (Fisher combination). With $\rho(x^{(b)})=\#\{b':x^{(b')}\le x^{(b)}\}/(B+1)$ and $F^{(b)}_g=\min\{\rho(M^{(b)}_g),\rho(S^{(b)}_g)\}$, the joint $p$-value of cell $g$ is $p^{\mathrm{in}}_g=\#\{b\le B:F^{(b)}_g\le F^{(0)}_g\}/(B+1)$.
\end{definition}

All permutation tests use $B=1{,}999$ Monte-Carlo draws shared by all channels and cells. Validity holds for any $B$ (Theorem~\ref{thm:validity}), while the sufficient condition of Theorem~\ref{thm:power} asks for more draws and is conservative. A cell enters the contract only if each of its strata has at least four calibration units. For every setting, the honest batches of every cell are re-split ten times at random into a calibration set of the original size and audit batches, and all rates are computed over the ten splits, since Theorem~\ref{thm:validity} is marginal over the calibration data. An audit with $K>1$ batches is a random set of $K$ batches of the same service within one cell. Every in-harness audit is paired with a distinct probe audit of the same service and cell (i.e., no probe audit is reused), so that the fused audits are disjoint.

\subsection{Baselines}

\noindent\textbf{MET}~\citep{gao2025met}: the official Wikipedia-continuation prompts with the published budget of 25 prompts (five per language) and 10 samples each, and the MMD test with the Hamming kernel and its permutation $p$-value from the official package. \textbf{LLMmap}~\citep{pasquini2025llmmap}: the official pretrained open-set model and its eight queries; every public model is enrolled with templates from 100 prompt configurations of the official training pool, and a claim is rejected if a template of another model is nearer than every template of the claimed model, with the audited substitute not enrolled. \textbf{RUT}~\citep{rut2026}: 100 WildChat prompts~\citep{zhao2024wildchat} with one response each, 100 local reference responses per prompt (temperature $0.5$, 30 tokens), randomized ranks of the log-probability and the Cram\'er--von Mises test against the uniform distribution. \textbf{KBF}~\citep{kbf2026}: the released code and probe sets with the reference self-calibration re-run against the claimed model in its default deployment, and the 99\% binomial bound of the paper. \textbf{IRIS}~\citep{iris2026}: the official live-audit protocol with three random-sequence probes over 80 queries, an honest window of 120 responses, the content-only random-forest tell score and the one-sided two-proportion test at the level of the paper. \textbf{AgentProv}~\citep{agentprov2026}: 20 tool-choice templates with 50 samples each and a permutation test on the categorical fingerprints. All requests go through the same provider shim as the audits of CertClaim.

\subsection{Evaluation Metrics}

The score of an audit is $-\log_{10}p$ for methods that output a $p$-value and the score of the method itself otherwise (i.e., the distance margin of LLMmap and the likelihood ratio of LIDAR). The pAUC is the standardized partial AUC of scikit-learn over $\mathrm{FPR}\in[0,0.05]$, as in LeaFBench~\citep{shao2025sok}. For MD, $p$-values are floored at $1/401$ so that the exact zeros of some methods do not dominate. ACC is the accuracy of the decision of each method at $\alpha$ over positives and negatives. When a setting has fewer than 100 honest audits, TPR@1\%FPR equals the TPR at zero FPR. In the breakdown by type of authorized deployment change, the honest audits of all settings in which the type occurs are pooled for validity, where the audits of a baseline shared by the settings of one claim (which do not depend on the harness or the task suite) are counted once.

\section{Additional Results}
\label{app:results}

\subsection{The Distribution of the $p$-values of Honest Services}

\begin{figure*}[!htbp]
\centering
\includegraphics[width=\textwidth]{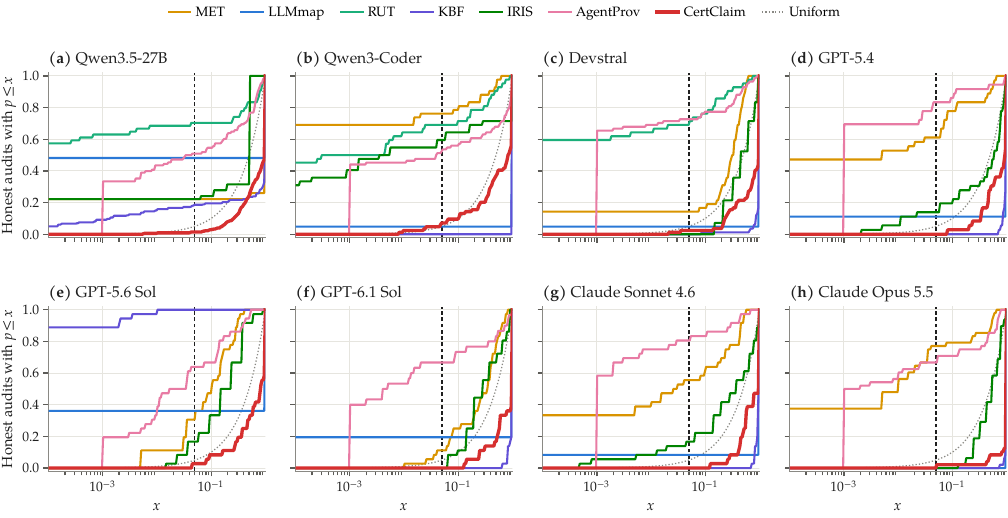}
\vspace{-1.4em}
\caption{The empirical cumulative distribution of the $p$-values of honest audits in all authorized deployments (authored task suite, mini-swe-agent). The dotted diagonal is the uniform distribution and the dashed line marks $\alpha=5\%$. LLMmap outputs a binary decision, which is shown as a $p$-value of 0 or 1.}
\label{fig:calibration}
\end{figure*}

Figure~\ref{fig:static}(a) reports the false-positive rate of every existing method in each authorized configuration of the Qwen3.5-27B claim. Figure~\ref{fig:calibration} further shows the empirical distribution of the $p$-values of the honest audits in all authorized deployments on the eight claims. A method is valid at $\alpha$ only if its curve at $x=\alpha$ stays below the diagonal up to sampling error. The curves of existing methods jump far above the diagonal on most claims, i.e., they reject a large proportion of honest services at any significance level, whereas the curve of CertClaim stays below the diagonal on every claim, since its $p$-value is the maximum of $p$-values that are valid within every cell (Theorem~\ref{thm:validity}).

\subsection{Results against Each Substitute}

\begin{figure*}[!htbp]
\centering
\includegraphics[width=\textwidth]{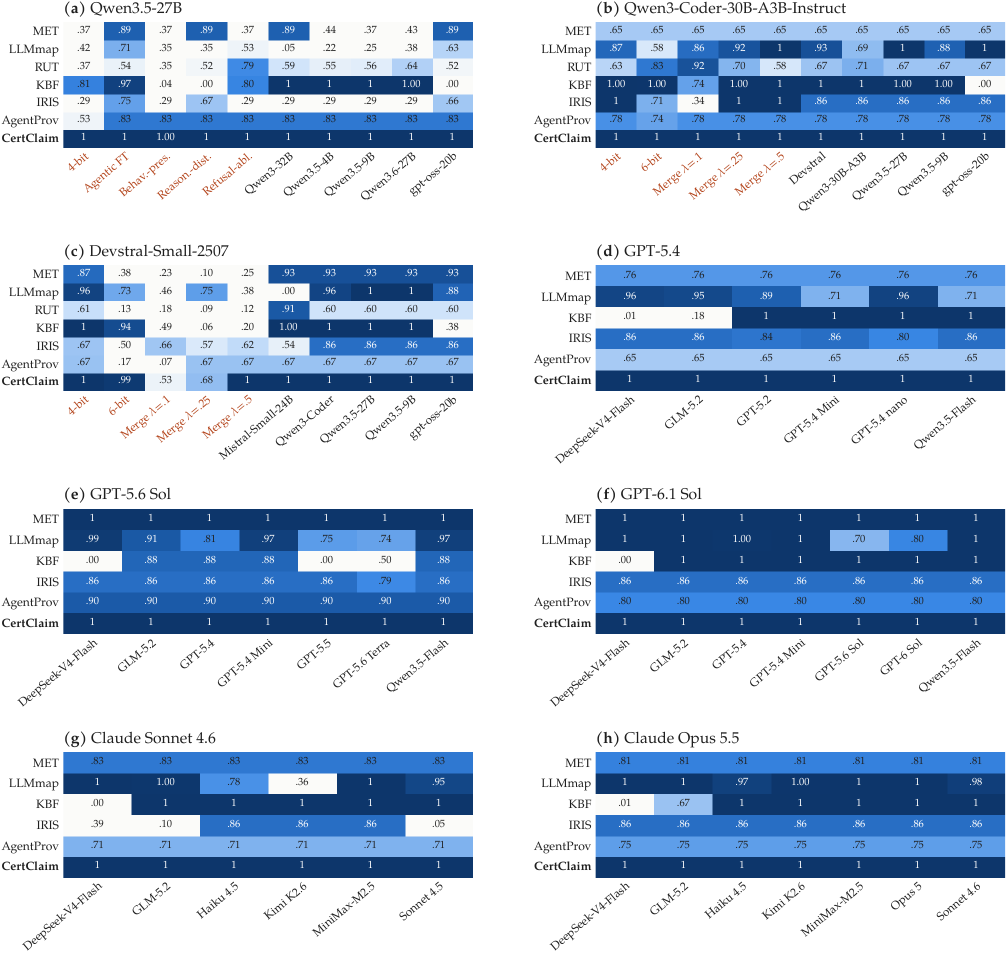}
\vspace{-1.4em}
\caption{The AUC of every method against each substitute (positives) and the honest services in all authorized deployments (negatives) on the eight claims (authored task suite, mini-swe-agent). Derivatives of the claimed weights are labeled in orange; white cells are at or below chance level.}
\label{fig:per-substitute}
\end{figure*}

Figure~\ref{fig:per-substitute} reports the AUC of every method against each substitute of the eight claims. Existing methods separate some substitutes well on some claims, but each of them fails on some derivative of the claimed weights (e.g., MET, RUT and IRIS on the 4-bit and the behavior-preserving derivatives of Qwen3.5-27B), whereas CertClaim achieves an AUC of at least 0.98 against every substitute except the merges of Devstral-Small-2507 with its predecessor at $\lambda\le0.25$.

\subsection{Breakdown by Type of Deployment Change and Substitution}

Table~\ref{tab:breakdown-full} extends Table~\ref{tab:breakdown-leaf} with the pAUC.

\input{tables/appendix_breakdown}

\subsection{Results on the EvalPlus and Real-issue Suites}

As shown in Table~\ref{tab:suites-leaf}, CertClaim achieves an AUC of 1.000 in all seven settings of the EvalPlus and real-issue suites, with an FPR of 1.7\%--6.0\%, whereas no existing method is valid on the Qwen3.5-27B claim and the best valid existing method achieves an AUC of at most 0.900 on the other claims.

\input{tables/per_setting_suites}

\begin{figure*}[!htbp]
\centering
\includegraphics[width=\textwidth]{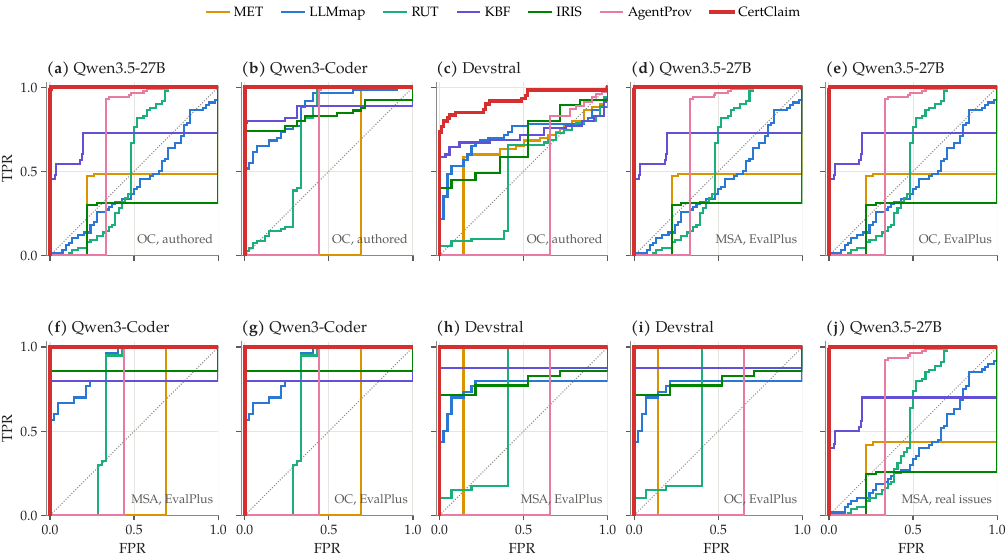}
\vspace{-1.4em}
\caption{ROC curves of model-claim verification on the remaining ten settings (harness and task suite in the lower right corner of each panel). The audits of existing methods do not depend on the harness or the task suite, so their curves are shared by the settings of the same claim.}
\label{fig:roc-more}
\end{figure*}

Figure~\ref{fig:roc-more} shows the ROC curves of the remaining ten settings, i.e., the open-weight claims under OpenCode on the authored suite, all open-weight settings on the EvalPlus suite and the Qwen3.5-27B claim on the real-issue suite.

\subsection{The Operating Point at the Significance Level}

Table~\ref{tab:operating} reports the false-positive rate and the mean detection rate over the whole-model substitutes of every method at $\alpha$. Existing methods that detect most substitutes (e.g., RUT and AgentProv) falsely accuse 50.9\%--83.3\% of the honest services, whereas the valid ones (e.g., KBF on the coding claims) detect 53.8\%--87.5\% of the substitute audits. In contrast, CertClaim detects 100\% of the substitute audits in all settings except those of the Devstral claim on the authored suite (80.0\%, where the misses are its merges at $\lambda\le0.25$ and one audit of its 6-bit derivative), with an FPR of 0\%--7.1\%.

\input{tables/appendix_operating}

\subsection{Effects of the Probe Budget}

On the Devstral claim, we vary the number of prompts of the token-rank probe R from 400 to 800 and 1{,}600 (with the banks of every cell drawn with one sample per request). The false-positive rate of channel R alone on the honest audits is 6.0\%, 2.4\% and 1.2\%, and its detection rate is 8.3\%, 33.3\% and 91.7\% on the 6-bit derivative and 100\% on the 4-bit derivative for every budget. In contrast, its detection rate on the merges at $\lambda=0.1$, $0.25$ and $0.5$ stays at most 8.3\%, 0\% and 41.7\% at every budget, which is consistent with their small separation from the claimed model. We use 1{,}600 prompts for the Devstral claim and 400 prompts for the other open-weight claims.

\subsection{Adaptive Attacks}

\noindent\textbf{Contract-boundary Substitution and Mixed Routing.} Table~\ref{tab:adaptive-all} reports the results of every setting. On the Qwen3.5-27B claim, the only contract-boundary substitute that is not always detected is the behavior-preserving derivative (66.7\% in four settings), and on the Devstral claim the misses are its merged derivatives. The hardest case of mixed routing is routing 30\% of the calls of GPT-5.6 Sol to GPT-5.4 Mini (50.0\% with one or four batches of tasks).

\input{tables/appendix_adaptive}

\vspace{0.3em}
\noindent\textbf{Probe-aware Routing.} Table~\ref{tab:probe-aware} reports the sweep over the fraction $\rho$ of the recognized probe requests, simulated per prompt on the stored audits of channel R (channel K is left out, since its answers are not stored per request). The substitutes that the in-harness channels separate from the claimed model are detected at every $\rho$, whereas the 4-bit and the behavior-preserving derivatives evade the verifier once most of the probe requests are recognized.

\input{tables/appendix_probe_aware}

\vspace{0.3em}
\noindent\textbf{Audit Recognition.} We train a logistic regression on character $n$-grams of the task text to separate audit tasks from real requests (20{,}532 SWE-bench issues and 4{,}012 WildChat coding requests), given $10$, $100$ or $1{,}000$ leaked audit sessions. On the authored and EvalPlus suites, with ten leaked audits, the classifier recognizes 99.6\%--100\% of the audits at 1\% FPR, even for audits drawn from task schemas that it has never seen. For the real-issue suite, the negatives are the other 1{,}599 issues of the same 11 projects: with 10, 100 and 300 leaked audits, the AUC of the classifier is 0.518, 0.528 and 0.536 and its TPR@1\%FPR is at most 1.1\%. With issues of other projects as negatives, the classifier achieves an AUC of 0.81--0.98, i.e., it recognizes the vocabulary of a project rather than the audit. Accordingly, an auditor should draw new issues from the projects of the audited traffic, rather than reuse a fixed public list.

\subsection{The Certified Detection Region in Other Settings}

\begin{figure*}[!htbp]
\centering
\includegraphics[width=\textwidth]{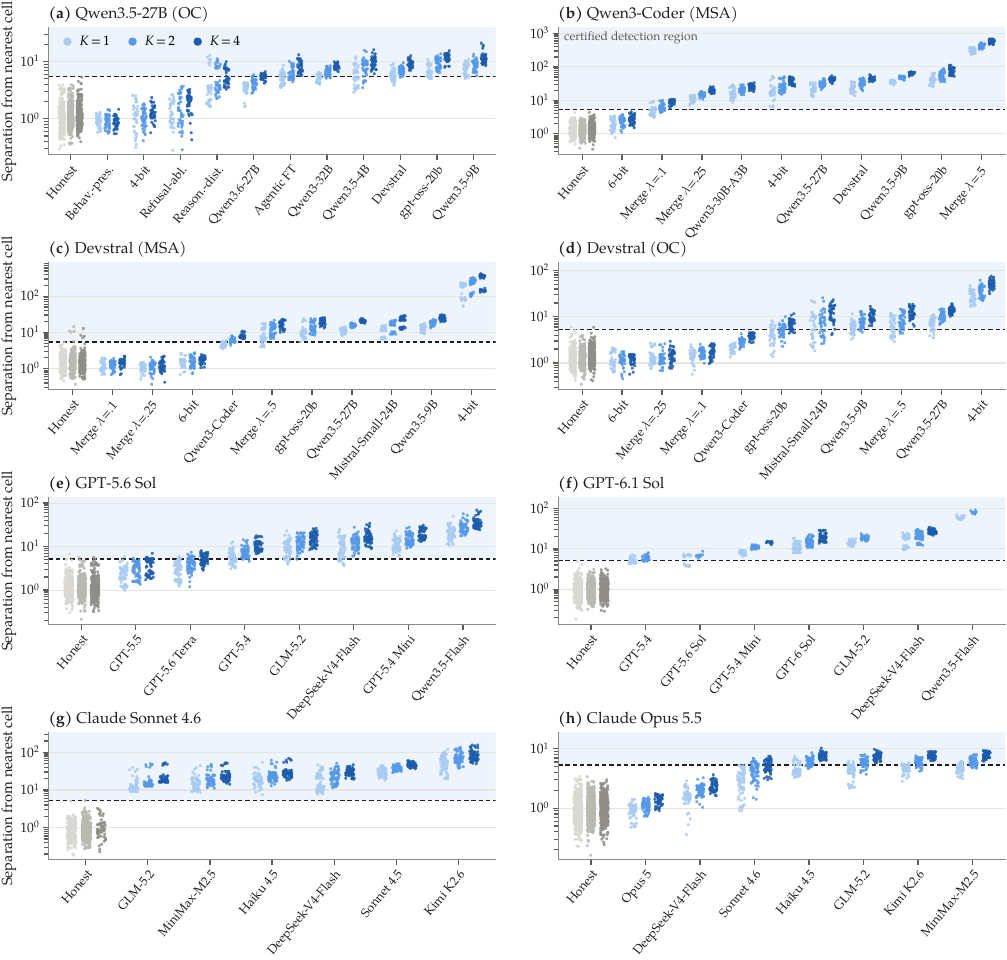}
\vspace{-1.4em}
\caption{The certified detection region of CertClaim on the in-harness channels in the other eight settings of the authored suite (as in Figure~\ref{fig:certified}).}
\label{fig:certified-more}
\end{figure*}

Figure~\ref{fig:certified-more} extends Figure~\ref{fig:certified} to the other eight settings of the authored suite. In every setting, the separation of honest services stays around one, while that of most substitutes grows with the audit budget and exceeds the threshold. The exceptions are the nearest derivatives of the open-weight claims (e.g., the 4-bit and the behavior-preserving derivatives of Qwen3.5-27B, and the 6-bit derivatives and the merges at $\lambda\le0.25$ of the two coding claims) and the closest siblings of the frontier claims (e.g., GPT-5.5 and GPT-5.6 Terra under GPT-5.6 Sol, and Claude Opus~5 under Claude Opus~5.5). The probe channels detect most of them in most audits, except for the merges of Devstral-Small-2507 at $\lambda\le0.25$ (Section~\ref{sec:exp-main}).

\subsection{Visualizing the In-harness Evidence of Other Claims}

\begin{figure*}[!htbp]
\centering
\includegraphics[width=\textwidth]{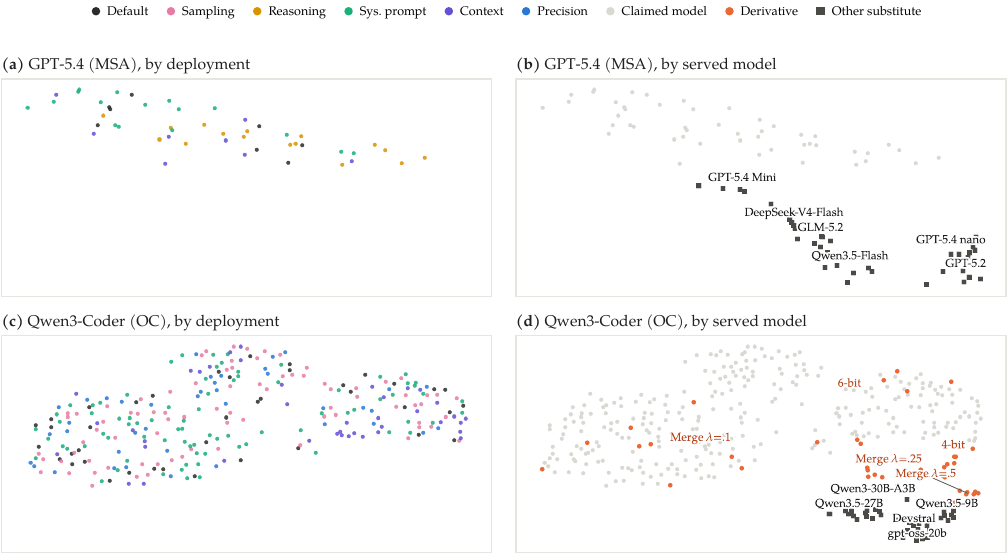}
\vspace{-1.4em}
\caption{The t-SNE visualization of the in-harness evidence on the GPT-5.4 claim (mini-swe-agent) and on the Qwen3-Coder-30B-A3B-Instruct claim (OpenCode), as in Figure~\ref{fig:tsne}.}
\label{fig:tsne-more}
\end{figure*}

Figure~\ref{fig:tsne-more} repeats the visualization of Figure~\ref{fig:tsne} on the GPT-5.4 claim under mini-swe-agent and on the Qwen3-Coder-30B-A3B-Instruct claim under OpenCode. On both claims, the substitutes of other models form their own clusters, whereas the nearest derivatives of Qwen3-Coder-30B-A3B-Instruct (e.g., its 6-bit derivative and its merge at $\lambda=0.1$) lie inside the cloud of honest batches, as on Qwen3.5-27B.

\subsection{Re-implementation of the Concurrent LIDAR Fingerprint}
\label{app:lidar}

LIDAR~\citep{lidar2026} is a concurrent work without released code, which identifies the model behind a coding harness among a set of enrolled models rather than verifying a claim. For reference, we re-implement its 958 behavioral fields and its Jeffreys classifier on our paired tasks, enroll every public model under its default deployment (with the audited substitute not enrolled), and score an audit by the likelihood ratio between the best-matching enrolled model and the claimed model. With its native decision (i.e., rejecting a claim when the predicted model is not the claimed one), LIDAR falsely accuses up to 92.7\% of the honest audits (e.g., 51.1\% on Qwen3.5-27B under mini-swe-agent), since it confuses authorized deployments with derivatives in its registry. With a threshold at the $(1-\alpha)$ quantile of the honest audits (an oracle that is tuned on the test audits, dubbed LIDAR$^{\dagger}$ in Tables~\ref{tab:app-leaf-qwen35-27b}--\ref{tab:app-leaf-qwen35-27b-swe}), it achieves an average AUC of 0.813, pAUC of 0.714 and TPR@1\%FPR of 0.398 over the 18 settings, compared with 0.992, 0.987 and 0.970 of CertClaim. Our identity channel L uses the per-episode part of these features inside the certified test, so that its false-accusation rate is controlled.

\subsection{Results of Every Method in Every Setting}

Tables~\ref{tab:app-leaf-qwen35-27b}--\ref{tab:app-leaf-qwen35-27b-swe} report all metrics of every method in every setting, including LIDAR$^{\dagger}$, the contract-calibrated versions of existing methods, our single channels and the variants of CertClaim.

\clearpage
\input{tables/appendix_settings_leaf}

\clearpage

\section{Discussions on the Closest Prior Work}
\label{app:relation}

\noindent\textbf{Fingerprints of Coding Agents.} LIDAR~\citep{lidar2026} answers the question of which enrolled model is behind a harness by closed-set classification with one enrollment per model, whereas CertClaim answers the question of whether the claimed model, in any authorized deployment, is behind the service, with a controlled false-accusation rate. We reuse behavioral fields of this kind as one channel. The differences are the composite null hypothesis over a contract, the stratified test of paired units that makes calibration feasible with a few batches per cell, and the additional channels that reach derivatives which behave like the claimed model.

\vspace{0.3em}
\noindent\textbf{Robust Probes.} KBF, AgentProv and IRIS~\citep{kbf2026,agentprov2026,iris2026} obtain robustness from the choice of evidence. CertClaim obtains validity from calibration, for any evidence, and can therefore absorb their probes as contract-calibrated channels (Theorem~\ref{thm:fusion}).

\vspace{0.3em}
\noindent\textbf{Equality Tests.} MET and RUT~\citep{gao2025met,rut2026} test equality with one reference configuration. Our token-rank channel uses a rank statistic of the same kind, calibrated per cell, with two client temperatures and four times as many prompts. Proposition~\ref{prop:paradox} explains why the uncalibrated tests accuse honest providers.

\vspace{0.3em}
\noindent\textbf{Certified Dataset Watermarks.} CertDW~\citep{qiao2025certdw} certifies dataset-ownership verification against bounded perturbations of the verification samples via conformal prediction. We share the logic of a certificate that must survive a specified set of benign changes. The object differs (a model claim instead of a watermark), the set of benign changes is a deployment contract instead of a norm ball, and the test is a composite-null randomization test with a detection condition and a matching indistinguishability bound.

\section{Potential Limitations and Future Directions}
\label{app:limitations}

Firstly, the validity of CertClaim covers the deployments listed in the contract and the deployments close to them (Corollary~\ref{cor:region}), so a provider that uses an unlisted configuration (e.g., an unauthorized temperature) may be flagged until the contract is extended. The contract therefore has to be agreed upon between the provider and its users, e.g., in a service agreement. Secondly, Theorem~\ref{thm:validity} requires the calibration data of every cell to be exchangeable with the honest audits in that cell. In a few cells, this premise appears to be violated, e.g., 25.0\% of the honest audits of the official FP8 cell of Qwen3-Coder under mini-swe-agent and, with four batches, 17.5\% of those of the context cell of Qwen3.5-27B on the EvalPlus suite under mini-swe-agent are rejected, although every setting remains valid. This is probably because the serving stack of these cells differs slightly between their calibration and audit batches; drift of the serving stack over time has the same effect, so calibration data should be refreshed together with the deployment. For closed claims, the reference is the first-party endpoint, so CertClaim detects deviations of a reseller from that endpoint rather than changes that the first party makes itself.

Thirdly, the nearest derivatives of a claim (e.g., its 4-bit quantization and a behavior-preserving edit) are separated from it only by the probe channels, so a provider that recognizes most of the probe requests and answers them with the claimed model evades CertClaim on these derivatives (Table~\ref{tab:probe-aware}). Our simulation assumes that the provider answers recognized probes in the default cell; in an exploratory analysis, a provider that instead answers them in the cell that best explains the in-harness behavior of its substitute reduces the detection rate of Qwen3.6-27B to 45\% with one batch, which is close to that of the in-harness channels alone. Fourthly, a provider may fine-tune a cheaper model on trajectories of the claimed model. In an exploratory experiment, a Qwen3.5-9B student fine-tuned on 8{,}000 trajectories of Qwen3.5-27B (35 minutes on eight GPUs) is detected by the in-harness channels in only 2.5\% and 75.0\% of the audits with one and four batches; it is still detected by the probe channels, but combined with probe forwarding in the best-matching cell, the detection rate of the fused verifier falls to 2.5\% and 35.0\%. Fifthly, no method in our experiments detects the merges of Devstral-Small-2507 with its predecessor at $\lambda\le0.25$, which lie below the radius of Theorem~\ref{thm:power} in every channel and may be indistinguishable on our workload (Theorem~\ref{thm:impossibility}). Making probes indistinguishable from agent traffic (e.g., by embedding them in the sessions of the harness, as the in-session probe C does), adding evidence channels that are sensitive to the weights of the claimed model in closed claims, and auditing with workloads that are closer to the traffic of real users are interesting directions for future work.

\section{Reproducibility Statement}
\label{app:repro}

The detailed settings of our experiments are in Section~\ref{sec:exp-setup} and Appendix~\ref{app:details}. Every agent episode and probe audit is collected from plan files that list its model, cell, role, task schema and seed. All tables are generated by scripts from the stored trajectories, probe audits and baseline audits (e.g., the metric files of every setting and the scripts that turn them into the tables of the LeaFBench metrics, the ablation and the adaptive attacks). We will release our code, the plan files and the task suites upon the acceptance of this paper. Open-weight checkpoints are public, and the results of the frontier claims depend on proprietary endpoints as served in October 2026.

\section{Ethics Statement}
\label{app:ethics}

Our audits query models with synthetic coding tasks, public problems of EvalPlus, public issues of SWE-bench and public WildChat prompts, and involve no personal data. The refusal-ablated and edited checkpoints are public models, which we use only as substitutes to be detected. API services were accessed through a paid commercial gateway under its terms of use, and every substitution in this paper was staged by us on our own infrastructure, so no real service is accused of substitution. A verifier of model claims can wrongly accuse honest providers, which is exactly the harm that this work quantifies and controls. We recommend that an accusation be reported together with its contract, significance level and audit budget.

%% file: sections/appendix_proofs.tex
\section{Proofs}
\label{app:proofs}

Throughout, $g^\star$ denotes the cell under which an honest service operates, $\mathcal{D}_{\mathrm{cal}}$ the
calibration data, and $\omega$ the auxiliary randomness of the verifier (the Monte-Carlo permutations). The audit
data are independent of $(\mathcal{D}_{\mathrm{cal}},\omega)$.

\subsection{Auxiliary Lemmas}

\begin{lemma}[Monte-Carlo randomization test]
\label{lem:perm}
Let $Z$ be a random element whose law is invariant under a finite group $\Pi$, let $D$ be any real statistic, and
let $\pi_1,\dots,\pi_B$ be i.i.d.\ uniform on $\Pi$, independent of $Z$. Then
$p=\bigl(1+\sum_{b}\mathbb{1}\{D(\pi_bZ)\ge D(Z)\}\bigr)/(B+1)$ satisfies $\Pr(p\le u)\le u$ for all $u\in[0,1]$.
\end{lemma}
\begin{proof}
Let $\pi_0$ be uniform on $\Pi$ and independent of everything else, and put $Z'=\pi_0Z$, which has the law of $Z$.
The vector $\bigl(D(Z'),D(\pi_1\pi_0^{-1}Z'),\dots,D(\pi_B\pi_0^{-1}Z')\bigr)$ equals
$\bigl(D(\pi_0Z),D(\pi_1Z),\dots,D(\pi_BZ)\bigr)$, whose coordinates are exchangeable because
$\pi_0,\dots,\pi_B$ are i.i.d. Since
\[(Z',\pi_1\pi_0^{-1},\dots,\pi_B\pi_0^{-1})\overset{d}{=}(Z,\pi_1,\dots,\pi_B)\]
by invariance of the law of $Z$ and of the uniform law on $\Pi$, the vector $\bigl(D(Z),D(\pi_1Z),\dots,D(\pi_BZ)\bigr)$
is exchangeable as well. The rank of its first coordinate among $B+1$ exchangeable values (ties counted against
it) is therefore stochastically no smaller than a uniform rank on $\{1,\dots,B+1\}$, which is the claim.
\end{proof}

\begin{lemma}[Serfling]
\label{lem:serfling}
Let $x_1,\dots,x_N\in[a,b]$ have mean $\bar x$ and let $\bar X_K$ be the mean of $K$ draws without replacement.
Then for all $\lambda\in\mathbb{R}$,
$\mathbb{E}\exp\{\lambda(\bar X_K-\bar x)\}\le\exp\bigl\{\lambda^2(b-a)^2\bigl(1-\tfrac{K-1}{N}\bigr)/(8K)\bigr\}$.
\end{lemma}
\noindent This is the moment-generating-function bound underlying Serfling's inequality (Serfling, 1974, Cor.~1.1).

\begin{lemma}[Joint randomization test]
\label{lem:joint}
In the setting of Lemma~\ref{lem:perm} put $\pi_0=\mathrm{id}$ and let $W_b=\pi_bZ$, $b=0,\dots,B$. Then
$(W_0,\dots,W_B)$ is exchangeable. Consequently, for any statistics $D_1,\dots,D_J$, the joint $p$-value of
Definition~\ref{def:joint} satisfies $\Pr(p^{\mathrm{in}}\le u)\le u$ for all $u\in[0,1]$, and
$p^{\mathrm{in}}\le2J\min_jr^{(0)}_j$.
\end{lemma}
\begin{proof}
With $\pi'_0$ uniform and independent of everything else, $(\pi'_0Z,\pi_1Z,\dots,\pi_BZ)$ is exchangeable, because
given $Z$ its entries are images of $Z$ under i.i.d.\ uniform group elements. Write $Z'=\pi'_0Z$ and
$\sigma_b=\pi_b(\pi'_0)^{-1}$; then $(\pi'_0Z,\pi_1Z,\dots,\pi_BZ)=(Z',\sigma_1Z',\dots,\sigma_BZ')$. Given $\pi'_0$ and
$Z$, the $\sigma_b$ are i.i.d.\ uniform, so they are independent of $Z'$, and $Z'\overset{d}{=}Z$ by invariance. Hence
$(Z',\sigma_1Z',\dots,\sigma_BZ')\overset{d}{=}(Z,\pi_1Z,\dots,\pi_BZ)=(W_0,\dots,W_B)$, which is therefore exchangeable.
Each of $r^{(b)}_j$, $M^{(b)}$, $S^{(b)}$ and $F^{(b)}$ is a function of $W_b$ and of the multiset
$\{W_0,\dots,W_B\}$. The vector $(F^{(0)},\dots,F^{(B)})$ is therefore exchangeable, and $p^{\mathrm{in}}$, the rank of
$F^{(0)}$ with ties counted against it, is super-uniform.
For the second claim we use the fact that for any reals $x_0,\dots,x_B$ and integer $k$, at most $k$ indices $b$
satisfy $\#\{b':x_{b'}\le x_b\}\le k$ (and likewise with $\ge$). Since $F^{(b)}\le F^{(0)}$ requires
$\rho(M^{(b)})\le F^{(0)}$ or $\rho(S^{(b)})\le F^{(0)}$, and $(B+1)F^{(0)}$ is an integer,
$\#\{b:F^{(b)}\le F^{(0)}\}\le2(B+1)F^{(0)}\le2(B+1)\rho(M^{(0)})$. By the same fact applied to each channel,
$\#\{b:M^{(b)}\le M^{(0)}\}\le\sum_j\#\{b:r^{(b)}_j\le M^{(0)}\}\le J(B+1)M^{(0)}$, i.e.\ $\rho(M^{(0)})\le JM^{(0)}$.
Together, $p^{\mathrm{in}}\le2JM^{(0)}=2J\min_jr^{(0)}_j$; with the min-$p$ part alone the factor $2$ disappears.
\end{proof}

\subsection{Proof of Theorem~\ref{thm:validity} and its Corollaries}

\begin{proof}[Proof of Theorem~\ref{thm:validity}]
Consider the true cell $g^\star$. In each stratum $s$ the $K$ audit units and the $n_{g^\star,s}$ calibration
units are i.i.d.\ draws from $P^{s}_{M_c,g^\star}$, and the score functions are fixed independently of both
samples. The pooled units are therefore exchangeable within the stratum, and the strata are independent, so the
joint law of all pooled units (each carrying its vector of channel scores) is invariant under the product group
$\Pi$ of Definition~\ref{def:pvalue}. Lemma~\ref{lem:joint} gives $\Pr(p^{\mathrm{in}}_{g^\star}\le u)\le u$, and by
assumption $\Pr(p^{(e)}_{g^\star}\le u)\le u$ for every probe channel. A union bound over the $1+E$ terms gives
$\Pr(p_{g^\star}\le\alpha)\le\Pr(p^{\mathrm{in}}_{g^\star}\le\tfrac{\alpha}{1+E})+\sum_e\Pr(p^{(e)}_{g^\star}\le\tfrac{\alpha}{1+E})\le\alpha$,
and since $\max_gp_g\ge p_{g^\star}$ the verifier rejects with probability at most $\alpha$. No assumption on the
dependence between channels, or between the in-harness and probe audits, is used.
\end{proof}

\begin{proof}[Proof of Corollary~\ref{cor:region}]
Let $g\in\mathcal{G}$ with $\lVert\theta-g\rVert\le h$. The rejection event is a measurable function of the audit
data, $\mathcal{D}_{\mathrm{cal}}$ and $\omega$, and the latter two have the same law whether the service runs under
$\theta$ or $g$. Hence
$\Pr_{\theta}(\text{reject})\le\Pr_{g}(\text{reject})+\mathrm{TV}\bigl(\bigotimes_{k,s}P^{s}_{M_c,\theta},\bigotimes_{k,s}P^{s}_{M_c,g}\bigr)$.
The first term is at most $\alpha$ by Theorem~\ref{thm:validity}. By additivity of the Kullback--Leibler divergence
over the $K|\mathcal{S}|$ independent units and Pinsker's inequality, the second is at most
$\sqrt{K|\mathcal{S}|\cdot2L^2h^2/2}=\sqrt{K|\mathcal{S}|}\,Lh$.
\end{proof}

\begin{proof}[Proof of Corollary~\ref{cor:mondrian}]
Under an honest service in cell $g^\star$, $\kappa(\text{audit})=\kappa(g^\star)$ almost surely, so $g^\star$ belongs
to the restricted index set and the restricted maximum is still at least $p_{g^\star}$. The proof of
Theorem~\ref{thm:validity} applies unchanged.
\end{proof}

\begin{proof}[Proof of Proposition~\ref{prop:coherent}]
(1) is the second claim of Lemma~\ref{lem:joint}, applied to cell $g$.
(2) For every $g$ and $c$,
\[\min_{c'}q_{g,c'}/w_{c'}\le q_{g,c}/w_c\le\max_{g'}q_{g',c}/w_c;\]
take the maximum over $g$ on the left and then the minimum over $c$ on the right. With the stated weights, (1) gives
$p_g\le\min\{1,\min_c p_{g,c}/w_c\}$ where $c$ ranges over in-harness channels (with $p_{g,c}=p_{g,j}$) and probe
channels (with $p_{g,c}=p^{(e)}_g$). So
\[\max_gp_g\le\max_g\min_cp_{g,c}/w_c\le\min_c\max_gp_{g,c}/w_c,\]
and the channel-first rule rejecting means the right-hand side is at most $\alpha$.
(3) Let $p_{g,1}\le\epsilon$ for $g\ne g_1$, $p_{g,2}\le\epsilon$ for $g\ne g_2$, and $p_{g_1,1}=p_{g_2,2}=1$. Then
$\max_gp_{g,c}=1$ for both channels, so the channel-first rule never rejects. Every cell has
$\min_cp_{g,c}\le\epsilon$, so the cell-coherent rule rejects as soon as $\epsilon\le w\alpha$.
\end{proof}

\subsection{Proof of Proposition~\ref{prop:paradox}}

\begin{proof}
(1) $P_{M_c,g}$ is an alternative to $H_0$, and consistency against it means $\Pr_{M_c,g}(\psi_N=1)\to1$.
(2) The decision is a function of $\rho(\text{audit})$, $\mathcal{D}_{\mathrm{cal}}$ and $\omega$. By assumption
$\rho(\text{audit})$ has the same law under $(S,\eta)$ as under $(M_c,g)$, and it is independent of
$(\mathcal{D}_{\mathrm{cal}},\omega)$ in both cases. So the rejection probability under $(S,\eta)$ equals that under
the honest service $(M_c,g)$, which is at most $\alpha$.
\end{proof}

\subsection{Proof of Theorem~\ref{thm:power}}

\begin{proof}
For every cell $g$ fix a channel $j(g)$ with $\Delta_{j(g)}(g)\ge r_{j(g)}(K,n)$. By
Proposition~\ref{prop:coherent}(1), $p_g\le(1+E)p^{\mathrm{in}}_g\le J'p_{g,j(g)}$, so it suffices to show that with
probability at least $1-\beta$ every cell has $p_{g,j(g)}\le\alpha/J'$. Write $S=|\mathcal{S}|$ and $c=c_{j(g)}$ when the
cell is clear from the context.

\emph{Step 1: the observed statistics.}
For a channel $j$, $\sum_s(\overline{T^s_j}-\mu^s_{S,j})$ is a sum of $KS$ independent centered terms, each ranging in an
interval of length $c_j/K$. Hoeffding's inequality gives, with probability at least $1-\beta/(4J)$,
$\bigl|\sum_s(\overline{T^s_j}-\mu^s_{S,j})\bigr|\le c_j\sqrt{S\log(8J/\beta)/(2K)}$; by a union bound this holds for all
(at most $J$) channels in $\{j(g)\}$ with probability at least $1-\beta/4$.
Likewise, for each cell $g$, $\sum_s(\overline{C^{g,s}_{j(g)}}-\mu^s_{g,j(g)})$ is a sum of independent centered terms with
ranges at most $c/n$, so with probability at least $1-\beta/(4m)$ it is at most $c\sqrt{S\log(8m/\beta)/(2n)}$ in absolute
value; by a union bound this holds for all $m$ cells with probability at least $1-\beta/4$. On the intersection
$\mathcal{E}$ of these events (probability at least $1-\beta/2$), for every $g$,
\begin{equation}
\label{eq:obs}
D_{g,j(g)}\ \ge\ \Delta_{j(g)}(g)-c\sqrt{\tfrac{S}{2K}\log\tfrac{8J}\beta}-c\sqrt{\tfrac{S}{2n}\log\tfrac{8m}\beta}.
\end{equation}

\emph{Step 2: the permutation distribution.}
Condition on the pooled scores. For a uniform $\pi\in\Pi$ and a stratum $s$ with $N_s=K+n_{g,s}$ pooled values of mean
$\bar x_s$, the permuted audit mean $\overline{T^{s,\pi}}$ is the mean of $K$ draws without replacement, and
$\overline{T^{s,\pi}}-\overline{C^{g,s,\pi}}=\frac{N_s}{n_{g,s}}(\overline{T^{s,\pi}}-\bar x_s)$. By
Lemma~\ref{lem:serfling} this difference is sub-Gaussian with variance proxy
\[
\frac{N_s^2}{n_{g,s}^2}\cdot\frac{c^2}{4K}\cdot\frac{n_{g,s}+1}{N_s}
=\frac{c^2}{4}\Bigl(\frac1K+\frac1{n_{g,s}}\Bigr)\Bigl(1+\frac1{n_{g,s}}\Bigr)\ \le\ \frac{c^2}{4}\Bigl(\frac1K+\frac1n\Bigr)\Bigl(1+\frac1n\Bigr)=:v .
\]
Permutations act independently across strata, so $\sum_s(\overline{T^{s,\pi}}-\overline{C^{g,s,\pi}})$ is sub-Gaussian
with proxy $Sv$ and
$\Pr_\pi\bigl(D^{\pi}_{g,j(g)}\ge t\bigr)\le2\exp\{-t^2/(2Sv)\}$. Taking
\[t_\alpha=\sqrt{2Sv\log(8J'/\alpha)}=c\sqrt{S}\sqrt{\tfrac12(\tfrac1K+\tfrac1n)(1+\tfrac1n)\log\tfrac{8J'}{\alpha}},\]
the exact
permutation tail probability $q_g=\Pr_\pi(D^\pi_{g,j(g)}\ge D_{g,j(g)})$ satisfies $q_g\le\alpha/(4J')$ whenever
$D_{g,j(g)}\ge t_\alpha$. By \eqref{eq:obs} and the assumption $\Delta_{j(g)}(g)\ge r_{j(g)}(K,n)$, this holds for all $g$
on $\mathcal{E}$.

\emph{Step 3: the Monte-Carlo $p$-values.}
Given the data, $W_g=\sum_b\mathbb{1}\{D^{\pi_b}_{g,j(g)}\ge D_{g,j(g)}\}\sim\mathrm{Bin}(B,q_g)$ and
$p_{g,j(g)}\le\alpha/J'$ iff $W_g\le(B+1)\alpha/J'-1$. Put $\mu=B\alpha/(4J')\ge\mathbb{E}W_g$. The condition on $B$ gives
$\mu\ge2$ and $\mu\ge\frac34\log\frac{4m}{\beta}$. From the first, $(B+1)\alpha/J'-1\ge4\mu-1\ge3.5\mu$, and the
multiplicative Chernoff bound $\Pr(W\ge(1+\delta)\mu)\le\exp\{-\mu\delta^2/(2+\delta)\}$ with $\delta=2.5$ yields,
using the second,
$\Pr(p_{g,j(g)}>\alpha/J')\le\exp\{-1.38\mu\}\le\exp\{-\log(4m/\beta)\}=\beta/(4m)$.
A union bound over the $m$ cells shows that, on $\mathcal{E}$, every cell has $p_{g,j(g)}\le\alpha/J'$ with conditional
probability at least $1-\beta/4$. Altogether $V_\alpha$ rejects with probability at least $1-\beta/2-\beta/4\ge1-\beta$.
\end{proof}

\subsection{Proofs of Theorems~\ref{thm:impossibility} and~\ref{thm:fusion}}

\begin{proof}[Proof of Theorem~\ref{thm:impossibility}]
Let $R$ be the rejection event, a measurable function of the audit data, $\mathcal{D}_{\mathrm{cal}}$ and $\omega$.
For any $g\in\mathcal{G}$, conditioning on $(\mathcal{D}_{\mathrm{cal}},\omega)$, which has the same law in both
situations, gives
$\Pr_{S,\eta}(R)\le\Pr_{M_c,g}(R)+\mathrm{TV}(P_{S,\eta}^{\otimes N},P_{M_c,g}^{\otimes N})\le\alpha+\mathrm{TV}(P_{S,\eta}^{\otimes N},P_{M_c,g}^{\otimes N})$,
where the last step uses that the verifier is certified. Minimizing over $g$ gives the first inequality. The second
follows from sub-additivity of total variation over product measures and from Pinsker's inequality with additivity
of the Kullback--Leibler divergence. Solving $1-\beta\le\alpha+\sqrt{N\delta_{\mathrm{KL}}/2}$ for $N$ gives the budget
bound, and enlarging $\mathcal{G}$ can only decrease the minimum.
\end{proof}

\begin{proof}[Proof of Theorem~\ref{thm:fusion}]
Validity is Theorem~\ref{thm:validity}. For power, by Proposition~\ref{prop:coherent},
\[\max_gp_g\le\max_g\min\{J'p_{g,j_0},(1+E)p^{(e_0)}_g\}\le\min\{J'\max_gp_{g,j_0},(1+E)\max_gp^{(e_0)}_g\}\]
for every in-harness channel $j_0$ and probe channel $e_0$, so the rejection event contains each of the events on the right-hand
side of the claim. A channel used alone corresponds to $J'=1$ in Theorem~\ref{thm:power}; inside the fusion only the
term $\log(8J'/\alpha)$ of its radius changes.
\end{proof}

\subsection{Proof of Theorem~\ref{thm:pivot}}

\begin{proof}
(1) Local simulation produces i.i.d.\ draws from $P^{s}_{M_c,g}$, which is all that the proof of
Theorem~\ref{thm:validity} uses. For the rank statistic, under an honest service in cell $g^\star$ the scores
$t,t_1,\dots,t_m$ are i.i.d., hence exchangeable. Order the $m+1$ values and let a tie group of size $e+1$ occupy
positions $a+1,\dots,a+e+1$; the suspect score falls in this group with probability $(e+1)/(m+1)$, in which case
$a=\#\{i:t_i<t\}$, $e=\#\{i:t_i=t\}$ and the randomized rank is uniform on $\bigl(\frac{a}{m+1},\frac{a+e+1}{m+1}\bigr)$,
an interval of length $(e+1)/(m+1)$. The tie groups partition $(0,1)$ into such intervals, each selected with
probability equal to its length, so the randomized rank is uniform on $(0,1)$.

(2) Conditionally on the history $h_t$, $y_t$ has the discrete law $q_t$ with distribution function $Q_t$, and the
randomized probability integral transform of a discrete variable under its own law is uniform on $(0,1)$
(Brockwell, 2007). This conditional law does not depend on $h_t$, so $u_t$ is independent of
$\sigma(u_1,\dots,u_{t-1})\subseteq\sigma(h_t,V_{1:t-1})$, and by induction $u_1,\dots,u_T$ are i.i.d.\ uniform; note
that $h_t$ may contain tool outputs chosen adaptively by the environment. If the provider samples from $\tilde q_t$
with $\mathrm{TV}(\tilde q_t,q_t)\le\varepsilon_t$ for every history, the maximal coupling of the two token sequences
differs with probability at most $\sum_t\varepsilon_t$, so any event has probability within $\sum_t\varepsilon_t$ of
its value under exact sampling.
\end{proof}

\subsection{Proof of Theorem~\ref{thm:sequential}}

\begin{proof}
(1) The verifier's $p$-value $p^{(r)}=\max_gp^{(r)}_g$ is super-uniform given the past rounds, by the proof of
Theorem~\ref{thm:validity} applied to round $r$, whose audit and calibration data are independent of earlier rounds. The map
$u\mapsto\kappa u^{\kappa-1}$ is non-increasing on $(0,1]$ and integrates to one against the uniform law, so
$\mathbb{E}[\kappa(p^{(r)})^{\kappa-1}\mid\mathcal{F}_{r-1}]\le1$ and $E_t$ is a non-negative supermartingale with
$E_0=1$. Ville's inequality gives $\Pr(\sup_tE_t\ge1/\alpha)\le\alpha$.

(2) Under per-session routing every episode is served by the substitute with probability $\varepsilon$. A unit's
score is the mean of its episodes' scores, so by linearity its expectation in stratum $s$ and channel $j$ is
$\mu^s_{g^\star,j}+\varepsilon(\mu^s_{S,j}-\mu^s_{g^\star,j})$, where $\mu^s_{S,j}$ and $\mu^s_{g^\star,j}$ are the unit means under
the substitute and the claimed model. Its separation from cell $g^\star$ is $\varepsilon\Delta_j(g^\star)$, so
cell $g^\star$ is refuted only through a channel with $r_j(K,n)\le\varepsilon\Delta_j(g^\star)$; the budget follows by
inverting this condition. Under
per-call routing let $A$ be the event that no call is served by the substitute, $\Pr(A)=(1-\varepsilon)^N$. On $A$
the audit is distributed as under the honest service, so
$\Pr(R)\le\Pr(R\mid A)\Pr(A)+1-\Pr(A)\le\alpha+1-(1-\varepsilon)^N$.
\end{proof}

\subsection{Derivations for the Examples}

\paragraph{Example~\ref{ex:gaussian}.}
For Gaussian scores the difference of pooled stratum means between the audit and cell $g$ is
$\mathcal{N}\bigl(\mu_S-\mu_g,\sigma^2(\tfrac{1}{K|\mathcal{S}|}+\tfrac{1}{n|\mathcal{S}|})\bigr)$, and the permutation
critical value converges to the two-sided normal quantile $z_{1-\alpha/(2J)}$ times its standard deviation
(Lehmann and Romano, 2005, Thm.~15.2.3). Requiring power $1-\beta/m$ against each of the $m$ cells gives the stated
condition.

\paragraph{Example~\ref{ex:temperature}.}
Softmax distributions $p_\tau\propto\exp(z/\tau)$ form an exponential family in $\theta=1/\tau$ with log-partition
function $A$, $A''(\theta)=\mathrm{Var}_{p_\theta}(z)$. Hence
$\mathrm{KL}(p_\theta\Vert p_{\theta'})=A(\theta')-A(\theta)-(\theta'-\theta)A'(\theta)=\tfrac12(\theta'-\theta)^2A''(\xi)$
for some $\xi$ between $\theta$ and $\theta'$, and
$(\theta'-\theta)^2=(\tau-\tau')^2/(\tau\tau')^2\le(\tau-\tau')^2/\min(\tau,\tau')^4$. By the chain rule the divergence
between the laws of a unit with at most $T$ sampled tokens is at most $T$ times the per-token bound, i.e.\
$Tv(\tau-\tau')^2/(2\tau_{\min}^4)=2L^2(\tau-\tau')^2$ with $L=\sqrt{Tv}/(2\tau_{\min}^2)$.

%% file: sections/appendix_models.tex
\Needspace{12\baselineskip}
\subsection{Model References and Checkpoint Provenance}
\label{app:model-sources}

Table~\ref{tab:model-citations} gives the release-specific source references for every named model and the official low-precision checkpoints. The Qwen3 family is described in its technical report~\citep{qwen3technicalreport}. For a gateway-served model, a source reference documents the advertised upstream release; the gateway alias and the upstream checkpoint are distinct identifiers. In particular, \texttt{gpt-6.1-sol-coding} is the gateway alias used for the GPT-6.1 Sol claim.

\input{tables/appendix_model_sources}

\noindent\textbf{Low-precision releases and constructed derivatives.}
The official FP8 releases used by the Qwen3.5-27B and Qwen3-Coder contracts have their own model cards~\citep{qwen35_27b_fp8,qwen3coder30b_fp8}. The official 4-bit substitute of Qwen3.5-27B is Qwen3.5-27B-GPTQ-Int4~\citep{qwen35_27b_gptq}. These checkpoint references distinguish an official low-precision release from online quantization by the serving stack. Table~\ref{tab:derivative-sources} records the source models and construction of the 15 derivatives in the experimental inventory. A source-model or algorithm citation does not uniquely identify a separately released post-trained checkpoint.

\begin{longtable}{@{}p{0.37\linewidth}p{0.57\linewidth}@{}}
\caption{Provenance of the 15 derivatives. The four generically named public post-trained checkpoints require their exact repository identifiers; the available manuscript specifies only their roles and base model.}
\label{tab:derivative-sources}\\
\toprule
Derivative & Source or construction \\
\midrule
\endfirsthead
\toprule
Derivative & Source or construction \\
\midrule
\endhead
\bottomrule
\endfoot
Qwen3.5-27B: official 4-bit & Official GPTQ-Int4 checkpoint~\citep{qwen35_27b_gptq}. \\
Qwen3.5-27B: agentic fine-tune & Base model~\citep{qwen35_27b}; checkpoint ID not supplied. \\
Qwen3.5-27B: reasoning-distilled & Base model~\citep{qwen35_27b}; checkpoint ID not supplied. \\
Qwen3.5-27B: refusal-ablated & Base model~\citep{qwen35_27b}; method reference~\citep{arditi2024refusal}; checkpoint ID not supplied. \\
Qwen3.5-27B: behavior-preserving edit & Base model~\citep{qwen35_27b}; checkpoint ID not supplied. \\
Qwen3-Coder: 4-bit and 6-bit (2) & Round-to-nearest quantization by us of Qwen3-Coder-30B-A3B-Instruct~\citep{qwen3coder30b}. \\
Qwen3-Coder: merges (3) & $(1-\lambda)\theta_c+\lambda\theta'$ with Qwen3-30B-A3B-Instruct-2507~\citep{qwen3coder30b,qwen3_30b_2507}, $\lambda\in\{0.1,0.25,0.5\}$. \\
Devstral: 4-bit and 6-bit (2) & Round-to-nearest quantization by us of Devstral-Small-2507~\citep{devstral2507}. \\
Devstral: merges (3) & $(1-\lambda)\theta_c+\lambda\theta'$ with Devstral-Small-2505~\citep{devstral2507,devstral2505}, $\lambda\in\{0.1,0.25,0.5\}$. \\
\end{longtable}


\noindent\textbf{Versioned API identities.}
API names may outlive the release that they originally denoted. For example, DeepSeek's V4.1-Flash release notice states that \texttt{deepseek-v4-flash} temporarily routes to V4.1-Flash~\citep{deepseek2026aliasupdate}. Thus, a public model-card citation identifies a release but does not establish the backend served by a gateway at collection time. The October 2026 gateway results retain the model labels recorded in the experiments; interpreting these labels as exact upstream checkpoints additionally requires the gateway's version or routing record.

%% file: tables/appendix_model_sources.tex
{\small
\setlength{\tabcolsep}{3pt}
\begin{longtable}{@{}p{0.41\textwidth}p{0.055\textwidth}@{\hspace{0.03\textwidth}}p{0.41\textwidth}p{0.055\textwidth}@{}}
\caption{Named model releases and their source references. C: claim; S: substitute; P: merge parent; A: authorized precision variant. A release can have multiple roles. Author-constructed quantizations and merges inherit the source models cited here.}
\label{tab:model-citations}\\
\toprule
Model release (source) & Role & Model release (source) & Role \\
\midrule
\endfirsthead
\toprule
Model release (source) & Role & Model release (source) & Role \\
\midrule
\endhead
\bottomrule
\endfoot
Qwen3.5-27B~\citep{qwen35_27b} & C/S & GPT-5.5~\citep{openai_gpt55} & S \\
Qwen3.5-9B~\citep{qwen35_9b} & S & GPT-5.6 Terra~\citep{openai_gpt56terra} & S \\
Qwen3.5-4B~\citep{qwen35_4b} & S & GPT-6 Sol~\citep{openai_gpt6sol} & S \\
Qwen3.6-27B~\citep{qwen36_27b} & S & Claude Sonnet 4.6~\citep{anthropic_sonnet46} & C/S \\
Qwen3-32B~\citep{qwen3_32b} & S & Claude Opus 5.5~\citep{anthropic_opus55} & C \\
Qwen3-Coder-30B-A3B-Instruct~\citep{qwen3coder30b} & C/S/P & Claude Haiku 4.5~\citep{anthropic_haiku45} & S \\
Qwen3-30B-A3B-Instruct-2507~\citep{qwen3_30b_2507} & S/P & Claude Sonnet 4.5~\citep{anthropic_sonnet45} & S \\
Devstral-Small-2507~\citep{devstral2507} & C/S/P & Claude Opus 5~\citep{anthropic_opus5} & S \\
Devstral-Small-2505~\citep{devstral2505} & P & DeepSeek-V4-Flash~\citep{deepseek_v4flash} & S \\
Mistral-Small-24B-Instruct-2501~\citep{mistralsmall2501} & S & GLM-5.2~\citep{zai_glm52} & S \\
gpt-oss-20b~\citep{openai2025gptoss} & S & Qwen3.5-Flash~\citep{qwen35flash} & S \\
GPT-5.4~\citep{openai_gpt54} & C/S & Kimi K2.6~\citep{moonshot_kimi26} & S \\
GPT-5.6 Sol~\citep{openai_gpt56sol} & C/S & MiniMax-M2.5~\citep{minimax_m25} & S \\
GPT-6.1 Sol~\citep{openai_gpt61sol} & C & Qwen3.5-27B-FP8~\citep{qwen35_27b_fp8} & A \\
GPT-5.2~\citep{openai_gpt52} & S & Qwen3-Coder-30B-A3B-Instruct-FP8~\citep{qwen3coder30b_fp8} & A \\
GPT-5.4 Mini~\citep{openai_gpt54mini} & S & Qwen3.5-27B-GPTQ-Int4~\citep{qwen35_27b_gptq} & S \\
GPT-5.4 nano~\citep{openai_gpt54nano} & S &  &  \\
\end{longtable}
}

%% file: tables/appendix_breakdown.tex
\begin{table*}[t]
\centering
\caption{The AUC and pAUC by type of authorized deployment change (left: negatives restricted to honest services in that type of deployment; positives are all substitutes) and by type of substitution (right: positives restricted to that type; negatives are all honest services), averaged over the settings in which the type occurs. A cell on the left is red when the method is not valid at $\alpha=5\%$ on the pooled honest services of that type.}
\label{tab:breakdown-full}
\scriptsize
\setlength{\tabcolsep}{3.2pt}
\begin{tabular}{llccccccccccc}
\toprule
 &  & \multicolumn{6}{c}{Authorized deployment change (honest services)} & \multicolumn{5}{c}{Substitution (substitutes)} \\
\cmidrule(lr){3-8}\cmidrule(lr){9-13}
Method & Metric & Default & Sampling & Reasoning & \begin{tabular}[b]{@{}c@{}}Sys.\\prompt\end{tabular} & Context & Precision & \begin{tabular}[b]{@{}c@{}}Other\\family\end{tabular} & \begin{tabular}[b]{@{}c@{}}Same\\developer\end{tabular} & \begin{tabular}[b]{@{}c@{}}Post-\\trained\end{tabular} & Quantized & Merged \\
\midrule
\multirow{2}{*}{MET} & AUC$\uparrow$ & \textcolor{red}{0.898} & \textcolor{red}{0.730} & \textcolor{red}{0.486} & \textcolor{red}{0.728} & \textcolor{red}{0.734} & \textcolor{red}{0.736} & 0.844 & 0.745 & 0.630 & 0.489 & 0.425 \\
 & pAUC$\uparrow$ & \textcolor{red}{0.882} & \textcolor{red}{0.620} & \textcolor{red}{0.609} & \textcolor{red}{0.707} & \textcolor{red}{0.736} & \textcolor{red}{0.715} & 0.590 & 0.578 & 0.516 & 0.493 & 0.496 \\
\midrule
\multirow{2}{*}{LLMmap} & AUC$\uparrow$ & \textcolor{red}{0.684} & 0.687 & \textcolor{red}{0.696} & \textcolor{red}{0.724} & \textcolor{red}{0.742} & \textcolor{red}{0.659} & 0.877 & 0.497 & 0.488 & 0.581 & 0.728 \\
 & pAUC$\uparrow$ & \textcolor{red}{0.696} & 0.739 & \textcolor{red}{0.691} & \textcolor{red}{0.734} & \textcolor{red}{0.823} & \textcolor{red}{0.721} & 0.737 & 0.600 & 0.501 & 0.537 & 0.668 \\
\midrule
\multirow{2}{*}{RUT} & AUC$\uparrow$ & 0.966 & \textcolor{red}{0.506} & \textcolor{red}{0.600} & \textcolor{red}{0.307} & \textcolor{red}{0.264} & \textcolor{red}{0.955} & 0.588 & 0.713 & 0.549 & 0.447 & 0.433 \\
 & pAUC$\uparrow$ & 0.958 & \textcolor{red}{0.657} & \textcolor{red}{0.548} & \textcolor{red}{0.502} & \textcolor{red}{0.511} & \textcolor{red}{0.967} & 0.487 & 0.566 & 0.487 & 0.497 & 0.510 \\
\midrule
\multirow{2}{*}{KBF} & AUC$\uparrow$ & 0.765 & \textcolor{red}{0.733} & \textcolor{red}{0.681} & 0.784 & 0.764 & 0.769 & 0.552 & 0.976 & 0.454 & 0.889 & 0.581 \\
 & pAUC$\uparrow$ & 0.861 & \textcolor{red}{0.837} & \textcolor{red}{0.783} & 0.867 & 0.865 & 0.875 & 0.755 & 0.972 & 0.531 & 0.661 & 0.658 \\
\midrule
\multirow{2}{*}{IRIS} & AUC$\uparrow$ & 0.732 & 0.748 & \textcolor{red}{0.408} & \textcolor{red}{0.676} & \textcolor{red}{0.720} & \textcolor{red}{0.704} & 0.794 & 0.606 & 0.500 & 0.483 & 0.698 \\
 & pAUC$\uparrow$ & 0.802 & 0.783 & \textcolor{red}{0.673} & \textcolor{red}{0.789} & \textcolor{red}{0.797} & \textcolor{red}{0.772} & 0.797 & 0.698 & 0.497 & 0.557 & 0.668 \\
\midrule
\multirow{2}{*}{AgentProv} & AUC$\uparrow$ & \textcolor{red}{0.964} & \textcolor{red}{0.845} & \textcolor{red}{0.636} & \textcolor{red}{0.678} & \textcolor{red}{0.482} & \textcolor{red}{0.965} & 0.766 & 0.766 & 0.833 & 0.555 & 0.626 \\
 & pAUC$\uparrow$ & \textcolor{red}{0.937} & \textcolor{red}{0.793} & \textcolor{red}{0.626} & \textcolor{red}{0.659} & \textcolor{red}{0.500} & \textcolor{red}{0.865} & 0.518 & 0.518 & 0.526 & 0.495 & 0.508 \\
\midrule
\multirow{2}{*}{CertClaim} & AUC$\uparrow$ & \textbf{0.992} & \textbf{0.986} & \textbf{1.000} & \textbf{0.994} & \textbf{0.993} & \textbf{0.988} & \textbf{1.000} & \textbf{1.000} & \textbf{1.000} & \textbf{0.998} & \textbf{0.877} \\
 & pAUC$\uparrow$ & \textbf{0.990} & \textbf{0.982} & \textbf{1.000} & \textbf{0.989} & \textbf{0.991} & \textbf{0.987} & \textbf{1.000} & \textbf{1.000} & \textbf{0.998} & \textbf{0.984} & \textbf{0.837} \\
\bottomrule
\end{tabular}
\end{table*}

%% file: tables/per_setting_suites.tex
\begin{table*}[t]
\centering
\caption{The performance of model-claim verification on the EvalPlus and real-issue task suites. Metrics of LeaFBench: AUC, standardized pAUC over FPR $\in[0,0.05]$ and the Mahalanobis distance (MD) between the scores of substitutes (positives) and of honest services in all authorized deployments (negatives). The best result in each column is in boldface; results of methods that are not valid at $\alpha=5\%$ (i.e., the lower end of the 95\% Clopper--Pearson interval of their FPR exceeds $\alpha$) are in red.}
\label{tab:suites-leaf}
\scriptsize
\setlength{\tabcolsep}{3.0pt}
\begin{tabular}{llccccccc}
\toprule
 & & \multicolumn{2}{c}{Qwen3.5-27B} & \multicolumn{2}{c}{Qwen3-Coder} & \multicolumn{2}{c}{Devstral} & \multicolumn{1}{c}{Qwen3.5-27B} \\
\cmidrule(lr){3-4} \cmidrule(lr){5-6} \cmidrule(lr){7-8} \cmidrule(lr){9-9}
Method & Metric & MSA & OC & MSA & OC & MSA & OC & MSA \\
\midrule
\multirow{3}{*}{MET} & AUC$\uparrow$ & \textcolor{red}{0.618} & \textcolor{red}{0.618} & \textcolor{red}{0.655} & \textcolor{red}{0.655} & \textcolor{red}{0.929} & \textcolor{red}{0.929} & \textcolor{red}{0.591} \\
 & pAUC$\uparrow$ & \textcolor{red}{0.513} & \textcolor{red}{0.513} & \textcolor{red}{0.506} & \textcolor{red}{0.506} & \textcolor{red}{0.577} & \textcolor{red}{0.577} & \textcolor{red}{0.510} \\
 & MD$\uparrow$ & \textcolor{red}{0.51} & \textcolor{red}{0.51} & \textcolor{red}{0.88} & \textcolor{red}{0.88} & \textcolor{red}{3.35} & \textcolor{red}{3.35} & \textcolor{red}{0.39} \\
\midrule
\multirow{3}{*}{LLMmap} & AUC$\uparrow$ & \textcolor{red}{0.427} & \textcolor{red}{0.427} & 0.900 & 0.900 & 0.768 & 0.768 & \textcolor{red}{0.390} \\
 & pAUC$\uparrow$ & \textcolor{red}{0.492} & \textcolor{red}{0.492} & 0.788 & 0.788 & 0.730 & 0.730 & \textcolor{red}{0.493} \\
 & MD$\uparrow$ & \textcolor{red}{0.30} & \textcolor{red}{0.30} & 1.83 & 1.83 & 0.70 & 0.70 & \textcolor{red}{0.43} \\
\midrule
\multirow{3}{*}{RUT} & AUC$\uparrow$ & \textcolor{red}{0.539} & \textcolor{red}{0.539} & \textcolor{red}{0.677} & \textcolor{red}{0.677} & \textcolor{red}{0.658} & \textcolor{red}{0.658} & \textcolor{red}{0.541} \\
 & pAUC$\uparrow$ & \textcolor{red}{0.487} & \textcolor{red}{0.487} & \textcolor{red}{0.487} & \textcolor{red}{0.487} & \textcolor{red}{0.538} & \textcolor{red}{0.538} & \textcolor{red}{0.487} \\
 & MD$\uparrow$ & \textcolor{red}{0.82} & \textcolor{red}{0.82} & \textcolor{red}{1.11} & \textcolor{red}{1.11} & \textcolor{red}{1.00} & \textcolor{red}{1.00} & \textcolor{red}{0.80} \\
\midrule
\multirow{3}{*}{KBF} & AUC$\uparrow$ & \textcolor{red}{0.693} & \textcolor{red}{0.693} & 0.800 & 0.800 & 0.875 & 0.875 & \textcolor{red}{0.663} \\
 & pAUC$\uparrow$ & \textcolor{red}{0.735} & \textcolor{red}{0.735} & 0.897 & 0.897 & 0.936 & 0.936 & \textcolor{red}{0.709} \\
 & MD$\uparrow$ & \textcolor{red}{1.07} & \textcolor{red}{1.07} & 2.83 & 2.83 & 3.66 & 3.66 & \textcolor{red}{0.97} \\
\midrule
\multirow{3}{*}{IRIS} & AUC$\uparrow$ & \textcolor{red}{0.445} & \textcolor{red}{0.445} & \textcolor{red}{0.857} & \textcolor{red}{0.857} & 0.795 & 0.795 & \textcolor{red}{0.413} \\
 & pAUC$\uparrow$ & \textcolor{red}{0.495} & \textcolor{red}{0.495} & \textcolor{red}{0.927} & \textcolor{red}{0.927} & 0.853 & 0.853 & \textcolor{red}{0.491} \\
 & MD$\uparrow$ & \textcolor{red}{0.01} & \textcolor{red}{0.01} & \textcolor{red}{0.63} & \textcolor{red}{0.63} & 1.75 & 1.75 & \textcolor{red}{0.17} \\
\midrule
\multirow{3}{*}{AgentProv} & AUC$\uparrow$ & \textcolor{red}{0.805} & \textcolor{red}{0.805} & \textcolor{red}{0.780} & \textcolor{red}{0.780} & \textcolor{red}{0.673} & \textcolor{red}{0.673} & \textcolor{red}{0.803} \\
 & pAUC$\uparrow$ & \textcolor{red}{0.522} & \textcolor{red}{0.522} & \textcolor{red}{0.516} & \textcolor{red}{0.516} & \textcolor{red}{0.507} & \textcolor{red}{0.507} & \textcolor{red}{0.522} \\
 & MD$\uparrow$ & \textcolor{red}{1.40} & \textcolor{red}{1.40} & \textcolor{red}{1.45} & \textcolor{red}{1.45} & \textcolor{red}{0.92} & \textcolor{red}{0.92} & \textcolor{red}{1.38} \\
\midrule
\multirow{3}{*}{CertClaim} & AUC$\uparrow$ & \textbf{1.000} & \textbf{1.000} & \textbf{1.000} & \textbf{1.000} & \textbf{1.000} & \textbf{1.000} & \textbf{1.000} \\
 & pAUC$\uparrow$ & \textbf{1.000} & \textbf{1.000} & \textbf{1.000} & \textbf{1.000} & \textbf{1.000} & \textbf{1.000} & \textbf{1.000} \\
 & MD$\uparrow$ & \textbf{10.29} & \textbf{10.29} & \textbf{7.72} & \textbf{7.72} & \textbf{9.68} & \textbf{9.23} & \textbf{9.79} \\
\bottomrule
\end{tabular}
\end{table*}

%% file: tables/appendix_operating.tex
\begin{table*}[t]
\centering
\caption{The operating point of every method at $\alpha=5\%$: the false-positive rate on honest services in all authorized deployments and the mean detection rate over the whole-model substitutes (\%, FPR / detection rate), with one batch of tasks for CertClaim. Methods that are not valid at $\alpha$ are in red.}
\label{tab:operating}
\scriptsize
\setlength{\tabcolsep}{3.0pt}
\begin{tabular}{lccccccc}
\toprule
Setting & MET & LLMmap & RUT & KBF & IRIS & AgentProv & CertClaim \\
\midrule
Qwen3.5-27B, MSA, authored & \textcolor{red}{22.2 / 40.0} & \textcolor{red}{48.1 / 30.0} & \textcolor{red}{70.4 / 97.5} & \textcolor{red}{18.6 / 52.5} & \textcolor{red}{22.2 / 24.3} & \textcolor{red}{50.9 / 96.2} & 1.4 / 100.0 \\
Qwen3.5-27B, OC, authored & \textcolor{red}{22.2 / 45.5} & \textcolor{red}{48.1 / 36.4} & \textcolor{red}{70.4 / 97.7} & \textcolor{red}{18.6 / 56.8} & \textcolor{red}{22.2 / 29.9} & \textcolor{red}{50.9 / 96.6} & 3.3 / 100.0 \\
Qwen3-Coder, MSA, authored & \textcolor{red}{76.2 / 100.0} & 4.8 / 60.0 & \textcolor{red}{69.0 / 100.0} & 0.0 / 60.0 & \textcolor{red}{59.5 / 84.5} & \textcolor{red}{53.6 / 100.0} & 7.1 / 100.0 \\
Qwen3-Coder, OC, authored & \textcolor{red}{76.2 / 100.0} & 4.8 / 60.0 & \textcolor{red}{69.0 / 100.0} & 0.0 / 60.0 & \textcolor{red}{59.5 / 84.5} & \textcolor{red}{53.6 / 100.0} & 3.6 / 100.0 \\
Devstral, MSA, authored & \textcolor{red}{14.3 / 58.3} & 4.8 / 43.3 & \textcolor{red}{71.4 / 68.3} & 0.0 / 53.8 & 0.0 / 31.4 & \textcolor{red}{72.6 / 80.0} & 2.4 / 80.0 \\
Devstral, OC, authored & \textcolor{red}{14.3 / 58.3} & 4.8 / 43.3 & \textcolor{red}{71.4 / 68.3} & 0.0 / 53.8 & 0.0 / 31.4 & \textcolor{red}{72.6 / 80.0} & 2.4 / 80.0 \\
GPT-5.4, MSA, authored & \textcolor{red}{61.1 / 100.0} & 11.1 / 69.4 & -- & 0.0 / 69.4 & 13.9 / 85.7 & \textcolor{red}{83.3 / 100.0} & 0.0 / 100.0 \\
GPT-5.6 Sol, MSA, authored & \textcolor{red}{30.6 / 100.0} & \textcolor{red}{36.1 / 92.9} & -- & \textcolor{red}{100.0 / 76.2} & \textcolor{red}{16.7 / 83.7} & \textcolor{red}{63.9 / 100.0} & 2.8 / 100.0 \\
GPT-6.1 Sol, MSA, authored & 11.1 / 100.0 & \textcolor{red}{19.4 / 81.0} & -- & 0.0 / 85.7 & 0.0 / 85.7 & \textcolor{red}{66.7 / 100.0} & 0.0 / 100.0 \\
Sonnet 4.6, MSA, authored & \textcolor{red}{55.6 / 100.0} & 8.3 / 69.4 & -- & 0.0 / 83.3 & \textcolor{red}{16.7 / 42.9} & \textcolor{red}{80.6 / 100.0} & 0.0 / 100.0 \\
Opus 5.5, MSA, authored & \textcolor{red}{77.1 / 100.0} & 0.0 / 54.8 & -- & 0.0 / 81.0 & 0.0 / 85.7 & \textcolor{red}{66.7 / 100.0} & 2.1 / 100.0 \\
Qwen3.5-27B, MSA, EvalPlus & \textcolor{red}{22.2 / 45.5} & \textcolor{red}{48.1 / 36.4} & \textcolor{red}{70.4 / 97.7} & \textcolor{red}{18.6 / 56.8} & \textcolor{red}{22.2 / 29.9} & \textcolor{red}{50.9 / 96.6} & 1.7 / 100.0 \\
Qwen3.5-27B, OC, EvalPlus & \textcolor{red}{22.2 / 45.5} & \textcolor{red}{48.1 / 36.4} & \textcolor{red}{70.4 / 97.7} & \textcolor{red}{18.6 / 56.8} & \textcolor{red}{22.2 / 29.9} & \textcolor{red}{50.9 / 96.6} & 1.7 / 100.0 \\
Qwen3-Coder, MSA, EvalPlus & \textcolor{red}{76.2 / 100.0} & 4.8 / 66.7 & \textcolor{red}{69.0 / 100.0} & 0.0 / 80.0 & \textcolor{red}{59.5 / 85.7} & \textcolor{red}{53.6 / 100.0} & 4.8 / 100.0 \\
Qwen3-Coder, OC, EvalPlus & \textcolor{red}{76.2 / 100.0} & 4.8 / 66.7 & \textcolor{red}{69.0 / 100.0} & 0.0 / 80.0 & \textcolor{red}{59.5 / 85.7} & \textcolor{red}{53.6 / 100.0} & 6.0 / 100.0 \\
Devstral, MSA, EvalPlus & \textcolor{red}{14.3 / 100.0} & 4.8 / 53.3 & \textcolor{red}{71.4 / 100.0} & 0.0 / 87.5 & 0.0 / 62.9 & \textcolor{red}{72.6 / 100.0} & 2.4 / 100.0 \\
Devstral, OC, EvalPlus & \textcolor{red}{14.3 / 100.0} & 4.8 / 53.3 & \textcolor{red}{71.4 / 100.0} & 0.0 / 87.5 & 0.0 / 62.9 & \textcolor{red}{72.6 / 100.0} & 3.6 / 100.0 \\
Qwen3.5-27B, MSA, real issues & \textcolor{red}{22.2 / 40.0} & \textcolor{red}{48.1 / 30.0} & \textcolor{red}{70.4 / 97.5} & \textcolor{red}{18.6 / 52.5} & \textcolor{red}{22.2 / 24.3} & \textcolor{red}{50.9 / 96.2} & 1.9 / 100.0 \\
\bottomrule
\end{tabular}
\end{table*}

%% file: tables/appendix_adaptive.tex
\begin{table}[t]
\centering
\caption{The detection rate (\%) of CertClaim at $\alpha=5\%$ under adaptive attacks in every setting. Boundary: the mean (worst) detection rate over the substitutes served in an authorized configuration of the claim, with one batch of tasks. Routing: $\varepsilon=0.25$ (0.3 for the frontier claims) and $\varepsilon=0.5$, with $K=1$ / $K=4$ batches of tasks.}
\label{tab:adaptive-all}
\scriptsize
\setlength{\tabcolsep}{3.0pt}
\begin{tabular}{lccc}
\toprule
Setting & Boundary & Routing, $\varepsilon=0.25$ (0.3) & Routing, $\varepsilon=0.5$ \\
\midrule
Qwen3.5-27B MSA (authored) & 100.0 (100.0) & 91.7 / 91.7 & 100.0 / 100.0 \\
Qwen3.5-27B OC (authored) & 97.0 (66.7) & 66.7 / 66.7 & 91.7 / 100.0 \\
Qwen3-Coder MSA (authored) & 100.0 (100.0) & 75.0 / 100.0 & 100.0 / 100.0 \\
Qwen3-Coder OC (authored) & 100.0 (100.0) & 66.7 / 100.0 & 100.0 / 100.0 \\
Devstral MSA (authored) & 76.7 (0.0) & 100.0 / 100.0 & 100.0 / 100.0 \\
Devstral OC (authored) & 83.3 (0.0) & 100.0 / 100.0 & 100.0 / 100.0 \\
GPT-5.4 & -- & 100.0 / 100.0 & 100.0 / 100.0 \\
GPT-5.6 Sol & -- & 50.0 / 50.0 & 100.0 / 100.0 \\
GPT-6.1 Sol & -- & 100.0 / 100.0 & 100.0 / 100.0 \\
Sonnet 4.6 & -- & 100.0 / 100.0 & 100.0 / 100.0 \\
Opus 5.5 & -- & 100.0 / 100.0 & 100.0 / 100.0 \\
Qwen3.5-27B MSA (EvalPlus) & 97.0 (66.7) & 83.3 / 91.7 & 83.3 / 100.0 \\
Qwen3.5-27B OC (EvalPlus) & 97.0 (66.7) & 66.7 / 83.3 & 83.3 / 100.0 \\
Qwen3-Coder MSA (EvalPlus) & 100.0 (100.0) & 66.7 / 91.7 & 100.0 / 100.0 \\
Qwen3-Coder OC (EvalPlus) & 100.0 (100.0) & 75.0 / 75.0 & 100.0 / 100.0 \\
Devstral MSA (EvalPlus) & 100.0 (100.0) & 100.0 / 100.0 & 100.0 / 100.0 \\
Devstral OC (EvalPlus) & 100.0 (100.0) & 100.0 / 100.0 & 100.0 / 100.0 \\
Qwen3.5-27B MSA (real issues) & 96.7 (66.7) & 75.0 / 100.0 & 83.3 / 100.0 \\
\bottomrule
\end{tabular}
\end{table}

%% file: tables/appendix_probe_aware.tex
\begin{table}[t]
\centering
\caption{The detection rate (\%) of CertClaim at $\alpha=5\%$ under probe-aware routing on the Qwen3.5-27B claim (mini-swe-agent, one batch of tasks), where a fraction $\rho$ of the probe requests is answered by the claimed model in its default cell. The in-harness channels are fused with channel R. Last column: the in-harness channels alone.}
\label{tab:probe-aware}
\scriptsize
\setlength{\tabcolsep}{3.0pt}
\begin{tabular}{lccccccc}
\toprule
Substitute & $\rho=0$ & $\rho=0.25$ & $\rho=0.5$ & $\rho=0.75$ & $\rho=0.9$ & $\rho=1$ & In-harness \\
\midrule
4-bit (INT4) & 100 & 75 & 15 & 5 & 2 & 5 & 0 \\
Behavior-preserving edit & 100 & 72 & 10 & 0 & 0 & 5 & 0 \\
Refusal-ablated & 100 & 100 & 100 & 85 & 32 & 25 & 22 \\
Agentic fine-tune & 100 & 100 & 100 & 100 & 100 & 100 & 100 \\
Reasoning-distilled & 100 & 100 & 100 & 100 & 100 & 100 & 100 \\
Qwen3.5-9B & 100 & 100 & 100 & 100 & 100 & 100 & 100 \\
Qwen3.5-4B & 100 & 100 & 100 & 100 & 100 & 100 & 100 \\
Qwen3-32B & 100 & 100 & 100 & 100 & 100 & 100 & 100 \\
gpt-oss-20b & 100 & 100 & 100 & 100 & 100 & 100 & 100 \\
Qwen3.6-27B & 100 & 100 & 100 & 100 & 100 & 100 & 52 \\
\bottomrule
\end{tabular}
\end{table}

%% file: tables/appendix_settings_leaf.tex
\begin{table}[h]
\centering
\caption{LeaFBench metrics and false-positive rate (\%, with 95\% Clopper--Pearson interval) on the setting Qwen3.5-27B, mini-swe-agent, authored task suite. Positives: whole-model substitutes; negatives: honest services in every authorized deployment. Invalid FPR (lower end of the interval above $\alpha=5\%$) in red. In-harness only and CertClaim-S use every in-harness audit, whereas CertClaim and the ablation in Table~\ref{tab:ablation} use the in-harness audits that are paired with probe audits.}
\label{tab:app-leaf-qwen35-27b}
\scriptsize
\setlength{\tabcolsep}{3.0pt}
\begin{tabular}{lccccccc}
\toprule
Method & AUC$\uparrow$ & pAUC$\uparrow$ & MD$\uparrow$ & ACC$\uparrow$ & TPR@1\%FPR$\uparrow$ & FPR (\%) & \#pos / \#neg \\
\midrule
MET & 0.591 & 0.510 & 0.39 & 0.579 & 0.000 & \textcolor{red}{22.2} (12.0--35.6) & 60 / 54 \\
LLMmap & 0.390 & 0.493 & 0.43 & 0.404 & 0.000 & \textcolor{red}{48.1} (34.3--62.2) & 60 / 54 \\
RUT & 0.541 & 0.487 & 0.80 & 0.701 & 0.000 & \textcolor{red}{70.4} (56.4--82.0) & 80 / 54 \\
KBF & 0.663 & 0.709 & 0.97 & 0.761 & 0.400 & \textcolor{red}{18.6} (14.7--23.0) & 80 / 360 \\
IRIS & 0.413 & 0.491 & 0.17 & 0.476 & 0.000 & \textcolor{red}{22.2} (12.0--35.6) & 70 / 54 \\
AgentProv & 0.803 & 0.522 & 1.38 & 0.691 & 0.000 & \textcolor{red}{50.9} (41.1--60.7) & 80 / 108 \\
LIDAR$^{\dagger}$ & 0.654 & 0.538 & 0.55 & 0.459 & 0.069 & 4.5 (1.3--11.2) & 130 / 88 \\
MET-C & 0.635 & 0.692 & 1.04 & 0.684 & 0.400 & 0.0 (0.0--6.6) & 60 / 54 \\
LLMmap-C & 0.957 & 0.949 & 2.53 & 0.821 & 0.900 & 0.0 (0.0--6.6) & 80 / 54 \\
KBF-C & 0.852 & 0.759 & 1.40 & 0.900 & 0.500 & 0.0 (0.0--1.0) & 80 / 360 \\
IRIS-C & 0.498 & 0.501 & 0.01 & 0.474 & 0.000 & 0.0 (0.0--6.6) & 60 / 54 \\
AgentProv-C & 1.000 & 0.998 & 7.91 & 0.989 & 0.988 & 0.9 (0.0--5.1) & 80 / 108 \\
R alone & 1.000 & 1.000 & 8.73 & 0.984 & 1.000 & 1.9 (0.8--4.0) & 70 / 360 \\
K alone & 0.866 & 0.781 & 1.54 & 0.902 & 0.545 & 0.0 (0.0--1.0) & 88 / 360 \\
In-harness only & 0.902 & 0.846 & 2.16 & 0.826 & 0.713 & 0.6 (0.1--2.0) & 400 / 360 \\
CertClaim-S (static) & 0.830 & 0.749 & 0.97 & 0.675 & 0.463 & \textcolor{red}{68.6} (63.5--73.4) & 400 / 360 \\
CertClaim ($K{=}4$) & 1.000 & 1.000 & 10.44 & 0.983 & 1.000 & 1.9 (0.8--4.0) & 62 / 360 \\
CertClaim & 1.000 & 0.999 & 10.32 & 0.988 & 1.000 & 1.4 (0.5--3.2) & 62 / 360 \\
\bottomrule
\end{tabular}
\end{table}

\begin{table}[h]
\centering
\caption{LeaFBench metrics and false-positive rate (\%, with 95\% Clopper--Pearson interval) on the setting Qwen3.5-27B, OpenCode, authored task suite. Positives: whole-model substitutes; negatives: honest services in every authorized deployment. Invalid FPR (lower end of the interval above $\alpha=5\%$) in red. In-harness only and CertClaim-S use every in-harness audit, whereas CertClaim and the ablation in Table~\ref{tab:ablation} use the in-harness audits that are paired with probe audits.}
\label{tab:app-leaf-qwen35-27b-opencode}
\scriptsize
\setlength{\tabcolsep}{3.0pt}
\begin{tabular}{lccccccc}
\toprule
Method & AUC$\uparrow$ & pAUC$\uparrow$ & MD$\uparrow$ & ACC$\uparrow$ & TPR@1\%FPR$\uparrow$ & FPR (\%) & \#pos / \#neg \\
\midrule
MET & 0.618 & 0.513 & 0.51 & 0.600 & 0.000 & \textcolor{red}{22.2} (12.0--35.6) & 66 / 54 \\
LLMmap & 0.427 & 0.492 & 0.30 & 0.433 & 0.000 & \textcolor{red}{48.1} (34.3--62.2) & 66 / 54 \\
RUT & 0.539 & 0.487 & 0.82 & 0.718 & 0.000 & \textcolor{red}{70.4} (56.4--82.0) & 88 / 54 \\
KBF & 0.693 & 0.735 & 1.07 & 0.766 & 0.455 & \textcolor{red}{18.6} (14.7--23.0) & 88 / 360 \\
IRIS & 0.445 & 0.495 & 0.01 & 0.496 & 0.000 & \textcolor{red}{22.2} (12.0--35.6) & 77 / 54 \\
AgentProv & 0.805 & 0.522 & 1.40 & 0.704 & 0.000 & \textcolor{red}{50.9} (41.1--60.7) & 88 / 108 \\
LIDAR$^{\dagger}$ & 0.625 & 0.627 & 0.67 & 0.512 & 0.261 & 4.9 (0.6--16.5) & 88 / 41 \\
MET-C & 0.669 & 0.720 & 1.17 & 0.700 & 0.455 & 0.0 (0.0--6.6) & 66 / 54 \\
LLMmap-C & 0.961 & 0.953 & 2.68 & 0.831 & 0.909 & 0.0 (0.0--6.6) & 88 / 54 \\
KBF-C & 0.866 & 0.781 & 1.54 & 0.902 & 0.545 & 0.0 (0.0--1.0) & 88 / 360 \\
IRIS-C & 0.531 & 0.504 & 0.13 & 0.450 & 0.000 & 0.0 (0.0--6.6) & 66 / 54 \\
AgentProv-C & 1.000 & 0.998 & 8.09 & 0.990 & 0.989 & 0.9 (0.0--5.1) & 88 / 108 \\
R alone & 1.000 & 1.000 & 8.73 & 0.984 & 1.000 & 1.9 (0.8--4.0) & 70 / 360 \\
K alone & 0.866 & 0.781 & 1.54 & 0.902 & 0.545 & 0.0 (0.0--1.0) & 88 / 360 \\
In-harness only & 0.852 & 0.825 & 1.82 & 0.782 & 0.641 & 0.8 (0.2--2.4) & 440 / 360 \\
CertClaim-S (static) & 0.846 & 0.772 & 0.97 & 0.694 & 0.511 & \textcolor{red}{68.1} (63.0--72.8) & 440 / 360 \\
CertClaim ($K{=}4$) & 1.000 & 0.999 & 8.06 & 0.970 & 1.000 & 3.6 (1.9--6.1) & 68 / 360 \\
CertClaim & 1.000 & 0.999 & 9.13 & 0.972 & 1.000 & 3.3 (1.7--5.8) & 68 / 360 \\
\bottomrule
\end{tabular}
\end{table}

\begin{table}[h]
\centering
\caption{LeaFBench metrics and false-positive rate (\%, with 95\% Clopper--Pearson interval) on the setting Qwen3-Coder, mini-swe-agent, authored task suite. Positives: whole-model substitutes; negatives: honest services in every authorized deployment. Invalid FPR (lower end of the interval above $\alpha=5\%$) in red. In-harness only and CertClaim-S use every in-harness audit, whereas CertClaim and the ablation in Table~\ref{tab:ablation} use the in-harness audits that are paired with probe audits.}
\label{tab:app-leaf-qwen3-coder-mini-swe-agent}
\scriptsize
\setlength{\tabcolsep}{3.0pt}
\begin{tabular}{lccccccc}
\toprule
Method & AUC$\uparrow$ & pAUC$\uparrow$ & MD$\uparrow$ & ACC$\uparrow$ & TPR@1\%FPR$\uparrow$ & FPR (\%) & \#pos / \#neg \\
\midrule
MET & 0.655 & 0.506 & 0.88 & 0.686 & 0.000 & \textcolor{red}{76.2} (60.5--87.9) & 60 / 42 \\
LLMmap & 0.873 & 0.763 & 1.65 & 0.745 & 0.517 & 4.8 (0.6--16.2) & 60 / 42 \\
RUT & 0.701 & 0.506 & 1.11 & 0.741 & 0.029 & \textcolor{red}{69.0} (52.9--82.4) & 70 / 42 \\
KBF & 0.864 & 0.896 & 1.97 & 0.831 & 0.786 & 0.0 (0.0--4.3) & 70 / 84 \\
IRIS & 0.835 & 0.866 & 0.65 & 0.673 & 0.738 & \textcolor{red}{59.5} (43.3--74.4) & 65 / 42 \\
AgentProv & 0.777 & 0.516 & 1.45 & 0.708 & 0.000 & \textcolor{red}{53.6} (42.4--64.5) & 70 / 84 \\
LIDAR$^{\dagger}$ & 0.932 & 0.856 & 2.26 & 0.852 & 0.718 & 4.8 (0.6--16.2) & 39 / 42 \\
MET-C & 0.988 & 0.878 & 5.82 & 0.971 & 0.000 & 7.1 (1.5--19.5) & 60 / 42 \\
LLMmap-C & 1.000 & 1.000 & 3.85 & 0.893 & 1.000 & 0.0 (0.0--8.4) & 70 / 42 \\
KBF-C & 0.978 & 0.954 & 3.37 & 0.922 & 0.900 & 0.0 (0.0--4.3) & 70 / 84 \\
IRIS-C & 0.913 & 0.897 & 2.75 & 0.882 & 0.800 & 0.0 (0.0--8.4) & 60 / 42 \\
AgentProv-C & 1.000 & 1.000 & 8.76 & 0.974 & 1.000 & 4.8 (1.3--11.7) & 70 / 84 \\
R alone & 1.000 & 1.000 & 7.79 & 0.979 & 1.000 & 3.6 (0.7--10.1) & 60 / 84 \\
K alone & 0.978 & 0.954 & 3.37 & 0.922 & 0.900 & 0.0 (0.0--4.3) & 70 / 84 \\
In-harness only & 0.972 & 0.961 & 4.27 & 0.951 & 0.920 & 0.0 (0.0--1.3) & 400 / 280 \\
CertClaim-S (static) & 0.731 & 0.758 & 0.56 & 0.643 & 0.517 & \textcolor{red}{86.8} (82.2--90.5) & 400 / 280 \\
CertClaim ($K{=}4$) & 1.000 & 1.000 & 7.69 & 0.951 & 1.000 & 8.3 (3.4--16.4) & 60 / 84 \\
CertClaim & 1.000 & 1.000 & 6.64 & 0.958 & 1.000 & 7.1 (2.7--14.9) & 60 / 84 \\
\bottomrule
\end{tabular}
\end{table}

\begin{table}[h]
\centering
\caption{LeaFBench metrics and false-positive rate (\%, with 95\% Clopper--Pearson interval) on the setting Qwen3-Coder, OpenCode, authored task suite. Positives: whole-model substitutes; negatives: honest services in every authorized deployment. Invalid FPR (lower end of the interval above $\alpha=5\%$) in red. In-harness only and CertClaim-S use every in-harness audit, whereas CertClaim and the ablation in Table~\ref{tab:ablation} use the in-harness audits that are paired with probe audits.}
\label{tab:app-leaf-qwen3-coder-opencode}
\scriptsize
\setlength{\tabcolsep}{3.0pt}
\begin{tabular}{lccccccc}
\toprule
Method & AUC$\uparrow$ & pAUC$\uparrow$ & MD$\uparrow$ & ACC$\uparrow$ & TPR@1\%FPR$\uparrow$ & FPR (\%) & \#pos / \#neg \\
\midrule
MET & 0.655 & 0.506 & 0.88 & 0.686 & 0.000 & \textcolor{red}{76.2} (60.5--87.9) & 60 / 42 \\
LLMmap & 0.873 & 0.763 & 1.65 & 0.745 & 0.517 & 4.8 (0.6--16.2) & 60 / 42 \\
RUT & 0.701 & 0.506 & 1.11 & 0.741 & 0.029 & \textcolor{red}{69.0} (52.9--82.4) & 70 / 42 \\
KBF & 0.864 & 0.896 & 1.97 & 0.831 & 0.786 & 0.0 (0.0--4.3) & 70 / 84 \\
IRIS & 0.835 & 0.866 & 0.65 & 0.673 & 0.738 & \textcolor{red}{59.5} (43.3--74.4) & 65 / 42 \\
AgentProv & 0.777 & 0.516 & 1.45 & 0.708 & 0.000 & \textcolor{red}{53.6} (42.4--64.5) & 70 / 84 \\
LIDAR$^{\dagger}$ & 0.875 & 0.786 & 1.77 & 0.841 & 0.475 & 4.8 (0.6--16.2) & 40 / 42 \\
MET-C & 0.988 & 0.878 & 5.82 & 0.971 & 0.000 & 7.1 (1.5--19.5) & 60 / 42 \\
LLMmap-C & 1.000 & 1.000 & 3.85 & 0.893 & 1.000 & 0.0 (0.0--8.4) & 70 / 42 \\
KBF-C & 0.978 & 0.954 & 3.37 & 0.922 & 0.900 & 0.0 (0.0--4.3) & 70 / 84 \\
IRIS-C & 0.913 & 0.897 & 2.75 & 0.882 & 0.800 & 0.0 (0.0--8.4) & 60 / 42 \\
AgentProv-C & 1.000 & 1.000 & 8.76 & 0.974 & 1.000 & 4.8 (1.3--11.7) & 70 / 84 \\
R alone & 1.000 & 1.000 & 7.79 & 0.979 & 1.000 & 3.6 (0.7--10.1) & 60 / 84 \\
K alone & 0.978 & 0.954 & 3.37 & 0.922 & 0.900 & 0.0 (0.0--4.3) & 70 / 84 \\
In-harness only & 0.952 & 0.922 & 3.24 & 0.894 & 0.835 & 0.4 (0.0--2.0) & 400 / 280 \\
CertClaim-S (static) & 0.731 & 0.758 & 0.57 & 0.647 & 0.517 & \textcolor{red}{85.7} (81.1--89.6) & 400 / 280 \\
CertClaim ($K{=}4$) & 1.000 & 1.000 & 9.02 & 0.979 & 1.000 & 3.6 (0.7--10.1) & 60 / 84 \\
CertClaim & 1.000 & 1.000 & 9.41 & 0.979 & 1.000 & 3.6 (0.7--10.1) & 60 / 84 \\
\bottomrule
\end{tabular}
\end{table}

\begin{table}[h]
\centering
\caption{LeaFBench metrics and false-positive rate (\%, with 95\% Clopper--Pearson interval) on the setting Devstral, mini-swe-agent, authored task suite. Positives: whole-model substitutes; negatives: honest services in every authorized deployment. Invalid FPR (lower end of the interval above $\alpha=5\%$) in red. In-harness only and CertClaim-S use every in-harness audit, whereas CertClaim and the ablation in Table~\ref{tab:ablation} use the in-harness audits that are paired with probe audits.}
\label{tab:app-leaf-devstral-mini-swe-agent}
\scriptsize
\setlength{\tabcolsep}{3.0pt}
\begin{tabular}{lccccccc}
\toprule
Method & AUC$\uparrow$ & pAUC$\uparrow$ & MD$\uparrow$ & ACC$\uparrow$ & TPR@1\%FPR$\uparrow$ & FPR (\%) & \#pos / \#neg \\
\midrule
MET & 0.647 & 0.534 & 0.87 & 0.696 & 0.000 & \textcolor{red}{14.3} (5.4--28.5) & 60 / 42 \\
LLMmap & 0.712 & 0.637 & 0.63 & 0.647 & 0.217 & 4.8 (0.6--16.2) & 60 / 42 \\
RUT & 0.472 & 0.516 & 0.03 & 0.562 & 0.057 & \textcolor{red}{71.4} (55.4--84.3) & 70 / 42 \\
KBF & 0.730 & 0.792 & 1.67 & 0.812 & 0.586 & 0.0 (0.0--4.3) & 70 / 84 \\
IRIS & 0.706 & 0.701 & 1.03 & 0.598 & 0.400 & 0.0 (0.0--8.4) & 65 / 42 \\
AgentProv & 0.578 & 0.503 & 0.27 & 0.526 & 0.000 & \textcolor{red}{72.6} (61.8--81.8) & 70 / 84 \\
LIDAR$^{\dagger}$ & 0.900 & 0.636 & 1.83 & 0.780 & 0.025 & 4.8 (0.6--16.2) & 40 / 42 \\
MET-C & 0.775 & 0.787 & 1.52 & 0.706 & 0.567 & 0.0 (0.0--8.4) & 60 / 42 \\
LLMmap-C & 0.871 & 0.824 & 1.95 & 0.786 & 0.657 & 0.0 (0.0--8.4) & 70 / 42 \\
KBF-C & 0.801 & 0.826 & 1.92 & 0.838 & 0.657 & 1.2 (0.0--6.5) & 70 / 84 \\
IRIS-C & 0.708 & 0.675 & 0.95 & 0.588 & 0.367 & 0.0 (0.0--8.4) & 60 / 42 \\
AgentProv-C & 0.940 & 0.760 & 2.79 & 0.890 & 0.000 & 7.1 (2.7--14.9) & 70 / 84 \\
R alone & 0.858 & 0.810 & 1.82 & 0.816 & 0.533 & 1.2 (0.0--6.5) & 90 / 84 \\
K alone & 0.801 & 0.826 & 1.92 & 0.838 & 0.657 & 1.2 (0.0--6.5) & 70 / 84 \\
In-harness only & 0.833 & 0.831 & 1.84 & 0.801 & 0.677 & 2.1 (0.8--4.6) & 400 / 280 \\
CertClaim-S (static) & 0.618 & 0.753 & 0.28 & 0.509 & 0.517 & \textcolor{red}{86.4} (81.9--90.2) & 400 / 280 \\
CertClaim ($K{=}4$) & 0.931 & 0.867 & 2.55 & 0.889 & 0.717 & 4.8 (1.3--11.7) & 60 / 84 \\
CertClaim & 0.920 & 0.890 & 2.61 & 0.903 & 0.733 & 2.4 (0.3--8.3) & 60 / 84 \\
\bottomrule
\end{tabular}
\end{table}

\begin{table}[h]
\centering
\caption{LeaFBench metrics and false-positive rate (\%, with 95\% Clopper--Pearson interval) on the setting Devstral, OpenCode, authored task suite. Positives: whole-model substitutes; negatives: honest services in every authorized deployment. Invalid FPR (lower end of the interval above $\alpha=5\%$) in red. In-harness only and CertClaim-S use every in-harness audit, whereas CertClaim and the ablation in Table~\ref{tab:ablation} use the in-harness audits that are paired with probe audits.}
\label{tab:app-leaf-devstral-opencode}
\scriptsize
\setlength{\tabcolsep}{3.0pt}
\begin{tabular}{lccccccc}
\toprule
Method & AUC$\uparrow$ & pAUC$\uparrow$ & MD$\uparrow$ & ACC$\uparrow$ & TPR@1\%FPR$\uparrow$ & FPR (\%) & \#pos / \#neg \\
\midrule
MET & 0.647 & 0.534 & 0.87 & 0.696 & 0.000 & \textcolor{red}{14.3} (5.4--28.5) & 60 / 42 \\
LLMmap & 0.712 & 0.637 & 0.63 & 0.647 & 0.217 & 4.8 (0.6--16.2) & 60 / 42 \\
RUT & 0.472 & 0.516 & 0.03 & 0.562 & 0.057 & \textcolor{red}{71.4} (55.4--84.3) & 70 / 42 \\
KBF & 0.730 & 0.792 & 1.67 & 0.812 & 0.586 & 0.0 (0.0--4.3) & 70 / 84 \\
IRIS & 0.706 & 0.701 & 1.03 & 0.598 & 0.400 & 0.0 (0.0--8.4) & 65 / 42 \\
AgentProv & 0.578 & 0.503 & 0.27 & 0.526 & 0.000 & \textcolor{red}{72.6} (61.8--81.8) & 70 / 84 \\
LIDAR$^{\dagger}$ & 0.796 & 0.684 & 1.20 & 0.790 & 0.359 & 4.8 (0.6--16.2) & 39 / 42 \\
MET-C & 0.775 & 0.787 & 1.52 & 0.706 & 0.567 & 0.0 (0.0--8.4) & 60 / 42 \\
LLMmap-C & 0.871 & 0.824 & 1.95 & 0.786 & 0.657 & 0.0 (0.0--8.4) & 70 / 42 \\
KBF-C & 0.801 & 0.826 & 1.92 & 0.838 & 0.657 & 1.2 (0.0--6.5) & 70 / 84 \\
IRIS-C & 0.708 & 0.675 & 0.95 & 0.588 & 0.367 & 0.0 (0.0--8.4) & 60 / 42 \\
AgentProv-C & 0.940 & 0.760 & 2.79 & 0.890 & 0.000 & 7.1 (2.7--14.9) & 70 / 84 \\
R alone & 0.858 & 0.810 & 1.82 & 0.816 & 0.533 & 1.2 (0.0--6.5) & 90 / 84 \\
K alone & 0.801 & 0.826 & 1.92 & 0.838 & 0.657 & 1.2 (0.0--6.5) & 70 / 84 \\
In-harness only & 0.836 & 0.784 & 1.63 & 0.740 & 0.562 & 0.7 (0.1--2.6) & 400 / 280 \\
CertClaim-S (static) & 0.620 & 0.753 & 0.25 & 0.519 & 0.517 & \textcolor{red}{85.7} (81.1--89.6) & 400 / 280 \\
CertClaim ($K{=}4$) & 0.947 & 0.894 & 2.86 & 0.903 & 0.750 & 3.6 (0.7--10.1) & 60 / 84 \\
CertClaim & 0.929 & 0.886 & 2.62 & 0.903 & 0.733 & 2.4 (0.3--8.3) & 60 / 84 \\
\bottomrule
\end{tabular}
\end{table}

\begin{table}[h]
\centering
\caption{LeaFBench metrics and false-positive rate (\%, with 95\% Clopper--Pearson interval) on the setting GPT-5.4, mini-swe-agent, authored task suite. Positives: whole-model substitutes; negatives: honest services in every authorized deployment. Invalid FPR (lower end of the interval above $\alpha=5\%$) in red. In-harness only and CertClaim-S use every in-harness audit, whereas CertClaim and the ablation in Table~\ref{tab:ablation} use the in-harness audits that are paired with probe audits.}
\label{tab:app-leaf-api-gpt-54}
\scriptsize
\setlength{\tabcolsep}{3.0pt}
\begin{tabular}{lccccccc}
\toprule
Method & AUC$\uparrow$ & pAUC$\uparrow$ & MD$\uparrow$ & ACC$\uparrow$ & TPR@1\%FPR$\uparrow$ & FPR (\%) & \#pos / \#neg \\
\midrule
MET & 0.764 & 0.514 & 1.28 & 0.694 & 0.000 & \textcolor{red}{61.1} (43.5--76.9) & 36 / 36 \\
LLMmap & 0.863 & 0.601 & 1.63 & 0.792 & 0.111 & 11.1 (3.1--26.1) & 36 / 36 \\
KBF & 0.699 & 0.843 & 2.11 & 0.847 & 0.694 & 0.0 (0.0--9.7) & 36 / 36 \\
IRIS & 0.845 & 0.841 & 1.94 & 0.859 & 0.690 & 13.9 (4.7--29.5) & 42 / 36 \\
AgentProv & 0.653 & 0.506 & 0.87 & 0.583 & 0.000 & \textcolor{red}{83.3} (67.2--93.6) & 36 / 36 \\
LIDAR$^{\dagger}$ & 0.984 & 0.943 & 3.07 & 0.917 & 0.889 & 4.2 (0.1--21.1) & 36 / 24 \\
MET-C & 1.000 & 1.000 & 10.73 & 1.000 & 1.000 & 0.0 (0.0--9.7) & 36 / 36 \\
LLMmap-C & 0.919 & 0.564 & 3.32 & 0.931 & 0.000 & 13.9 (4.7--29.5) & 36 / 36 \\
KBF-C & 0.895 & 0.915 & 3.00 & 0.917 & 0.833 & 0.0 (0.0--9.7) & 36 / 36 \\
IRIS-C & 0.819 & 0.523 & 1.56 & 0.500 & 0.000 & 0.0 (0.0--9.7) & 36 / 36 \\
AgentProv-C & 0.681 & 0.507 & 0.84 & 0.569 & 0.000 & \textcolor{red}{86.1} (70.5--95.3) & 36 / 36 \\
K alone & 0.895 & 0.915 & 3.00 & 0.917 & 0.833 & 0.0 (0.0--9.7) & 36 / 36 \\
In-harness only & 1.000 & 1.000 & 11.26 & 0.997 & 1.000 & 0.0 (0.0--1.5) & 360 / 240 \\
CertClaim-S (static) & 0.723 & 0.510 & 1.15 & 0.730 & 0.000 & \textcolor{red}{67.5} (61.2--73.4) & 360 / 240 \\
CertClaim ($K{=}4$) & 1.000 & 1.000 & 12.71 & 1.000 & 1.000 & 0.0 (0.0--9.7) & 36 / 36 \\
CertClaim & 1.000 & 1.000 & 13.60 & 1.000 & 1.000 & 0.0 (0.0--9.7) & 36 / 36 \\
\bottomrule
\end{tabular}
\end{table}

\begin{table}[h]
\centering
\caption{LeaFBench metrics and false-positive rate (\%, with 95\% Clopper--Pearson interval) on the setting GPT-5.6 Sol, mini-swe-agent, authored task suite. Positives: whole-model substitutes; negatives: honest services in every authorized deployment. Invalid FPR (lower end of the interval above $\alpha=5\%$) in red. In-harness only and CertClaim-S use every in-harness audit, whereas CertClaim and the ablation in Table~\ref{tab:ablation} use the in-harness audits that are paired with probe audits.}
\label{tab:app-leaf-api-gpt-56-sol}
\scriptsize
\setlength{\tabcolsep}{3.0pt}
\begin{tabular}{lccccccc}
\toprule
Method & AUC$\uparrow$ & pAUC$\uparrow$ & MD$\uparrow$ & ACC$\uparrow$ & TPR@1\%FPR$\uparrow$ & FPR (\%) & \#pos / \#neg \\
\midrule
MET & 1.000 & 1.000 & 3.91 & 0.859 & 1.000 & \textcolor{red}{30.6} (16.3--48.1) & 42 / 36 \\
LLMmap & 0.877 & 0.654 & 1.58 & 0.795 & 0.071 & \textcolor{red}{36.1} (20.8--53.8) & 42 / 36 \\
KBF & 0.571 & 0.516 & 0.81 & 0.410 & 0.000 & \textcolor{red}{100.0} (90.3--100.0) & 42 / 36 \\
IRIS & 0.848 & 0.893 & 1.83 & 0.835 & 0.755 & \textcolor{red}{16.7} (6.4--32.8) & 49 / 36 \\
AgentProv & 0.903 & 0.553 & 1.84 & 0.705 & 0.000 & \textcolor{red}{63.9} (46.2--79.2) & 42 / 36 \\
LIDAR$^{\dagger}$ & 0.925 & 0.710 & 2.10 & 0.717 & 0.339 & 2.8 (0.1--14.5) & 56 / 36 \\
MET-C & 1.000 & 1.000 & 30.20 & 1.000 & 1.000 & 0.0 (0.0--9.7) & 42 / 36 \\
LLMmap-C & 1.000 & 1.000 & 5.95 & 0.949 & 1.000 & 0.0 (0.0--9.7) & 42 / 36 \\
KBF-C & 0.647 & 0.723 & 1.19 & 0.718 & 0.429 & 2.8 (0.1--14.5) & 42 / 36 \\
IRIS-C & 0.995 & 0.981 & 4.59 & 0.936 & 0.952 & 0.0 (0.0--9.7) & 42 / 36 \\
AgentProv-C & 1.000 & 1.000 & 7.90 & 1.000 & 1.000 & 0.0 (0.0--9.7) & 42 / 36 \\
K alone & 0.647 & 0.723 & 1.19 & 0.718 & 0.429 & 2.8 (0.1--14.5) & 42 / 36 \\
In-harness only & 0.966 & 0.865 & 3.01 & 0.892 & 0.640 & 2.8 (1.3--5.0) & 420 / 360 \\
CertClaim-S (static) & 1.000 & 1.000 & 5.46 & 0.937 & 1.000 & \textcolor{red}{13.6} (10.2--17.6) & 420 / 360 \\
CertClaim ($K{=}4$) & 1.000 & 1.000 & 11.61 & 0.987 & 1.000 & 2.8 (0.1--14.5) & 42 / 36 \\
CertClaim & 1.000 & 1.000 & 9.63 & 0.987 & 1.000 & 2.8 (0.1--14.5) & 42 / 36 \\
\bottomrule
\end{tabular}
\end{table}

\begin{table}[h]
\centering
\caption{LeaFBench metrics and false-positive rate (\%, with 95\% Clopper--Pearson interval) on the setting GPT-6.1 Sol, mini-swe-agent, authored task suite. Positives: whole-model substitutes; negatives: honest services in every authorized deployment. Invalid FPR (lower end of the interval above $\alpha=5\%$) in red. In-harness only and CertClaim-S use every in-harness audit, whereas CertClaim and the ablation in Table~\ref{tab:ablation} use the in-harness audits that are paired with probe audits.}
\label{tab:app-leaf-api-gpt-61-coding}
\scriptsize
\setlength{\tabcolsep}{3.0pt}
\begin{tabular}{lccccccc}
\toprule
Method & AUC$\uparrow$ & pAUC$\uparrow$ & MD$\uparrow$ & ACC$\uparrow$ & TPR@1\%FPR$\uparrow$ & FPR (\%) & \#pos / \#neg \\
\midrule
MET & 1.000 & 1.000 & 5.94 & 0.949 & 1.000 & 11.1 (3.1--26.1) & 42 / 36 \\
LLMmap & 0.928 & 0.871 & 2.10 & 0.808 & 0.738 & \textcolor{red}{19.4} (8.2--36.0) & 42 / 36 \\
KBF & 0.857 & 0.927 & 3.43 & 0.923 & 0.857 & 0.0 (0.0--9.7) & 42 / 36 \\
IRIS & 0.857 & 0.927 & 2.46 & 0.918 & 0.857 & 0.0 (0.0--9.7) & 49 / 36 \\
AgentProv & 0.800 & 0.519 & 1.21 & 0.718 & 0.000 & \textcolor{red}{66.7} (47.2--82.7) & 41 / 30 \\
LIDAR$^{\dagger}$ & 0.742 & 0.753 & 1.23 & 0.778 & 0.519 & 2.8 (0.1--14.5) & 27 / 36 \\
MET-C & 1.000 & 1.000 & 26.28 & 1.000 & 1.000 & 0.0 (0.0--9.7) & 42 / 36 \\
LLMmap-C & 1.000 & 1.000 & 5.83 & 0.930 & 1.000 & 0.0 (0.0--10.6) & 38 / 33 \\
KBF-C & 1.000 & 1.000 & 7.25 & 0.962 & 1.000 & 0.0 (0.0--9.7) & 42 / 36 \\
IRIS-C & 1.000 & 1.000 & 13.28 & 1.000 & 1.000 & 0.0 (0.0--9.7) & 42 / 36 \\
AgentProv-C & 0.983 & 0.829 & 5.48 & 0.958 & 0.000 & 10.0 (2.1--26.5) & 41 / 30 \\
K alone & 1.000 & 1.000 & 7.25 & 0.962 & 1.000 & 0.0 (0.0--9.7) & 42 / 36 \\
In-harness only & 1.000 & 0.998 & 7.15 & 0.987 & 1.000 & 0.3 (0.0--1.5) & 250 / 360 \\
CertClaim-S (static) & 1.000 & 1.000 & 6.94 & 0.967 & 1.000 & 5.6 (3.4--8.4) & 250 / 360 \\
CertClaim ($K{=}4$) & 1.000 & 1.000 & 11.00 & 1.000 & 1.000 & 0.0 (0.0--9.7) & 18 / 36 \\
CertClaim & 1.000 & 1.000 & 14.04 & 1.000 & 1.000 & 0.0 (0.0--9.7) & 42 / 36 \\
\bottomrule
\end{tabular}
\end{table}

\begin{table}[h]
\centering
\caption{LeaFBench metrics and false-positive rate (\%, with 95\% Clopper--Pearson interval) on the setting Sonnet 4.6, mini-swe-agent, authored task suite. Positives: whole-model substitutes; negatives: honest services in every authorized deployment. Invalid FPR (lower end of the interval above $\alpha=5\%$) in red. In-harness only and CertClaim-S use every in-harness audit, whereas CertClaim and the ablation in Table~\ref{tab:ablation} use the in-harness audits that are paired with probe audits.}
\label{tab:app-leaf-api-claude-sonnet-46}
\scriptsize
\setlength{\tabcolsep}{3.0pt}
\begin{tabular}{lccccccc}
\toprule
Method & AUC$\uparrow$ & pAUC$\uparrow$ & MD$\uparrow$ & ACC$\uparrow$ & TPR@1\%FPR$\uparrow$ & FPR (\%) & \#pos / \#neg \\
\midrule
MET & 0.833 & 0.526 & 1.62 & 0.722 & 0.000 & \textcolor{red}{55.6} (38.1--72.1) & 36 / 36 \\
LLMmap & 0.848 & 0.804 & 1.34 & 0.806 & 0.556 & 8.3 (1.8--22.5) & 36 / 36 \\
KBF & 0.833 & 0.915 & 3.12 & 0.917 & 0.833 & 0.0 (0.0--9.7) & 36 / 36 \\
IRIS & 0.519 & 0.707 & 0.48 & 0.615 & 0.429 & \textcolor{red}{16.7} (6.4--32.8) & 42 / 36 \\
AgentProv & 0.708 & 0.509 & 0.83 & 0.597 & 0.000 & \textcolor{red}{80.6} (64.0--91.8) & 36 / 36 \\
LIDAR$^{\dagger}$ & 0.977 & 0.972 & 3.72 & 0.950 & 0.944 & 4.2 (0.1--21.1) & 36 / 24 \\
MET-C & 1.000 & 1.000 & 12.47 & 0.986 & 1.000 & 2.8 (0.1--14.5) & 36 / 36 \\
LLMmap-C & 0.828 & 0.525 & 1.79 & 0.806 & 0.000 & \textcolor{red}{38.9} (23.1--56.5) & 36 / 36 \\
KBF-C & 0.852 & 0.915 & 2.89 & 0.917 & 0.833 & 0.0 (0.0--9.7) & 36 / 36 \\
IRIS-C & 0.518 & 0.508 & 0.16 & 0.500 & 0.000 & 0.0 (0.0--9.7) & 36 / 36 \\
AgentProv-C & 0.681 & 0.507 & 0.95 & 0.611 & 0.000 & \textcolor{red}{77.8} (60.8--89.9) & 36 / 36 \\
K alone & 0.852 & 0.915 & 2.89 & 0.917 & 0.833 & 0.0 (0.0--9.7) & 36 / 36 \\
In-harness only & 1.000 & 1.000 & 12.84 & 1.000 & 1.000 & 0.0 (0.0--1.5) & 360 / 240 \\
CertClaim-S (static) & 0.792 & 0.518 & 1.54 & 0.802 & 0.000 & \textcolor{red}{49.6} (43.1--56.1) & 360 / 240 \\
CertClaim ($K{=}4$) & 1.000 & 1.000 & 13.43 & 1.000 & 1.000 & 0.0 (0.0--9.7) & 36 / 36 \\
CertClaim & 1.000 & 1.000 & 16.04 & 1.000 & 1.000 & 0.0 (0.0--9.7) & 36 / 36 \\
\bottomrule
\end{tabular}
\end{table}

\begin{table}[h]
\centering
\caption{LeaFBench metrics and false-positive rate (\%, with 95\% Clopper--Pearson interval) on the setting Opus 5.5, mini-swe-agent, authored task suite. Positives: whole-model substitutes; negatives: honest services in every authorized deployment. Invalid FPR (lower end of the interval above $\alpha=5\%$) in red. In-harness only and CertClaim-S use every in-harness audit, whereas CertClaim and the ablation in Table~\ref{tab:ablation} use the in-harness audits that are paired with probe audits.}
\label{tab:app-leaf-api-claude-opus-55}
\scriptsize
\setlength{\tabcolsep}{3.0pt}
\begin{tabular}{lccccccc}
\toprule
Method & AUC$\uparrow$ & pAUC$\uparrow$ & MD$\uparrow$ & ACC$\uparrow$ & TPR@1\%FPR$\uparrow$ & FPR (\%) & \#pos / \#neg \\
\midrule
MET & 0.812 & 0.521 & 1.31 & 0.589 & 0.000 & \textcolor{red}{77.1} (62.7--88.0) & 42 / 48 \\
LLMmap & 0.992 & 0.941 & 2.76 & 0.789 & 0.857 & 0.0 (0.0--7.4) & 42 / 48 \\
KBF & 0.812 & 0.902 & 2.91 & 0.911 & 0.810 & 0.0 (0.0--7.4) & 42 / 48 \\
IRIS & 0.857 & 0.927 & 2.89 & 0.928 & 0.857 & 0.0 (0.0--7.4) & 49 / 48 \\
AgentProv & 0.750 & 0.513 & 1.14 & 0.644 & 0.000 & \textcolor{red}{66.7} (51.6--79.6) & 42 / 48 \\
LIDAR$^{\dagger}$ & 0.811 & 0.701 & 1.29 & 0.731 & 0.375 & 4.2 (0.5--14.3) & 56 / 48 \\
MET-C & 1.000 & 1.000 & 85.13 & 1.000 & 1.000 & 0.0 (0.0--7.4) & 42 / 48 \\
LLMmap-C & 1.000 & 1.000 & 10.45 & 1.000 & 1.000 & 0.0 (0.0--7.4) & 42 / 48 \\
KBF-C & 0.998 & 0.988 & 6.60 & 0.978 & 0.976 & 0.0 (0.0--7.4) & 42 / 48 \\
IRIS-C & 1.000 & 1.000 & 13.49 & 1.000 & 1.000 & 0.0 (0.0--7.4) & 42 / 48 \\
AgentProv-C & 0.875 & 0.538 & 1.96 & 0.767 & 0.000 & \textcolor{red}{43.8} (29.5--58.8) & 42 / 48 \\
K alone & 0.998 & 0.988 & 6.60 & 0.978 & 0.976 & 0.0 (0.0--7.4) & 42 / 48 \\
In-harness only & 0.893 & 0.838 & 2.01 & 0.824 & 0.674 & 0.2 (0.0--1.2) & 420 / 480 \\
CertClaim-S (static) & 1.000 & 1.000 & 7.11 & 0.967 & 1.000 & 6.2 (4.3--8.8) & 420 / 480 \\
CertClaim ($K{=}4$) & 1.000 & 1.000 & 40.44 & 1.000 & 1.000 & 0.0 (0.0--7.4) & 42 / 48 \\
CertClaim & 1.000 & 1.000 & 17.94 & 0.989 & 1.000 & 2.1 (0.1--11.1) & 42 / 48 \\
\bottomrule
\end{tabular}
\end{table}

\begin{table}[h]
\centering
\caption{LeaFBench metrics and false-positive rate (\%, with 95\% Clopper--Pearson interval) on the setting Qwen3.5-27B, mini-swe-agent, EvalPlus task suite. Positives: whole-model substitutes; negatives: honest services in every authorized deployment. Invalid FPR (lower end of the interval above $\alpha=5\%$) in red. In-harness only and CertClaim-S use every in-harness audit, whereas CertClaim and the ablation in Table~\ref{tab:ablation} use the in-harness audits that are paired with probe audits.}
\label{tab:app-leaf-qwen35-27b-evalplus-mini-swe-agent}
\scriptsize
\setlength{\tabcolsep}{3.0pt}
\begin{tabular}{lccccccc}
\toprule
Method & AUC$\uparrow$ & pAUC$\uparrow$ & MD$\uparrow$ & ACC$\uparrow$ & TPR@1\%FPR$\uparrow$ & FPR (\%) & \#pos / \#neg \\
\midrule
MET & 0.618 & 0.513 & 0.51 & 0.600 & 0.000 & \textcolor{red}{22.2} (12.0--35.6) & 66 / 54 \\
LLMmap & 0.427 & 0.492 & 0.30 & 0.433 & 0.000 & \textcolor{red}{48.1} (34.3--62.2) & 66 / 54 \\
RUT & 0.539 & 0.487 & 0.82 & 0.718 & 0.000 & \textcolor{red}{70.4} (56.4--82.0) & 88 / 54 \\
KBF & 0.693 & 0.735 & 1.07 & 0.766 & 0.455 & \textcolor{red}{18.6} (14.7--23.0) & 88 / 360 \\
IRIS & 0.445 & 0.495 & 0.01 & 0.496 & 0.000 & \textcolor{red}{22.2} (12.0--35.6) & 77 / 54 \\
AgentProv & 0.805 & 0.522 & 1.40 & 0.704 & 0.000 & \textcolor{red}{50.9} (41.1--60.7) & 88 / 108 \\
LIDAR$^{\dagger}$ & 0.611 & 0.530 & 0.40 & 0.426 & 0.069 & 3.7 (0.5--12.7) & 87 / 54 \\
MET-C & 0.669 & 0.720 & 1.17 & 0.700 & 0.455 & 0.0 (0.0--6.6) & 66 / 54 \\
LLMmap-C & 0.961 & 0.953 & 2.68 & 0.831 & 0.909 & 0.0 (0.0--6.6) & 88 / 54 \\
KBF-C & 0.866 & 0.781 & 1.54 & 0.902 & 0.545 & 0.0 (0.0--1.0) & 88 / 360 \\
IRIS-C & 0.531 & 0.504 & 0.13 & 0.450 & 0.000 & 0.0 (0.0--6.6) & 66 / 54 \\
AgentProv-C & 1.000 & 0.998 & 8.09 & 0.990 & 0.989 & 0.9 (0.0--5.1) & 88 / 108 \\
R alone & 1.000 & 1.000 & 8.73 & 0.984 & 1.000 & 1.9 (0.8--4.0) & 70 / 360 \\
K alone & 0.866 & 0.781 & 1.54 & 0.902 & 0.545 & 0.0 (0.0--1.0) & 88 / 360 \\
In-harness only & 0.908 & 0.844 & 2.22 & 0.821 & 0.657 & 1.9 (0.8--4.0) & 440 / 360 \\
CertClaim-S (static) & 0.846 & 0.772 & 0.97 & 0.694 & 0.511 & \textcolor{red}{68.1} (63.0--72.8) & 440 / 360 \\
CertClaim ($K{=}4$) & 1.000 & 0.999 & 8.56 & 0.977 & 1.000 & 2.8 (1.3--5.0) & 68 / 360 \\
CertClaim & 1.000 & 1.000 & 10.29 & 0.986 & 1.000 & 1.7 (0.6--3.6) & 68 / 360 \\
\bottomrule
\end{tabular}
\end{table}

\begin{table}[h]
\centering
\caption{LeaFBench metrics and false-positive rate (\%, with 95\% Clopper--Pearson interval) on the setting Qwen3.5-27B, OpenCode, EvalPlus task suite. Positives: whole-model substitutes; negatives: honest services in every authorized deployment. Invalid FPR (lower end of the interval above $\alpha=5\%$) in red. In-harness only and CertClaim-S use every in-harness audit, whereas CertClaim and the ablation in Table~\ref{tab:ablation} use the in-harness audits that are paired with probe audits.}
\label{tab:app-leaf-qwen35-27b-evalplus-opencode}
\scriptsize
\setlength{\tabcolsep}{3.0pt}
\begin{tabular}{lccccccc}
\toprule
Method & AUC$\uparrow$ & pAUC$\uparrow$ & MD$\uparrow$ & ACC$\uparrow$ & TPR@1\%FPR$\uparrow$ & FPR (\%) & \#pos / \#neg \\
\midrule
MET & 0.618 & 0.513 & 0.51 & 0.600 & 0.000 & \textcolor{red}{22.2} (12.0--35.6) & 66 / 54 \\
LLMmap & 0.427 & 0.492 & 0.30 & 0.433 & 0.000 & \textcolor{red}{48.1} (34.3--62.2) & 66 / 54 \\
RUT & 0.539 & 0.487 & 0.82 & 0.718 & 0.000 & \textcolor{red}{70.4} (56.4--82.0) & 88 / 54 \\
KBF & 0.693 & 0.735 & 1.07 & 0.766 & 0.455 & \textcolor{red}{18.6} (14.7--23.0) & 88 / 360 \\
IRIS & 0.445 & 0.495 & 0.01 & 0.496 & 0.000 & \textcolor{red}{22.2} (12.0--35.6) & 77 / 54 \\
AgentProv & 0.805 & 0.522 & 1.40 & 0.704 & 0.000 & \textcolor{red}{50.9} (41.1--60.7) & 88 / 108 \\
LIDAR$^{\dagger}$ & 0.824 & 0.700 & 1.23 & 0.662 & 0.330 & 3.7 (0.5--12.7) & 88 / 54 \\
MET-C & 0.669 & 0.720 & 1.17 & 0.700 & 0.455 & 0.0 (0.0--6.6) & 66 / 54 \\
LLMmap-C & 0.961 & 0.953 & 2.68 & 0.831 & 0.909 & 0.0 (0.0--6.6) & 88 / 54 \\
KBF-C & 0.866 & 0.781 & 1.54 & 0.902 & 0.545 & 0.0 (0.0--1.0) & 88 / 360 \\
IRIS-C & 0.531 & 0.504 & 0.13 & 0.450 & 0.000 & 0.0 (0.0--6.6) & 66 / 54 \\
AgentProv-C & 1.000 & 0.998 & 8.09 & 0.990 & 0.989 & 0.9 (0.0--5.1) & 88 / 108 \\
R alone & 1.000 & 1.000 & 8.73 & 0.984 & 1.000 & 1.9 (0.8--4.0) & 70 / 360 \\
K alone & 0.866 & 0.781 & 1.54 & 0.902 & 0.545 & 0.0 (0.0--1.0) & 88 / 360 \\
In-harness only & 0.900 & 0.850 & 2.08 & 0.792 & 0.648 & 0.8 (0.2--2.4) & 440 / 360 \\
CertClaim-S (static) & 0.846 & 0.772 & 0.97 & 0.694 & 0.511 & \textcolor{red}{68.1} (63.0--72.8) & 440 / 360 \\
CertClaim ($K{=}4$) & 1.000 & 1.000 & 10.77 & 0.986 & 1.000 & 1.7 (0.6--3.6) & 68 / 360 \\
CertClaim & 1.000 & 1.000 & 10.29 & 0.986 & 1.000 & 1.7 (0.6--3.6) & 68 / 360 \\
\bottomrule
\end{tabular}
\end{table}

\begin{table}[h]
\centering
\caption{LeaFBench metrics and false-positive rate (\%, with 95\% Clopper--Pearson interval) on the setting Qwen3-Coder, mini-swe-agent, EvalPlus task suite. Positives: whole-model substitutes; negatives: honest services in every authorized deployment. Invalid FPR (lower end of the interval above $\alpha=5\%$) in red. In-harness only and CertClaim-S use every in-harness audit, whereas CertClaim and the ablation in Table~\ref{tab:ablation} use the in-harness audits that are paired with probe audits.}
\label{tab:app-leaf-qwen3-coder-evalplus-mini-swe-agent}
\scriptsize
\setlength{\tabcolsep}{3.0pt}
\begin{tabular}{lccccccc}
\toprule
Method & AUC$\uparrow$ & pAUC$\uparrow$ & MD$\uparrow$ & ACC$\uparrow$ & TPR@1\%FPR$\uparrow$ & FPR (\%) & \#pos / \#neg \\
\midrule
MET & 0.655 & 0.506 & 0.88 & 0.556 & 0.000 & \textcolor{red}{76.2} (60.5--87.9) & 30 / 42 \\
LLMmap & 0.900 & 0.788 & 1.83 & 0.833 & 0.567 & 4.8 (0.6--16.2) & 30 / 42 \\
RUT & 0.677 & 0.487 & 1.11 & 0.646 & 0.000 & \textcolor{red}{69.0} (52.9--82.4) & 40 / 42 \\
KBF & 0.800 & 0.897 & 2.83 & 0.935 & 0.800 & 0.0 (0.0--4.3) & 40 / 84 \\
IRIS & 0.857 & 0.927 & 0.63 & 0.610 & 0.857 & \textcolor{red}{59.5} (43.3--74.4) & 35 / 42 \\
AgentProv & 0.780 & 0.516 & 1.45 & 0.637 & 0.000 & \textcolor{red}{53.6} (42.4--64.5) & 40 / 84 \\
LIDAR$^{\dagger}$ & 0.895 & 0.662 & 1.73 & 0.683 & 0.275 & 4.8 (0.6--16.2) & 40 / 42 \\
MET-C & 0.988 & 0.878 & 5.82 & 0.958 & 0.000 & 7.1 (1.5--19.5) & 30 / 42 \\
LLMmap-C & 1.000 & 1.000 & 1083.04 & 1.000 & 1.000 & 0.0 (0.0--8.4) & 40 / 42 \\
KBF-C & 1.000 & 1.000 & 17.62 & 1.000 & 1.000 & 0.0 (0.0--4.3) & 40 / 84 \\
IRIS-C & 1.000 & 1.000 & 28.24 & 1.000 & 1.000 & 0.0 (0.0--8.4) & 30 / 42 \\
AgentProv-C & 1.000 & 1.000 & 8.76 & 0.968 & 1.000 & 4.8 (1.3--11.7) & 40 / 84 \\
R alone & 1.000 & 1.000 & 7.79 & 0.979 & 1.000 & 3.6 (0.7--10.1) & 60 / 84 \\
K alone & 0.978 & 0.954 & 3.37 & 0.922 & 0.900 & 0.0 (0.0--4.3) & 70 / 84 \\
In-harness only & 1.000 & 1.000 & 9.07 & 1.000 & 1.000 & 0.0 (0.0--1.3) & 200 / 280 \\
CertClaim-S (static) & 0.747 & 0.795 & 0.57 & 0.498 & 0.600 & \textcolor{red}{86.1} (81.5--89.9) & 200 / 280 \\
CertClaim ($K{=}4$) & 1.000 & 1.000 & 10.56 & 0.982 & 1.000 & 2.4 (0.3--8.3) & 30 / 84 \\
CertClaim & 1.000 & 1.000 & 7.72 & 0.965 & 1.000 & 4.8 (1.3--11.7) & 30 / 84 \\
\bottomrule
\end{tabular}
\end{table}

\begin{table}[h]
\centering
\caption{LeaFBench metrics and false-positive rate (\%, with 95\% Clopper--Pearson interval) on the setting Qwen3-Coder, OpenCode, EvalPlus task suite. Positives: whole-model substitutes; negatives: honest services in every authorized deployment. Invalid FPR (lower end of the interval above $\alpha=5\%$) in red. In-harness only and CertClaim-S use every in-harness audit, whereas CertClaim and the ablation in Table~\ref{tab:ablation} use the in-harness audits that are paired with probe audits.}
\label{tab:app-leaf-qwen3-coder-evalplus-opencode}
\scriptsize
\setlength{\tabcolsep}{3.0pt}
\begin{tabular}{lccccccc}
\toprule
Method & AUC$\uparrow$ & pAUC$\uparrow$ & MD$\uparrow$ & ACC$\uparrow$ & TPR@1\%FPR$\uparrow$ & FPR (\%) & \#pos / \#neg \\
\midrule
MET & 0.655 & 0.506 & 0.88 & 0.556 & 0.000 & \textcolor{red}{76.2} (60.5--87.9) & 30 / 42 \\
LLMmap & 0.900 & 0.788 & 1.83 & 0.833 & 0.567 & 4.8 (0.6--16.2) & 30 / 42 \\
RUT & 0.677 & 0.487 & 1.11 & 0.646 & 0.000 & \textcolor{red}{69.0} (52.9--82.4) & 40 / 42 \\
KBF & 0.800 & 0.897 & 2.83 & 0.935 & 0.800 & 0.0 (0.0--4.3) & 40 / 84 \\
IRIS & 0.857 & 0.927 & 0.63 & 0.610 & 0.857 & \textcolor{red}{59.5} (43.3--74.4) & 35 / 42 \\
AgentProv & 0.780 & 0.516 & 1.45 & 0.637 & 0.000 & \textcolor{red}{53.6} (42.4--64.5) & 40 / 84 \\
LIDAR$^{\dagger}$ & 0.904 & 0.782 & 1.86 & 0.768 & 0.575 & 4.8 (0.6--16.2) & 40 / 42 \\
MET-C & 0.988 & 0.878 & 5.82 & 0.958 & 0.000 & 7.1 (1.5--19.5) & 30 / 42 \\
LLMmap-C & 1.000 & 1.000 & 1083.04 & 1.000 & 1.000 & 0.0 (0.0--8.4) & 40 / 42 \\
KBF-C & 1.000 & 1.000 & 17.62 & 1.000 & 1.000 & 0.0 (0.0--4.3) & 40 / 84 \\
IRIS-C & 1.000 & 1.000 & 28.24 & 1.000 & 1.000 & 0.0 (0.0--8.4) & 30 / 42 \\
AgentProv-C & 1.000 & 1.000 & 8.76 & 0.968 & 1.000 & 4.8 (1.3--11.7) & 40 / 84 \\
R alone & 1.000 & 1.000 & 7.79 & 0.979 & 1.000 & 3.6 (0.7--10.1) & 60 / 84 \\
K alone & 0.978 & 0.954 & 3.37 & 0.922 & 0.900 & 0.0 (0.0--4.3) & 70 / 84 \\
In-harness only & 1.000 & 1.000 & 11.27 & 0.998 & 1.000 & 0.0 (0.0--1.3) & 200 / 280 \\
CertClaim-S (static) & 0.747 & 0.795 & 0.56 & 0.487 & 0.600 & \textcolor{red}{87.9} (83.4--91.4) & 200 / 280 \\
CertClaim ($K{=}4$) & 1.000 & 1.000 & 8.82 & 0.974 & 1.000 & 3.6 (0.7--10.1) & 30 / 84 \\
CertClaim & 1.000 & 1.000 & 7.72 & 0.956 & 1.000 & 6.0 (2.0--13.3) & 30 / 84 \\
\bottomrule
\end{tabular}
\end{table}

\begin{table}[h]
\centering
\caption{LeaFBench metrics and false-positive rate (\%, with 95\% Clopper--Pearson interval) on the setting Devstral, mini-swe-agent, EvalPlus task suite. Positives: whole-model substitutes; negatives: honest services in every authorized deployment. Invalid FPR (lower end of the interval above $\alpha=5\%$) in red. In-harness only and CertClaim-S use every in-harness audit, whereas CertClaim and the ablation in Table~\ref{tab:ablation} use the in-harness audits that are paired with probe audits.}
\label{tab:app-leaf-devstral-evalplus-mini-swe-agent}
\scriptsize
\setlength{\tabcolsep}{3.0pt}
\begin{tabular}{lccccccc}
\toprule
Method & AUC$\uparrow$ & pAUC$\uparrow$ & MD$\uparrow$ & ACC$\uparrow$ & TPR@1\%FPR$\uparrow$ & FPR (\%) & \#pos / \#neg \\
\midrule
MET & 0.929 & 0.577 & 3.35 & 0.917 & 0.000 & \textcolor{red}{14.3} (5.4--28.5) & 30 / 42 \\
LLMmap & 0.768 & 0.730 & 0.70 & 0.778 & 0.433 & 4.8 (0.6--16.2) & 30 / 42 \\
RUT & 0.658 & 0.538 & 1.00 & 0.634 & 0.100 & \textcolor{red}{71.4} (55.4--84.3) & 40 / 42 \\
KBF & 0.875 & 0.936 & 3.66 & 0.960 & 0.875 & 0.0 (0.0--4.3) & 40 / 84 \\
IRIS & 0.795 & 0.853 & 1.75 & 0.831 & 0.714 & 0.0 (0.0--8.4) & 35 / 42 \\
AgentProv & 0.673 & 0.507 & 0.92 & 0.508 & 0.000 & \textcolor{red}{72.6} (61.8--81.8) & 40 / 84 \\
LIDAR$^{\dagger}$ & 0.986 & 0.974 & 2.10 & 0.963 & 0.949 & 4.8 (0.6--16.2) & 39 / 42 \\
MET-C & 1.000 & 1.000 & 21.76 & 1.000 & 1.000 & 0.0 (0.0--8.4) & 30 / 42 \\
LLMmap-C & 1.000 & 1.000 & 82.63 & 1.000 & 1.000 & 0.0 (0.0--8.4) & 40 / 42 \\
KBF-C & 1.000 & 1.000 & 16.15 & 0.992 & 1.000 & 1.2 (0.0--6.5) & 40 / 84 \\
IRIS-C & 0.777 & 0.795 & 1.70 & 0.833 & 0.600 & 0.0 (0.0--8.4) & 30 / 42 \\
AgentProv-C & 0.982 & 0.817 & 5.38 & 0.952 & 0.000 & 7.1 (2.7--14.9) & 40 / 84 \\
R alone & 0.858 & 0.810 & 1.82 & 0.816 & 0.533 & 1.2 (0.0--6.5) & 90 / 84 \\
K alone & 0.801 & 0.826 & 1.92 & 0.838 & 0.657 & 1.2 (0.0--6.5) & 70 / 84 \\
In-harness only & 1.000 & 0.997 & 7.93 & 0.988 & 0.975 & 2.1 (0.8--4.6) & 200 / 280 \\
CertClaim-S (static) & 0.945 & 0.915 & 0.56 & 0.494 & 0.835 & \textcolor{red}{86.8} (82.2--90.5) & 200 / 280 \\
CertClaim ($K{=}4$) & 1.000 & 1.000 & 10.67 & 0.974 & 1.000 & 3.6 (0.7--10.1) & 30 / 84 \\
CertClaim & 1.000 & 1.000 & 9.68 & 0.982 & 1.000 & 2.4 (0.3--8.3) & 30 / 84 \\
\bottomrule
\end{tabular}
\end{table}

\begin{table}[h]
\centering
\caption{LeaFBench metrics and false-positive rate (\%, with 95\% Clopper--Pearson interval) on the setting Devstral, OpenCode, EvalPlus task suite. Positives: whole-model substitutes; negatives: honest services in every authorized deployment. Invalid FPR (lower end of the interval above $\alpha=5\%$) in red. In-harness only and CertClaim-S use every in-harness audit, whereas CertClaim and the ablation in Table~\ref{tab:ablation} use the in-harness audits that are paired with probe audits.}
\label{tab:app-leaf-devstral-evalplus-opencode}
\scriptsize
\setlength{\tabcolsep}{3.0pt}
\begin{tabular}{lccccccc}
\toprule
Method & AUC$\uparrow$ & pAUC$\uparrow$ & MD$\uparrow$ & ACC$\uparrow$ & TPR@1\%FPR$\uparrow$ & FPR (\%) & \#pos / \#neg \\
\midrule
MET & 0.929 & 0.577 & 3.35 & 0.917 & 0.000 & \textcolor{red}{14.3} (5.4--28.5) & 30 / 42 \\
LLMmap & 0.768 & 0.730 & 0.70 & 0.778 & 0.433 & 4.8 (0.6--16.2) & 30 / 42 \\
RUT & 0.658 & 0.538 & 1.00 & 0.634 & 0.100 & \textcolor{red}{71.4} (55.4--84.3) & 40 / 42 \\
KBF & 0.875 & 0.936 & 3.66 & 0.960 & 0.875 & 0.0 (0.0--4.3) & 40 / 84 \\
IRIS & 0.795 & 0.853 & 1.75 & 0.831 & 0.714 & 0.0 (0.0--8.4) & 35 / 42 \\
AgentProv & 0.673 & 0.507 & 0.92 & 0.508 & 0.000 & \textcolor{red}{72.6} (61.8--81.8) & 40 / 84 \\
LIDAR$^{\dagger}$ & 0.651 & 0.487 & 0.39 & 0.488 & 0.000 & 4.8 (0.6--16.2) & 40 / 42 \\
MET-C & 1.000 & 1.000 & 21.76 & 1.000 & 1.000 & 0.0 (0.0--8.4) & 30 / 42 \\
LLMmap-C & 1.000 & 1.000 & 82.63 & 1.000 & 1.000 & 0.0 (0.0--8.4) & 40 / 42 \\
KBF-C & 1.000 & 1.000 & 16.15 & 0.992 & 1.000 & 1.2 (0.0--6.5) & 40 / 84 \\
IRIS-C & 0.777 & 0.795 & 1.70 & 0.833 & 0.600 & 0.0 (0.0--8.4) & 30 / 42 \\
AgentProv-C & 0.982 & 0.817 & 5.38 & 0.952 & 0.000 & 7.1 (2.7--14.9) & 40 / 84 \\
R alone & 0.858 & 0.810 & 1.82 & 0.816 & 0.533 & 1.2 (0.0--6.5) & 90 / 84 \\
K alone & 0.801 & 0.826 & 1.92 & 0.838 & 0.657 & 1.2 (0.0--6.5) & 70 / 84 \\
In-harness only & 0.919 & 0.767 & 1.96 & 0.819 & 0.435 & 3.9 (2.0--6.9) & 200 / 280 \\
CertClaim-S (static) & 0.945 & 0.915 & 0.55 & 0.492 & 0.835 & \textcolor{red}{87.1} (82.6--90.8) & 200 / 280 \\
CertClaim ($K{=}4$) & 1.000 & 1.000 & 7.87 & 0.974 & 1.000 & 3.6 (0.7--10.1) & 30 / 84 \\
CertClaim & 1.000 & 1.000 & 9.23 & 0.974 & 1.000 & 3.6 (0.7--10.1) & 30 / 84 \\
\bottomrule
\end{tabular}
\end{table}

\begin{table}[h]
\centering
\caption{LeaFBench metrics and false-positive rate (\%, with 95\% Clopper--Pearson interval) on the setting Qwen3.5-27B, mini-swe-agent, real issues task suite. Positives: whole-model substitutes; negatives: honest services in every authorized deployment. Invalid FPR (lower end of the interval above $\alpha=5\%$) in red. In-harness only and CertClaim-S use every in-harness audit, whereas CertClaim and the ablation in Table~\ref{tab:ablation} use the in-harness audits that are paired with probe audits.}
\label{tab:app-leaf-qwen35-27b-swe}
\scriptsize
\setlength{\tabcolsep}{3.0pt}
\begin{tabular}{lccccccc}
\toprule
Method & AUC$\uparrow$ & pAUC$\uparrow$ & MD$\uparrow$ & ACC$\uparrow$ & TPR@1\%FPR$\uparrow$ & FPR (\%) & \#pos / \#neg \\
\midrule
MET & 0.591 & 0.510 & 0.39 & 0.579 & 0.000 & \textcolor{red}{22.2} (12.0--35.6) & 60 / 54 \\
LLMmap & 0.390 & 0.493 & 0.43 & 0.404 & 0.000 & \textcolor{red}{48.1} (34.3--62.2) & 60 / 54 \\
RUT & 0.541 & 0.487 & 0.80 & 0.701 & 0.000 & \textcolor{red}{70.4} (56.4--82.0) & 80 / 54 \\
KBF & 0.663 & 0.709 & 0.97 & 0.761 & 0.400 & \textcolor{red}{18.6} (14.7--23.0) & 80 / 360 \\
IRIS & 0.413 & 0.491 & 0.17 & 0.476 & 0.000 & \textcolor{red}{22.2} (12.0--35.6) & 70 / 54 \\
AgentProv & 0.803 & 0.522 & 1.38 & 0.691 & 0.000 & \textcolor{red}{50.9} (41.1--60.7) & 80 / 108 \\
LIDAR$^{\dagger}$ & 0.547 & 0.504 & 0.13 & 0.425 & 0.000 & 3.8 (0.5--13.2) & 75 / 52 \\
MET-C & 0.635 & 0.692 & 1.04 & 0.684 & 0.400 & 0.0 (0.0--6.6) & 60 / 54 \\
LLMmap-C & 0.957 & 0.949 & 2.53 & 0.821 & 0.900 & 0.0 (0.0--6.6) & 80 / 54 \\
KBF-C & 0.852 & 0.759 & 1.40 & 0.900 & 0.500 & 0.0 (0.0--1.0) & 80 / 360 \\
IRIS-C & 0.498 & 0.501 & 0.01 & 0.474 & 0.000 & 0.0 (0.0--6.6) & 60 / 54 \\
AgentProv-C & 1.000 & 0.998 & 7.91 & 0.989 & 0.988 & 0.9 (0.0--5.1) & 80 / 108 \\
R alone & 1.000 & 1.000 & 8.73 & 0.984 & 1.000 & 1.9 (0.8--4.0) & 70 / 360 \\
K alone & 0.866 & 0.781 & 1.54 & 0.902 & 0.545 & 0.0 (0.0--1.0) & 88 / 360 \\
In-harness only & 0.890 & 0.755 & 1.76 & 0.762 & 0.440 & 2.5 (1.1--4.7) & 400 / 360 \\
CertClaim-S (static) & 0.830 & 0.749 & 0.94 & 0.662 & 0.463 & \textcolor{red}{71.4} (66.4--76.0) & 400 / 360 \\
CertClaim ($K{=}4$) & 1.000 & 0.999 & 8.79 & 0.976 & 1.000 & 2.8 (1.3--5.0) & 62 / 360 \\
CertClaim & 1.000 & 1.000 & 9.79 & 0.983 & 1.000 & 1.9 (0.8--4.0) & 62 / 360 \\
\bottomrule
\end{tabular}
\end{table}

%% file: main.bbl
\begin{thebibliography}{66}
\providecommand{\natexlab}[1]{#1}
\providecommand{\url}[1]{\texttt{#1}}
\expandafter\ifx\csname urlstyle\endcsname\relax
  \providecommand{\doi}[1]{doi: #1}\else
  \providecommand{\doi}{doi: \begingroup \urlstyle{rm}\Url}\fi

\bibitem[{Alibaba Cloud}(2026)]{qwen35flash}
{Alibaba Cloud}.
\newblock {Qwen3.5-Flash: Official Model Documentation}, 2026.
\newblock URL \url{https://www.alibabacloud.com/help/en/model-studio/qwen3-5-flash}.
\newblock Accessed November 1, 2026.

\bibitem[{Anthropic}(2025{\natexlab{a}})]{anthropic_haiku45}
{Anthropic}.
\newblock {Introducing Claude Haiku 4.5}, 2025{\natexlab{a}}.
\newblock URL \url{https://www.anthropic.com/news/claude-haiku-4-5}.
\newblock Accessed November 1, 2026.

\bibitem[{Anthropic}(2025{\natexlab{b}})]{anthropic_sonnet45}
{Anthropic}.
\newblock {Introducing Claude Sonnet 4.5}, 2025{\natexlab{b}}.
\newblock URL \url{https://www.anthropic.com/news/claude-sonnet-4-5}.
\newblock Accessed November 1, 2026.

\bibitem[{Anthropic}(2026{\natexlab{a}})]{anthropic_opus5}
{Anthropic}.
\newblock {Introducing Claude Opus 5}, 2026{\natexlab{a}}.
\newblock URL \url{https://www.anthropic.com/news/claude-opus-5}.
\newblock Accessed November 1, 2026.

\bibitem[{Anthropic}(2026{\natexlab{b}})]{anthropic_opus55}
{Anthropic}.
\newblock {Introducing Claude Opus 5.5}, 2026{\natexlab{b}}.
\newblock URL \url{https://www.anthropic.com/claude-opus-5-5}.
\newblock Accessed November 1, 2026.

\bibitem[{Anthropic}(2026{\natexlab{c}})]{anthropic_sonnet46}
{Anthropic}.
\newblock {Introducing Claude Sonnet 4.6}, 2026{\natexlab{c}}.
\newblock URL \url{https://www.anthropic.com/news/claude-sonnet-4-6}.
\newblock Accessed November 1, 2026.

\bibitem[Arditi et~al.(2024)Arditi, Obeso, Syed, Paleka, Panickssery, Gurnee, and Nanda]{arditi2024refusal}
Arditi, A., Obeso, O., Syed, A., Paleka, D., Panickssery, N., Gurnee, W., and Nanda, N.
\newblock Refusal in language models is mediated by a single direction.
\newblock In \emph{Advances in Neural Information Processing Systems}, 2024.

\bibitem[Bates et~al.(2023)Bates, Cand{\`e}s, Lei, Romano, and Sesia]{bates2023outliers}
Bates, S., Cand{\`e}s, E., Lei, L., Romano, Y., and Sesia, M.
\newblock Testing for outliers with conformal p-values.
\newblock \emph{The Annals of Statistics}, 51\penalty0 (1):\penalty0 149--178, 2023.

\bibitem[Cai et~al.(2025)Cai, Shi, Zhao, et~al.]{cai2025substitution}
Cai, W., Shi, T., Zhao, X., et~al.
\newblock Are you getting what you pay for? auditing model substitution in {LLM} {API}s.
\newblock \emph{arXiv preprint arXiv:2504.04715}, 2025.

\bibitem[Cohen et~al.(2019)Cohen, Rosenfeld, and Kolter]{cohen2019smoothing}
Cohen, J., Rosenfeld, E., and Kolter, Z.
\newblock Certified adversarial robustness via randomized smoothing.
\newblock In \emph{International Conference on Machine Learning}, 2019.

\bibitem[{DeepSeek-AI}(2026{\natexlab{a}})]{deepseek2026aliasupdate}
{DeepSeek-AI}.
\newblock {Introducing DeepSeek-V4.1-Flash: Smarter, Faster, More Efficient}, 2026{\natexlab{a}}.
\newblock URL \url{https://www.deepseek.com/en/news/deepseek-v4-1-flash/}.
\newblock Accessed November 1, 2026.

\bibitem[{DeepSeek-AI}(2026{\natexlab{b}})]{deepseek_v4flash}
{DeepSeek-AI}.
\newblock {DeepSeek-V4-Flash: Official Model Card}, 2026{\natexlab{b}}.
\newblock URL \url{https://huggingface.co/deepseek-ai/DeepSeek-V4-Flash}.
\newblock Accessed November 1, 2026.

\bibitem[Fang et~al.(2026)Fang, Feng, Li, et~al.]{kbf2026}
Fang, Y., Feng, Y., Li, B., et~al.
\newblock {KBF}: Knowledge boundary as fingerprint for language model and black-box {API} auditing.
\newblock \emph{arXiv preprint arXiv:2605.29524}, 2026.

\bibitem[Gao et~al.(2025)Gao, Liang, and Guestrin]{gao2025met}
Gao, I., Liang, P., and Guestrin, C.
\newblock Model equality testing: Which model is this {API} serving?
\newblock In \emph{International Conference on Learning Representations}, 2025.
\newblock arXiv:2410.20247.

\bibitem[Gupta et~al.(2026)Gupta, Chaturvedi, Huan, et~al.]{gupta2026dissecting}
Gupta, G., Chaturvedi, V., Huan, J., et~al.
\newblock Dissecting model behavior through agent trajectories.
\newblock \emph{arXiv preprint arXiv:2606.17454}, 2026.

\bibitem[Hemerik \& Goeman(2018)Hemerik and Goeman]{hemerik2018random}
Hemerik, J. and Goeman, J.
\newblock Exact testing with random permutations.
\newblock \emph{TEST}, 27\penalty0 (4):\penalty0 811--825, 2018.

\bibitem[Jimenez et~al.(2024)Jimenez, Yang, Wettig, Yao, Pei, Press, and Narasimhan]{jimenez2024swebench}
Jimenez, C.~E., Yang, J., Wettig, A., Yao, S., Pei, K., Press, O., and Narasimhan, K.
\newblock {SWE}-bench: Can language models resolve real-world {GitHub} issues?
\newblock In \emph{International Conference on Learning Representations}, 2024.

\bibitem[Kwon et~al.(2023)Kwon, Li, Zhuang, Sheng, Zheng, Yu, Gonzalez, Zhang, and Stoica]{kwon2023vllm}
Kwon, W., Li, Z., Zhuang, S., Sheng, Y., Zheng, L., Yu, C.~H., Gonzalez, J.~E., Zhang, H., and Stoica, I.
\newblock Efficient memory management for large language model serving with {PagedAttention}.
\newblock In \emph{Proceedings of the 29th Symposium on Operating Systems Principles}, 2023.
\newblock URL \url{https://arxiv.org/abs/2309.06180}.

\bibitem[Lin et~al.(2026)Lin, Wan, Pei, et~al.]{gatescope2026}
Lin, G., Wan, Y., Pei, S., et~al.
\newblock Behavioral consistency and transparency analysis on large language model {API} gateways.
\newblock In \emph{ACM Internet Measurement Conference}, 2026.
\newblock arXiv:2604.21083.

\bibitem[Liu et~al.(2023)Liu, Xia, Wang, and Zhang]{liu2023evalplus}
Liu, J., Xia, C.~S., Wang, Y., and Zhang, L.
\newblock Is your code generated by {ChatGPT} really correct? rigorous evaluation of large language models for code generation.
\newblock In \emph{Advances in Neural Information Processing Systems}, 2023.

\bibitem[Liu(2026)]{liu2026sae}
Liu, Z.
\newblock Committed {SAE}-feature traces for audited-session substitution detection in hosted {LLM}s.
\newblock \emph{arXiv preprint arXiv:2604.18179}, 2026.

\bibitem[Lugoloobi et~al.(2026)Lugoloobi, Marro, Magomere, et~al.]{lugoloobi2026actions}
Lugoloobi, W., Marro, S., Magomere, J., et~al.
\newblock Known by their actions: Fingerprinting {LLM} browser agents via {UI} traces.
\newblock \emph{arXiv preprint arXiv:2605.14786}, 2026.

\bibitem[{MiniMax}(2026)]{minimax_m25}
{MiniMax}.
\newblock {MiniMax M2.5: Built for Real-World Productivity}, 2026.
\newblock URL \url{https://www.minimax.io/news/minimax-m25}.
\newblock Accessed November 1, 2026.

\bibitem[{Mistral AI}(2025{\natexlab{a}})]{devstral2505}
{Mistral AI}.
\newblock {Devstral-Small-2505: Official Model Card}, 2025{\natexlab{a}}.
\newblock URL \url{https://huggingface.co/mistralai/Devstral-Small-2505}.
\newblock Accessed November 1, 2026.

\bibitem[{Mistral AI}(2025{\natexlab{b}})]{devstral2507}
{Mistral AI}.
\newblock {Devstral-Small-2507: Official Model Card}, 2025{\natexlab{b}}.
\newblock URL \url{https://huggingface.co/mistralai/Devstral-Small-2507}.
\newblock Accessed November 1, 2026.

\bibitem[{Mistral AI}(2025{\natexlab{c}})]{mistralsmall2501}
{Mistral AI}.
\newblock {Mistral-Small-24B-Instruct-2501: Official Model Card}, 2025{\natexlab{c}}.
\newblock URL \url{https://huggingface.co/mistralai/Mistral-Small-24B-Instruct-2501}.
\newblock Accessed November 1, 2026.

\bibitem[{Moonshot AI}(2026)]{moonshot_kimi26}
{Moonshot AI}.
\newblock {Kimi K2.6: Advancing Open-Source Coding}, 2026.
\newblock URL \url{https://www.kimi.com/en/blog/kimi-k2-6}.
\newblock Accessed November 1, 2026.

\bibitem[Oderinwale(2026)]{procgrep2026}
Oderinwale, H.
\newblock Agent trajectories as programs: Fingerprinting and programming coding-agent behavior.
\newblock \emph{arXiv preprint arXiv:2606.16988}, 2026.

\bibitem[Ong et~al.(2025)Ong, Di~Ferrante, Pazdera, et~al.]{toploc2025}
Ong, J.~M., Di~Ferrante, M., Pazdera, A., et~al.
\newblock {TOPLOC}: A locality sensitive hashing scheme for trustless verifiable inference.
\newblock \emph{arXiv preprint arXiv:2501.16007}, 2025.

\bibitem[{OpenAI}(2025{\natexlab{a}})]{openai2025gptoss}
{OpenAI}.
\newblock {gpt-oss-120b \& gpt-oss-20b Model Card}, 2025{\natexlab{a}}.
\newblock URL \url{https://arxiv.org/abs/2508.10925}.
\newblock Accessed November 1, 2026.

\bibitem[{OpenAI}(2025{\natexlab{b}})]{openai_gpt52}
{OpenAI}.
\newblock {GPT-5.2: Official Model Documentation}, 2025{\natexlab{b}}.
\newblock URL \url{https://developers.openai.com/api/docs/models/gpt-5.2}.
\newblock Accessed November 1, 2026.

\bibitem[{OpenAI}(2026{\natexlab{a}})]{openai_gpt54}
{OpenAI}.
\newblock {GPT-5.4: Official Model Documentation}, 2026{\natexlab{a}}.
\newblock URL \url{https://developers.openai.com/api/docs/models/gpt-5.4}.
\newblock Accessed November 1, 2026.

\bibitem[{OpenAI}(2026{\natexlab{b}})]{openai_gpt54mini}
{OpenAI}.
\newblock {GPT-5.4 Mini: Official Model Documentation}, 2026{\natexlab{b}}.
\newblock URL \url{https://developers.openai.com/api/docs/models/gpt-5.4-mini}.
\newblock Accessed November 1, 2026.

\bibitem[{OpenAI}(2026{\natexlab{c}})]{openai_gpt54nano}
{OpenAI}.
\newblock {GPT-5.4 nano: Official Model Documentation}, 2026{\natexlab{c}}.
\newblock URL \url{https://developers.openai.com/api/docs/models/gpt-5.4-nano}.
\newblock Accessed November 1, 2026.

\bibitem[{OpenAI}(2026{\natexlab{d}})]{openai_gpt55}
{OpenAI}.
\newblock {GPT-5.5: Official Model Documentation}, 2026{\natexlab{d}}.
\newblock URL \url{https://developers.openai.com/api/docs/models/gpt-5.5}.
\newblock Accessed November 1, 2026.

\bibitem[{OpenAI}(2026{\natexlab{e}})]{openai_gpt56sol}
{OpenAI}.
\newblock {GPT-5.6 Sol: Official Model Documentation}, 2026{\natexlab{e}}.
\newblock URL \url{https://developers.openai.com/api/docs/models/gpt-5.6-sol}.
\newblock Accessed November 1, 2026.

\bibitem[{OpenAI}(2026{\natexlab{f}})]{openai_gpt56terra}
{OpenAI}.
\newblock {GPT-5.6 Terra: Official Model Documentation}, 2026{\natexlab{f}}.
\newblock URL \url{https://developers.openai.com/api/docs/models/gpt-5.6-terra}.
\newblock Accessed November 1, 2026.

\bibitem[{OpenAI}(2026{\natexlab{g}})]{openai_gpt61sol}
{OpenAI}.
\newblock {GPT-6.1 Sol: Official Model Documentation}, 2026{\natexlab{g}}.
\newblock URL \url{https://developers.openai.com/api/docs/models/gpt-6.1-sol}.
\newblock Accessed November 1, 2026.

\bibitem[{OpenAI}(2026{\natexlab{h}})]{openai_gpt6sol}
{OpenAI}.
\newblock {GPT-6 Sol: Official Model Documentation}, 2026{\natexlab{h}}.
\newblock URL \url{https://developers.openai.com/api/docs/models/gpt-6-sol}.
\newblock Accessed November 1, 2026.

\bibitem[{OpenCode Team}(2026)]{opencode2026}
{OpenCode Team}.
\newblock {OpenCode}, 2026.
\newblock URL \url{https://opencode.ai/docs/}.
\newblock Software documentation. Accessed November 1, 2026.

\bibitem[Pasquini et~al.(2025)Pasquini, Kornaropoulos, and Ateniese]{pasquini2025llmmap}
Pasquini, D., Kornaropoulos, E.~M., and Ateniese, G.
\newblock {LLMmap}: Fingerprinting for large language models.
\newblock In \emph{USENIX Security Symposium}, 2025.

\bibitem[Qiao et~al.(2025)Qiao, Li, Li, Wang, Qi, Guo, Feng, and Tao]{qiao2025certdw}
Qiao, T., Li, Y., Li, J., Wang, Y., Qi, L., Guo, J., Feng, R., and Tao, D.
\newblock {CertDW}: Towards certified dataset ownership verification via conformal prediction.
\newblock \emph{arXiv preprint arXiv:2506.13160}, 2025.

\bibitem[{Qwen Team}(2025{\natexlab{a}})]{qwen3_30b_2507}
{Qwen Team}.
\newblock {Qwen3-30B-A3B-Instruct-2507: Official Model Card}, 2025{\natexlab{a}}.
\newblock URL \url{https://huggingface.co/Qwen/Qwen3-30B-A3B-Instruct-2507}.
\newblock Accessed November 1, 2026.

\bibitem[{Qwen Team}(2025{\natexlab{b}})]{qwen3_32b}
{Qwen Team}.
\newblock {Qwen3-32B: Official Model Card}, 2025{\natexlab{b}}.
\newblock URL \url{https://huggingface.co/Qwen/Qwen3-32B}.
\newblock Accessed November 1, 2026.

\bibitem[{Qwen Team}(2025{\natexlab{c}})]{qwen3coder30b}
{Qwen Team}.
\newblock {Qwen3-Coder-30B-A3B-Instruct: Official Model Card}, 2025{\natexlab{c}}.
\newblock URL \url{https://huggingface.co/Qwen/Qwen3-Coder-30B-A3B-Instruct}.
\newblock Accessed November 1, 2026.

\bibitem[{Qwen Team}(2025{\natexlab{d}})]{qwen3coder30b_fp8}
{Qwen Team}.
\newblock {Qwen3-Coder-30B-A3B-Instruct-FP8: Official Model Card}, 2025{\natexlab{d}}.
\newblock URL \url{https://huggingface.co/Qwen/Qwen3-Coder-30B-A3B-Instruct-FP8}.
\newblock Accessed November 1, 2026.

\bibitem[{Qwen Team}(2025{\natexlab{e}})]{qwen3technicalreport}
{Qwen Team}.
\newblock {Qwen3 Technical Report}, 2025{\natexlab{e}}.
\newblock URL \url{https://arxiv.org/abs/2505.09388}.

\bibitem[{Qwen Team}(2026{\natexlab{a}})]{qwen35_27b}
{Qwen Team}.
\newblock {Qwen3.5-27B: Official Model Card}, 2026{\natexlab{a}}.
\newblock URL \url{https://huggingface.co/Qwen/Qwen3.5-27B}.
\newblock Accessed November 1, 2026.

\bibitem[{Qwen Team}(2026{\natexlab{b}})]{qwen35_27b_fp8}
{Qwen Team}.
\newblock {Qwen3.5-27B-FP8: Official Model Card}, 2026{\natexlab{b}}.
\newblock URL \url{https://huggingface.co/Qwen/Qwen3.5-27B-FP8}.
\newblock Accessed November 1, 2026.

\bibitem[{Qwen Team}(2026{\natexlab{c}})]{qwen35_27b_gptq}
{Qwen Team}.
\newblock {Qwen3.5-27B-GPTQ-Int4: Official Model Card}, 2026{\natexlab{c}}.
\newblock URL \url{https://huggingface.co/Qwen/Qwen3.5-27B-GPTQ-Int4}.
\newblock Accessed November 1, 2026.

\bibitem[{Qwen Team}(2026{\natexlab{d}})]{qwen35_4b}
{Qwen Team}.
\newblock {Qwen3.5-4B: Official Model Card}, 2026{\natexlab{d}}.
\newblock URL \url{https://huggingface.co/Qwen/Qwen3.5-4B}.
\newblock Accessed November 1, 2026.

\bibitem[{Qwen Team}(2026{\natexlab{e}})]{qwen35_9b}
{Qwen Team}.
\newblock {Qwen3.5-9B: Official Model Card}, 2026{\natexlab{e}}.
\newblock URL \url{https://huggingface.co/Qwen/Qwen3.5-9B}.
\newblock Accessed November 1, 2026.

\bibitem[{Qwen Team}(2026{\natexlab{f}})]{qwen36_27b}
{Qwen Team}.
\newblock {Qwen3.6-27B: Official Model Card}, 2026{\natexlab{f}}.
\newblock URL \url{https://huggingface.co/Qwen/Qwen3.6-27B}.
\newblock Accessed November 1, 2026.

\bibitem[Shao et~al.(2025)Shao, Li, He, Yao, Yang, Tao, and Qin]{shao2025sok}
Shao, S., Li, Y., He, Y., Yao, H., Yang, W., Tao, D., and Qin, Z.
\newblock {SoK}: Large language model copyright auditing via fingerprinting.
\newblock \emph{arXiv preprint arXiv:2508.19843}, 2025.

\bibitem[{SWE-agent Team}(2025)]{minisweagent2025}
{SWE-agent Team}.
\newblock {mini-SWE-agent}, 2025.
\newblock URL \url{https://mini-swe-agent.com/latest/}.
\newblock Software documentation. Accessed November 1, 2026.

\bibitem[Vovk et~al.(2005)Vovk, Gammerman, and Shafer]{vovk2005alrw}
Vovk, V., Gammerman, A., and Shafer, G.
\newblock \emph{Algorithmic Learning in a Random World}.
\newblock Springer, 2005.

\bibitem[Wang et~al.(2026{\natexlab{a}})Wang, Xie, Wang, et~al.]{lidar2026}
Wang, C., Xie, X., Wang, T., et~al.
\newblock Who is behind the harness? fingerprinting {LLM}s through agentic behavior.
\newblock \emph{arXiv preprint arXiv:2609.28559}, 2026{\natexlab{a}}.

\bibitem[Wang et~al.(2026{\natexlab{b}})Wang, Zhao, Backes, et~al.]{agentprov2026}
Wang, X., Zhao, B., Backes, M., et~al.
\newblock {AgentProv}: Auditing agentic {LLM} {API} providers via tool-use policy probes.
\newblock \emph{arXiv preprint arXiv:2609.00052}, 2026{\natexlab{b}}.

\bibitem[White et~al.(2026)White, Jafari, and Berg-Kirkpatrick]{white2026forensics}
White, I., Jafari, Y., and Berg-Kirkpatrick, T.
\newblock Black-box forensics for conversational {LLM} agents.
\newblock \emph{arXiv preprint arXiv:2606.22698}, 2026.

\bibitem[Wu(2026)]{wu2026responsible}
Wu, C.
\newblock Responsible {AI} governance for {BNPL} consumer fintech lending: A risk-based framework.
\newblock In \emph{Proceedings of the 20th International Management Conference}, October 2026.
\newblock URL \url{https://conference.management.ase.ro/wp-content/uploads/2026/10/IMC2026_Conference_Programme_26-10-2026_v4.pdf}.
\newblock Listed in the official conference programme.

\bibitem[Xie et~al.(2026)Xie, Wu, Lu, et~al.]{xie2026tee}
Xie, S., Wu, Q., Lu, H., et~al.
\newblock The proxy knows too much: Sealing {LLM} {API} routers with attested {TEE}s.
\newblock \emph{arXiv preprint arXiv:2606.16358}, 2026.

\bibitem[Yang et~al.(2024)Yang, Jimenez, Wettig, Lieret, Yao, Narasimhan, and Press]{yang2024sweagent}
Yang, J., Jimenez, C.~E., Wettig, A., Lieret, K., Yao, S., Narasimhan, K., and Press, O.
\newblock {SWE-agent}: Agent-computer interfaces enable automated software engineering.
\newblock In \emph{Advances in Neural Information Processing Systems}, 2024.

\bibitem[{Z.ai}(2026)]{zai_glm52}
{Z.ai}.
\newblock {GLM-5.2: Built for Long-Horizon Tasks}, 2026.
\newblock URL \url{https://z.ai/blog/glm-5.2}.
\newblock Accessed November 1, 2026.

\bibitem[Zhang et~al.(2026)Zhang, Zhang, Qin, et~al.]{iris2026}
Zhang, Y., Zhang, Z.-H., Qin, H., et~al.
\newblock Which model is actually serving you? {IRIS}: Budgeted black-box auditing of model substitution and routing dilution.
\newblock \emph{arXiv preprint arXiv:2607.20860}, 2026.

\bibitem[Zhao et~al.(2024)Zhao, Ren, Hessel, Cardie, Choi, and Deng]{zhao2024wildchat}
Zhao, W., Ren, X., Hessel, J., Cardie, C., Choi, Y., and Deng, Y.
\newblock {WildChat}: 1m {ChatGPT} interaction logs in the wild.
\newblock In \emph{International Conference on Learning Representations}, 2024.
\newblock URL \url{https://arxiv.org/abs/2405.01470}.

\bibitem[Zhu et~al.(2026)Zhu, Ye, Qiu, et~al.]{rut2026}
Zhu, X., Ye, Y., Qiu, T., et~al.
\newblock Auditing black-box {LLM} {API}s with a rank-based uniformity test.
\newblock In \emph{International Conference on Learning Representations}, 2026.
\newblock arXiv:2506.06975.

\end{thebibliography}
